\documentclass[twocolumn,aps,prd,amsmath,amssymb,preprintnumbers,nofootinbib]{revtex4-1}

\usepackage{graphicx,caption}
\usepackage{tensor}
\usepackage{xcolor}

\usepackage{multirow}

\begin{document}

\setcounter{footnote}{0}

\title{Emergent electromagnetism}
\author{Erick I.\ Duque}
\email{eqd5272@psu.edu}
\affiliation{Institute for Gravitation and the Cosmos,
The Pennsylvania State University, 104 Davey Lab, University Park,
PA 16802, USA}

\begin{abstract}
We introduce the concept of \emph{emergent electric field}. This is distinguished from the fundamental one in that the emergent electric field directly appears in observations through the Lorentz force, while the latter enters the phase space as the canonical momentum of the electromagnetic field.
In Hamiltonian classical electromagnetism this concept naturally appears after introducing the topological $\theta$ term.
Furthermore, we show that in the spherically symmetric model the concept of emergent electric field allows us to formulate a modified theory of electromagnetism that is otherwise impossible.
The relation between the fundamental and the emergent electric fields is derived from the imposition of general covariance of the electromagnetic strength tensor, which is a nontrivial task in the canonical formulation the modified theory is based on.
We couple this theory to emergent modified gravity, where a similar distinction between spacetime and gravity is made such that the spacetime, which defines the observable geometry, is an emergent field composed of the fundamental gravitational field.
In this more encompassing emergent field theory coupling gravity and electromagnetism, we show that the spherically symmetric model contains a nonsingular black hole solution where not only modified gravity but also modified electromagnetism is crucial for a robust singularity resolution which, as a consequence, excludes the existence of (super)extremal black holes.
\end{abstract}

\maketitle

\section{Introduction}

The simplest and yet profound example discerning between the canonical and mechanical momenta is given by the nonrelativistic, single, charged particle in an external electromagnetic field.
This Lagrangian is given by
\begin{equation}
    L = \frac{m}{2} \dot{\vec{q}}\,^2 - e \phi + e \vec{A} \cdot \dot{\vec{q}}
    \,,
\end{equation}
where $\vec{q}$ denotes the particle's position, $m$ its mass, $e$ its electric charge, $\phi$ the electric potential, $\vec{A}$ the electromagnetic vector potential, and the dot denotes a derivative with respect to time.
The mechanical momentum is defined by
\begin{equation}\label{eq:Mechanical momentum - particle}
    \vec{p} = m \dot{\vec{q}}
    \,.
\end{equation}
On the other hand, the procedure of the Legendre transformation defines the canonical momentum
\begin{equation}\label{eq:Canonical momentum - particle}
    \vec{P} = \frac{\partial L}{\partial \dot{\vec q}} = m \dot{\vec q} + e \vec{A}
    \,,
\end{equation}
and the Hamiltonian
\begin{equation}\label{eq:Hamiltonian - particle}
    H = \vec{P} \cdot \dot{\vec{q}} - L = \frac{|\vec{P} - e \vec{A}|^2}{2 m} + e \phi
    \,.
\end{equation}
The mechanical momentum (\ref{eq:Mechanical momentum - particle}) differs from the canonical one (\ref{eq:Canonical momentum - particle}) in two important ways.
First, it is the canonical momentum that is conserved under time evolution, but it is the mechanical one which we associate with observations, and hence why magnetic fields can deviate the trajectory of particles without doing work.
The second important difference is that it is the canonical momentum that is directly used for quantization in promoting the phase space variables to operators and the Poisson brackets to commutators,
\begin{equation}
    \{ x^i , P_j \} = \delta^i_j \rightarrow [\hat{x}^i , \hat{P}_j] = i \hbar \delta^i_j
    \,,
\end{equation}
and hence the quantized spectrum is associated to the canonical momentum and not to the mechanical one.
This distinction between the momenta is crucial to explain for instance Landau levels, a phenomenon that describes the discrete energy levels of the electron cyclotron motion in a magnetic field \cite{Landau}.

This distinction between canonical and mechanical momenta is even more general.
Here, it is our purpose to extend this idea to the electromagnetic field itself.
In fact, an immediate example follows when including the topological $\theta$ term to the Maxwell action.
In flat spacetime, the total action in the four-dimensional region ${\cal R}$ reads
\begin{equation}\label{eq:Maxwell action - Minkowski}
    S_{\rm EM}^{\cal R} [A]
    = - \frac{1}{4} \int_{\cal R} {\rm d} x^4\ F_{\mu\nu} F^{\mu\nu}
    + S_\theta^{\cal R} [A]
    \,,
\end{equation}
where the $\theta$ term contribution is given by
\begin{equation}\label{eq:Theta term - Minkowski}
    S_\theta^{\cal R} [A]
    = \int_{\cal R} {\rm d} x^4\ \frac{\theta}{8} \epsilon^{\mu\nu\alpha\beta} F_{\mu\nu} F_{\alpha\beta}
    \,,
\end{equation}
with constant $\theta$.
The canonical analysis of the electromagnetic action shows that the component $A_t$ is nondynamical and hence a Lagrange multiplier related to the underlying U(1) symmetry of the action, while the canonical momentum of the electromagnetic field is given by
\begin{equation}\label{eq:Electric momentum - Minkowski}
    E^a (x) \equiv \frac{\partial \mathcal{L}_{\rm EM}}{ \partial \dot{A}_a (x)}
    = \tilde{E}^a - \theta B^a
    \,,
\end{equation}
where Latin indices indicate spatial components in contrast to the Greek indices for spacetime components.
The expressions in the right-hand-side of (\ref{eq:Electric momentum - Minkowski}) are the usual electric and magnetic fields directly related to the strength tensor components by
\begin{equation}
    \tilde{E}^a = \delta^{a b} F_{t b}
    \quad , \quad B^a = \frac{1}{2} \epsilon^{a b c} F_{b c}
    \,.
\end{equation}
It is the electric field $\tilde{E}^a$ that we directly observe in experiments because it is the strength tensor that couples to the charged particle's equation of motion through the Lorentz force.
This expression, however, differs from the canonical electric momentum (\ref{eq:Electric momentum - Minkowski}) by a contribution from the magnetic field for nonzero $\theta$.
It is the canonical momentum $E^a$ that must be used as a phase space variable and be promoted to a fundamental operator upon quantization.
However, most classical experiments are unable to distinguish between the two because the $\theta$ term contribution (\ref{eq:Theta term - Minkowski}) can be written as a total derivative and be integrated out to give the surface term
\begin{equation}\label{eq:Theta term - Minkowski - surface}
    S_\theta^{\partial {\cal R}} [A]
    = \int_{\partial {\cal R}} {\rm d} x^3\ \frac{\theta}{2} \hat{r}_\mu \epsilon^{\mu\nu\alpha\beta} A_\nu \partial_\alpha A_\beta
    \,,
\end{equation}
where $\hat{r}_\mu$ is the unit (co)normal to the three dimensional hypersurface $\partial {\cal R}$, which is the boundary of ${\cal R}$.
As a boundary term, the $\theta$ contribution to the action does not affect the bulk dynamics of the electromagnetic field and, therefore, it is difficult to distinguish $\tilde{E}^a$ from $E^a$ in experiments.
On the other hand, as a contribution to the action functional of electromagnetism, its variation with respect to $A_\nu$ generates a contribution to the electric current $J^\nu=\delta S [A]/\delta A^\nu$.
While the $\theta$ term is not associated to any bulk current, the variation of (\ref{eq:Theta term - Minkowski - surface}) with respect to the boundary variable $A_\nu |_{\partial {\cal R}}$ does generate the surface current contribution
\begin{equation}\label{eq:Theta current - Minkowski}
    J_{\theta,\partial {\cal R}}^\nu
    = \frac{\theta}{2} \hat{r}_\mu \epsilon^{\mu\nu\alpha\beta} F_{\alpha\beta}
    \,.
\end{equation}
It is therefore possible for the $\theta$ term to have boundary or topological effects.
In fact, picking the Cartesian coordinates $t,x,y,z$, if the hypersurface $\partial {\cal R}$ is spacelike such that $\hat{r}^\mu = \hat{z}^\mu = (0,0,0,1)$ and $\nu=x$, we obtain the current
\begin{equation}\label{eq:Theta current - Minkowski - simp}
    J^x_{\theta,\partial {\cal R}} = \theta \tilde{E}^y
    \,.
\end{equation}
This can be related to the surface (half-quantized) anomalous Hall effect in the context of topological (magnetic) insulators with a boundary \cite{Simon,AxionEM}.
This example shows that there are indeed some experimental observations with the potential to discriminate between the two electric fields, being $\tilde{E}^a$ the one appearing in the surface current (\ref{eq:Theta current - Minkowski - simp}), which also directly depends on the $\theta$ parameter.

Furthermore, we will be interested in the more general system of electromagnetism in curved spaces, i.e, coupled to the gravitational field.
Recently, a similar distinction between the fundamental variables composing the phase space and the fields concerning observations was explicitly applied to canonical gravity in \cite{EMG,EMGCov}.
In those works, the gravity was considered the fundamental field defining the phase space, while the spacetime is derived as a nontrivial function of the phase space and hence called emergent.
In this theory, named emergent modified gravity (EMG), the gravitational field and the emergent spacetime play the respective roles of canonical and mechanical momenta of the charged particle system above.
Applications of EMG in spherical symmetry have led to a nonsingular black hole solution in vacuum \cite{alonso2022nonsingular,Alonso_Bardaji_2022,ELBH}, the formation of wormholes or black-to-white-hole transitions as a result of gravitational collapse \cite{EMGPF,alonsobardaji2024nonsingular}, and a relativistic generalization of modified Newtonian dynamics (MOND) \cite{milgrom1983modification,banik2022galactic,MONDEMG}.
It has also been possible to couple scalar matter to EMG \cite{alonso2021anomaly,alonsobardaji2024spacetime,EMGscalar} and to apply the theory to Gowdy systems with a local, propagating gravitational degree of freedom \cite{EMGGowdy}.
One particular modification variable $\lambda$ in the spherically symmetric model of EMG is responsible for the singularity resolution of the vacuum black hole solution.
This parameter enters as a 'frequency' in trigonometric functions of the angular curvature in the Hamiltonian.
While this modification belongs to a general result of spherical EMG, these trigonometric functions have a natural interpretation as holonomy terms of the gravitational connection used for instance in loop quantum gravity (LQG) \cite{rovelli2004quantum,thiemann2008modern}, which are in turn based on Wilson loops, extensively used in nonperturbative lattice quantum field theory \cite{Wilson}.
EMG can then be used as an effective theory for LQG black holes, keeping in mind that the two theories are, in principle, independent from each other.

The emergence of the spacetime metric as a function of the fundamental gravitational field is derived from a subtle feature of canonical gravity.
Canonical gravity is a constrained gauge field theory where the gauge transformations, which are generated by the Hamiltonian constraint and the vector constraint, are equivalent to spacetime coordinate transformations only on shell, that is, when the constraints vanish.
The spacetime is foliated by space-like hypersurfaces, and the field content is given by a set of spatial tensors on these hypersurfaces with flow equations generated by the same constraints, determining the evolution of the fields between adjacent hypersurfaces.
These evolving spatial tensors can then reproduce the usual spacetime tensors of GR.
The action of the gauge generators can also be understood geometrically: the Hamiltonian constraint $H[N]$, smeared by a scalar lapse function $N$, generates a normal, infinitesimal hypersurface deformation with length $N$, while the vector constraint $\Vec{H}[\Vec{N}]$, smeared by the spatial shift vector $\Vec{N}$, generates a tangential hypersurface deformation with length $\Vec{N}$.

In ADM notation, the spacetime metric or line element is given by \cite{ADM,arnowitt2008republication}
\begin{equation}
  {\rm d} s^2 = - N^2 {\rm d} t^2 + q_{a b} ( {\rm d} x^a + N^a {\rm d} t )
  ( {\rm d} x^b + N^b {\rm d} t )
  \,,
  \label{eq:ADM line element}
\end{equation}
where $q_{ab}$ is the spatial metric of the hypersurfaces, and the functions $N$ and $N^a$ are the lapse and shift defining the observer's frame.
The vector and Hamiltonian constraints have the Poisson brackets
\begin{eqnarray}
    \{ \Vec{H} [ \Vec{N}_1] , \Vec{H} [ \Vec{N}_2 ] \} &=& - \vec{H} [\mathcal{L}_{\Vec{N}_2} \Vec{N}_1]
    \,,
    \label{eq:Hypersurface deformation algebra - HaHa}
    \\
    \{ H [ N ] , \Vec{H} [ \Vec{N}]\} &=& - H [ N^b \partial_b N ]
    \label{eq:Hypersurface deformation algebra - HHa}
    \,, \\
    \{ H [ N_1 ] , H [ N_2 ] \} &=& - \vec{H} [ q^{a b} ( N_2 \partial_b N_1 - N_1 \partial_b N_2 )]
    \,,\quad
    \label{eq:Hypersurface deformation algebra - HH}
\end{eqnarray}
depending not only on $\Vec{N}$ and $N$, but also on the inverse of the spatial metric $q^{a b}$ on a spatial hypersurface and is hence a structure function, rather than a typical structure constant of Lie algebras.
The early results of \cite{hojman1976geometrodynamics}, where the vacuum was considered with the spatial metric being the only configuration variable, state that the Hamiltonian constraint, at second order in spatial derivatives, is uniquely determined by the hypersurface deformation algebra (\ref{eq:Hypersurface deformation algebra - HaHa})-(\ref{eq:Hypersurface deformation algebra - HH}), and given by that of GR up to the choice of Newton's and the cosmological constants \cite{hojman1976geometrodynamics,kuchar1974geometrodynamics}, implying that no modifications are allowed.
A crucial ingredient to this conclusion lies in the common, and often only implicit, assumption that the spatial metric $q_{a b}$ is a configuration variable.
Physically, this is the assumption that the spacetime metric is gravity itself, that it is a fundamental field.
This assumption, however, is not necessary to obtain a field theory describing spacetime.
EMG postulates a set of fundamental fields composing the phase-space, and considers non-classical constraints (that is, different from those of GR) that still respect the form of the hypersurface deformation algebra (\ref{eq:Hypersurface deformation algebra - HaHa})-(\ref{eq:Hypersurface deformation algebra - HH}) up to a structure function $\tilde{q}^{a b}$ potentially differing from the classical one.
This new structure function, composed by the fundamental variables of the phase space but not identical to any one of them nor related by a simple canonical transformation, is interpreted as the inverse of the spatial metric.
This is an emergent spatial metric that, when embedded into a 4-dimensional manifold, gives rise to an emergent spacetime that is not gravity itself, but made of gravity.
Mathematically, this is a difference between the canonical variables and the mechanical or observed ones.
Adopting the terminology used in EMG, the canonical variables may be referred to as fundamental, while the mechanical variables directly related to observations and whose dependence on the canonical ones is derived by some regaining process may be referred to as emergent.

In the same spirit of EMG, here we will show that the imposition of covariance on the electromagnetic field similarly defines an electric field associated to the strength tensor, derived from the Hamiltonian, and which is generally different from the canonical electric momentum, and hence we refer to it as the \emph{emergent electric field}.
Similar to EMG, we expect that the introduction of this new concept will enable us to formulate new covariant theories of modified electromagnetism.
In particular, since spherical EMG contains modification terms that can be interpreted as gravitational holonomies, we are interested in obtaining similar terms that could be interpreted as the electromagnetic holonomies that appear in the Wilson action of lattice field theory in a covariant way.
The model of spherical EMG coupled to emergent electromagnetism presented here does have such a term.

It is important to note that EMG and its extension to emergent electromagnetism do not introduce the characteristic ingredients of the quantum theory\textemdash\,though some modifications may be motivated by it\textemdash\,, and are, therefore, strictly non-quantum; on the other hand, the theory is non-classical in the sense that it is not equivalent to GR or Maxwell's electrodynamics. In the following, we use the term \emph{classical} to refer to the latter theories. In particular, here we define the classical limit of EMG and emergent electromagnetism as the recovery of GR and Maxwell electrodynamics.

The organization of this work is as follows.
In Section~\ref{sec:Emergent modified gravity} we review the full theory of EMG emphasizing the covariance conditions.
We then focus on the spherically symmetric model in Section~\ref{sec:Spherical EMG Vacuum}.
In Section~\ref{sec:Symm red vs dim red} we discuss the larger freedom enjoyed by symmetry reduced models compared to the full four dimensional ones and how that can impact the allowed modifications of both gravity and electromagnetism.
In Section~\ref{sec:Classical electromagnetism} we review the canonical formulation of classical electromagnetism on a curved space, we include the $\theta$ term to understand how the emergent electric field is treated in the theory and how the symmetries of the system are affected by it, and we discuss how to formulate the covariance conditions of the electromagnetic field canonically.
In Section~\ref{sec: Emergent electromagnetism} we impose these electromagnetic covariance conditions in the full four dimensional theory starting from a general ansatz for a modified electromagnetic Hamiltonian, we find that these covariance conditions define how the emergent electric is derived from the Hamiltonian itself, and conclude that there is not much freedom for modifications beyond the $\theta$ term and the allowance for the structure function, and hence the spacetime metric, to depend on the electromagnetic phase space variables.
In Section~\ref{sec:Emergent electromagnetism: Spherical symmetry} we reformulate the electromagentic covariance conditions in the spherically symmetric model, which behaves as a $1+1$ dimensional theory, and find that it has a larger freedom for modifications that are possible only by the introduction of the concepts of emergent spacetime and electric fields.
We use this modification freedom in Section~\ref{eq:BH sol} to reproduce a simpler version of the Wilson action modification, which, together with a modified gravity term, leads to a nonsingular, electrically charged black hole solution.
Finally, we summarize the conclusions in Section~\ref{sec:Conclusions}, where we revisit the discussion of symmetry reduced models having a larger modification freedom than their full four dimensional counterparts with the retrospective insight of the obtained results; in particular, we point out that our results hint towards a more general emergent field theory where nonlocal modifications may be possible while preserving a generalized notion of covariance for both the spacetime and the electromagnetic force.

\section{Emergent modified gravity}
\label{sec:Emergent modified gravity}

We assume that the spacetime region of interest is globally hyperbolic, $M = \Sigma \times \mathbb{R}$, with a 3-dimensional spatial manifold $\Sigma$.
We parametrize the embeddings of $\Sigma$ by foliating $M$ into smooth families of spacelike hypersurfaces $\Sigma_t$, $t\in{\mathbb R}$.
We can then embed $\Sigma$ in $M$ as a constant-time hypersurface for any fixed $t_0$, $\Sigma\cong \Sigma_{t_0}\cong (\Sigma_{t_0},t_0)\hookrightarrow M$.

Given a foliation into spacelike hypersurfaces, the metric
$g_{\mu\nu}$ on $M$ defines the vector field $n^{\mu}$ unit normal to $\Sigma_{t_0}$, and induces the spatial metric $q_{ab}(t_0)$ on
$\Sigma_{t_0}$ by restricting the spacetime tensor $q_{\mu\nu}=g_{\mu\nu}+n_{\mu}n_{\nu}$ to $\Sigma_{t_0}$, such that $q_{\mu\nu}n^{\nu}=0$.
A family of spatial metrics $q_{ab}(t)$ is then reproduced by time-evolution between adjacent hypersurfaces, which requires the introduction of an observer's frame via a time-evolution vector field
\begin{equation}
  t^\mu = N n^\mu + N^a s_a^\mu\,,
  \label{eq:Time-evolution vector field}
\end{equation}
where $N$ is the lapse function and $N^a$ shift vector field
\cite{arnowitt2008republication}.
Here, $s_a^\mu$ are the basis vectors on the spatial hypersurfaces, $s_a^\mu(t_0): T\Sigma_{t_0}\to TM$, such that $g_{\mu \nu} n^\mu s^\mu_a = 0$.
The resulting spacetime line element is given by (\ref{eq:ADM line element}).

The Hamiltonian constraint $H$ and the vector constraint $H_a$ generate both time-evolution and gauge transformations via Poisson brackets by use of different smearing functions.
We will denote $H$ and $H_a$ as the local versions of the constraints, while we denote the smeared versions with square brackets enclosing the smearing function: $H[N]=\int {\rm d} x^3 H(x) N(x)$.
The infinitesimal gauge transformation of a phase-space function $\mathcal{O}$ is given by $\delta_\epsilon \mathcal{O} = \{ \mathcal{O} , H[\epsilon^0 , \epsilon^a]
\}$, where $H[\epsilon^0 , \epsilon^a] = H[\epsilon^0] + H_a [\epsilon^a]$, and $\epsilon^0$ and $\epsilon^a$ are the gauge parameters, while infinitesimal time-evolution is given by $\Dot{\mathcal{O}} = \delta_t \mathcal{O} = \{ \mathcal{O} , H[N , N^a] \}$, where lapse and shift play the role of gauge parameters.
As gauge generator, $H[\epsilon^0 , \epsilon^a]$ must vanish for all $\epsilon^0$ and $\epsilon^a$; therefore $H=0$ and $H_a=0$ on dynamical solutions and we say the system is "on shell".
The constraints must vanish consistently in all gauges, and hence the Poisson brackets of the constraints with themselves must vanish on shell.
This is indeed the case because the constraint algebra (\ref{eq:Hypersurface deformation algebra - HaHa})-(\ref{eq:Hypersurface deformation algebra - HH}) is first class and the right-hand-side vanishes on shell.

The Poisson brackets do not provide the gauge transformations of $N$ and $N^a$ because they do not have momenta:
They do not evolve dynamically because they specify the frame for time evolution.
The gauge transformations of the lapse and shift are instead derived from the condition that the equations of motion are gauge covariant, yielding \cite{pons1997gauge,salisbury1983realization,bojowald2018effective}
\begin{eqnarray}
    \delta_\epsilon N &=& \dot{\epsilon}^0 + \epsilon^a \partial_a N - N^a \partial_a \epsilon^0
    \,,
    \label{eq:Off-shell gauge transformations for lapse - intro}
    \\
    \delta_\epsilon N^a &=& \dot{\epsilon}^a + \epsilon^b \partial_b N^a - N^b \partial_b \epsilon^a
    \nonumber\\
    &&
    + q^{a b} \left(\epsilon^0 \partial_b N - N \partial_b \epsilon^0 \right)
    \,.
    \label{eq:Off-shell gauge transformations for shift - intro}
\end{eqnarray}

We say that the spacetime is covariant if
\begin{equation}
    \delta_\epsilon g_{\mu \nu} \big|_{\text{O.S.}} =
    \mathcal{L}_{\xi} g_{\mu \nu} \big|_{\text{O.S.}}
    \,,
    \label{eq:Covariance condition of spacetime}
\end{equation}
where we use "O.S." to indicate an evaluation on shell.
Equation (\ref{eq:Covariance condition of spacetime}) is the condition that the canonical gauge transformations with the gauge parameters $(\epsilon^0, \epsilon^a)$ is equivalent to a diffeomorphism of the spacetime metric generated by the 4-vector field $\xi^\mu$ with the componentes of the two generators related by
\begin{eqnarray}
    \xi^\mu &=& \epsilon^0 n^\mu + \epsilon^a s^\mu_a
    = \xi^t t^\mu + \xi^a s^\mu_a
    \,,
    \\
    \xi^t &=& \frac{\epsilon^0}{N}
    \,,
    \hspace{0.75cm}
    \xi^a = \epsilon^a - \frac{\epsilon^0}{N} N^a
    \,,
\label{eq:Diffeomorphism generator projection}
\end{eqnarray}
which is identified as a change of basis from the observer's frame to the Eulerian frame associated to the foliation.

The timelike components ($tt$ and $ta$) of the spacetime covariance condition (\ref{eq:Covariance condition of spacetime}) are automatically satisfied by the gauge transformations of the lapse and shift, (\ref{eq:Off-shell gauge transformations for lapse - intro}) and (\ref{eq:Off-shell gauge transformations for shift - intro}), provided the covariance condition of the 3-metric, $\delta_\epsilon q_{a b} |_{\rm O.S.} = \mathcal{L}_{\xi} q_{a b} |_{\rm O.S.}$, is satisfied too.
The latter can be simplified to the following series of conditions \cite{EMGCov}
\begin{equation}
    \frac{\partial (\delta_{\epsilon^0} q^{a b})}{\partial (\partial_c \epsilon^0)} \bigg|_{\text{O.S.}}
    = \frac{\partial (\delta_{\epsilon^0} q^{a b})}{\partial (\partial_c \partial_d \epsilon^0)} \bigg|_{\text{O.S.}}
    = \dotsi
    = 0
    \,,
    \label{eq:Covariance condition of 3-metric - reduced}
\end{equation}
terminating on the highest derivative order in the Hamiltonian constraint, here assumed to be finite, and hence local.

If one assumes that the spatial metric $q_{a b}$, being the inverse of the structure function, is a configuration variable with conjugate momentum $p^{a b}$, then, using $\{q_{a b} , H[\epsilon^0]\} = \delta H[\epsilon^0] / \delta p^{a b}$, the covariance condition (\ref{eq:Covariance condition of 3-metric - reduced}) implies that the Hamiltonian constraint cannot depend on spatial derivatives of $p^{a b}$.
If the Hamiltonian constraint contains only up to second-order spatial derivatives of $q_{ab}$, then it is uniquely determined by the constraint algebra (\ref{eq:Hypersurface deformation algebra - HaHa})-(\ref{eq:Hypersurface deformation algebra - HH}) up to the choice of Newton's and the cosmological
constant, and hence it is precisely the constraint of GR \cite{hojman1976geometrodynamics,kuchar1974geometrodynamics}.
Therefore, no generally covariant modifications are allowed.

This result is circumvented in EMG by dropping the assumption that the spatial metric $q_{ab}$ is a configuration variable.
Rather, we start with a phase-space composed of fundamental fields different from the metric, for which in vacuum only the gravitational field is considered, and the metric is an emergent object to be derived from imposing the preservation of the form of the constraint algebra as follows.
First, we distinguish the emergent spatial metric from its classical expression by writing it as $\tilde{q}_{ab}$.
We now consider a modified Hamiltonian constraint different from its classical expression and denote it by $\tilde{H}$, while we leave the vector constraint unmodified.
We then impose the preservation of the hypersurface deformation algebra of the modified constraints,
\begin{eqnarray}
    \{ \Vec{H} [ \Vec{N}_1] , \Vec{H} [ \Vec{N}_2 ] \} &=& - \vec{H} [\mathcal{L}_{\Vec{N}_2} \Vec{N}_1]
    \,,
    \label{eq:Hypersurface deformation algebra - HaHa - EMG}
    \\
    \{ \tilde{H} [ N ] , \Vec{H} [ \Vec{N}]\} &=& - \tilde{H} [ N^b \partial_b N ]
    \label{eq:Hypersurface deformation algebra - HHa - EMG}
    \,, \\
    \{ \tilde{H} [ N_1 ] , \tilde{H} [ N_2 ] \} &=& - \vec{H} [ \tilde{q}^{a b} ( N_2 \partial_b N_1 - N_1 \partial_b N_2 )]
    \,,\quad
    \label{eq:Hypersurface deformation algebra - HH - EMG}
\end{eqnarray}
up to a modified structure function $\tilde{q}^{ab}$ generally different from its classical expression.
It follows that the gauge transformations of the lapse and shift become
\begin{eqnarray}
    \delta_\epsilon N &=& \dot{\epsilon}^0 + \epsilon^a \partial_a N - N^a \partial_a \epsilon^0
    \,,
    \label{eq:Off-shell gauge transformations for lapse - EMG}
    \\
    \delta_\epsilon N^a &=& \dot{\epsilon}^a + \epsilon^b \partial_b N^a - N^b \partial_b \epsilon^a
    \nonumber\\
    &&
    + \tilde{q}^{a b} \left(\epsilon^0 \partial_b N - N \partial_b \epsilon^0 \right)
    \,.
    \label{eq:Off-shell gauge transformations for shift - EMG}
\end{eqnarray}
The spatial metric $\tilde{q}_{ab}$ is then defined as the inverse of this modified structure function, which in turn defines the emergent spacetime line element by
\begin{equation}
  {\rm d} s^2 = - N^2 {\rm d} t^2 + \tilde{q}_{a b} ( {\rm d} x^a + N^a {\rm d} t )
  ( {\rm d} x^b + N^b {\rm d} t )
  \,,
  \label{eq:ADM line element - EMG}
\end{equation}
whose components we denote as $\tilde{g}_{\mu\nu}$.
This candidate metric is not necessarily covariant and, therefore, we must impose that its gauge transformation be equivalent to infinitesimal coordinate transformations,
\begin{equation}
    \delta_\epsilon \tilde{g}_{\mu \nu} \big|_{\rm O.S.} =
    \mathcal{L}_{\xi} \tilde{g}_{\mu \nu} \big|_{\rm O.S.}
    \,,
    \label{eq:Covariance condition of spacetime - EMG}
\end{equation}
which can be reduced to \cite{EMGCov}
\begin{equation}
    \frac{\partial (\delta_{\epsilon^0} \tilde{q}^{a b})}{\partial (\partial_c \epsilon^0)} \bigg|_{\rm O.S.}
    = \frac{\partial (\delta_{\epsilon^0} \tilde{q}^{a b})}{\partial (\partial_c \partial_d \epsilon^0)} \bigg|_{\rm O.S.}
    = \dotsi
    = 0
    \,.
    \label{eq:Covariance condition of 3-metric - reduced - EMG}
\end{equation}
Implementing this condition, the gauge transformation of the structure function takes the form
\begin{eqnarray}\label{eq:Gauge transformation of emergent spatial metric}
    \delta_\epsilon \tilde{q}^{a b}
    = \tilde{Q}^{a b} \epsilon^0 + \mathcal{L}_{\vec{\epsilon}} \tilde{q}^{a b}
    \,,
\end{eqnarray}
\textcolor{red}{with some phase space function $\tilde{Q}^{a b}$.}

Once the emergent spacetime metric $\tilde{g}_{\mu \nu}$ has been found, there is a preferred derivative operator $\nabla_\mu$ associated to it by the usual relation $\nabla_\mu \tilde{g}_{\alpha \beta} = 0$.
It is this covariant derivative that must be used for instance for parallel transport and to compute geodesic motion.

It is important to note that the emergent spacetime volume $\sqrt{- \det \tilde{g}}=N\sqrt{\det \tilde{q}}$ is not the only covariant way to densitize objects.
Recall that the spacetime volume is used because its transformation under coordinate transformations, given by
\begin{eqnarray}
    &&\!\!\!\!\!\!
    \sqrt{- \det \tilde{g} (x)} \to
    \sqrt{- \det \tilde{g}'(x')}
    \nonumber\\
    &&\qquad\quad= \sqrt{- \det \tilde{g} (x)} + \mathcal{L}_\xi \sqrt{- \det \tilde{g} (x)}
    + O(\xi^2)
    \nonumber\\
    &&\qquad\quad=
    \sqrt{- \det \tilde{g} (x)}
    + \partial_\alpha (\xi^\alpha \sqrt{- \det \tilde{g}(x)})
    + O(\xi^2)
    \nonumber\\
    &&\qquad\quad=
    \sqrt{- \det \tilde{g}(x)} \det \left(\frac{\partial x^\alpha}{\partial x'^\beta}\right)
    + O(\xi^2)
    \,,
\end{eqnarray}
where $x'^\alpha=x^\alpha+\xi^\alpha$, can be used to counteract the Jacobian coming from the transformation of the coordinate volume
\begin{equation}\label{eq:Jacobian of coordinate volume}
    {\rm d}^4 x \to {\rm d}^4 x' = \det \left(\partial x'^\alpha/\partial x^\beta\right) {\rm d}^4 x
    \,,
\end{equation}
such that ${\rm d}^4 x \sqrt{- \det \tilde{g} (x)}$ remains invariant.
The above computation then implies that the canonical gauge transformation of the spacetime volume is given by
\begin{eqnarray}
    &&\!\!\!\!\!\!\!
    \delta_\epsilon (N\sqrt{\det \tilde{q}})
    =
    \partial_\alpha (\xi^\alpha N \sqrt{\det \tilde{q}})
    \nonumber\\
    &&\quad\quad=
    \partial_t ( \epsilon^0 \sqrt{\det \tilde{q}})
    + \partial_c \left( \left(N \epsilon^c - \epsilon^0 N^c \right) \sqrt{\det \tilde{q}}\right)
    \nonumber\\
    &&\quad\quad=
    \dot{\epsilon}^0 \sqrt{\det \tilde{q}}
    - \epsilon^0 \frac{\sqrt{\det \tilde{q}}}{2} \tilde{q}_{ab} \dot{\tilde{q}}^{a b}
    + \mathcal{L}_{\vec{\epsilon}} (N \sqrt{\det \tilde{q}})
    \nonumber\\
    &&\qquad\quad\qquad
    - \mathcal{L}_{\vec{N}} (\epsilon^0 \sqrt{\det \tilde{q}})
    \nonumber\\
    &&\quad\quad=
    \dot{\epsilon}^0 \sqrt{\det \tilde{q}}
    - N \epsilon^0 \frac{\sqrt{\det \tilde{q}}}{2} \tilde{q}_{ab} \tilde{Q}^{a b}
    + \epsilon^0 \mathcal{L}_{\vec{N}} \sqrt{\det \tilde{q}}
    \nonumber\\
    &&\qquad\qquad\quad
    + \mathcal{L}_{\vec{\epsilon}} (N \sqrt{\det \tilde{q}})
    - \mathcal{L}_{\vec{N}} (\epsilon^0 \sqrt{\det \tilde{q}})
    \nonumber\\
    &&\quad\quad=
    \dot{\epsilon}^0 \sqrt{\det \tilde{q}}
    - N \epsilon^0 \frac{\sqrt{\det \tilde{q}}}{2} \tilde{q}_{ab} \tilde{Q}^{a b}
    \nonumber\\
    &&\qquad\qquad\quad
    + \partial_c ( \epsilon^c N \sqrt{\det \tilde{q}})
    - \sqrt{\det \tilde{q}} N^c \partial_c \epsilon^0\,,
\end{eqnarray}
where we used (\ref{eq:Gauge transformation of emergent spatial metric}).

If one takes some spatial symmetric tensor $\bar{q}^{a b}$ different from the emergent structure function but with the same transformation properties, namely,
\begin{eqnarray}\label{eq:Gauge transformation of unnatural spatial metric}
    \delta_\epsilon \bar{q}^{a b}
    = \bar{Q}^{a b} \epsilon^0 + \mathcal{L}_{\vec{\epsilon}} \bar{q}^{a b}
    \,,
\end{eqnarray}
with some phase space function $\bar{Q}^{a b}$, then the function
\begin{eqnarray}\label{eq:Unnatural volume}
    N \sqrt{\det \bar{q}}
    \,,
\end{eqnarray}
where $\bar{q}$ stands for $\bar{q}_{ab}$ (the inverse of $\bar{q}^{a b}$), will have the gauge transformation
\begin{eqnarray}
    \delta_\epsilon (N\sqrt{\det \bar{q}})
    &=&
    \dot{\epsilon}^0 \sqrt{\det \bar{q}}
    - N \epsilon^0 \frac{\sqrt{\det \bar{q}}}{2} \bar{q}_{ab} \bar{Q}^{a b}
    \\
    &&
    + \partial_c ( \epsilon^c N \sqrt{\det \bar{q}})
    - \sqrt{\det \bar{q}} N^c \partial_c \epsilon^0\,,
    \nonumber
\end{eqnarray}
and hence
\begin{eqnarray}
    &&\!\!\!\!
    N\sqrt{\det \bar{q}} + \delta_\epsilon (N\sqrt{\det \bar{q}})
    \nonumber\\
    &&\qquad\qquad
    = N\sqrt{\det \bar{q}} \det \left(\frac{\partial x^\alpha}{\partial x'^\beta}\right)
    + O(\xi^2)
    \,.
\end{eqnarray}
Therefore, the \emph{unnatural} volume $N\sqrt{\det \bar{q}}$ can be used for densitization too because it can counteract the transformation of the coordinate volume (\ref{eq:Jacobian of coordinate volume}), such that ${\rm d}^4 x N\sqrt{\det \bar{q}}$ remains invariant.
Covariant spacetime integrations are therefore possible with use of an unnatural volume (\ref{eq:Unnatural volume}) replacing the \emph{natural} one, $\sqrt{-\det{\tilde{g}}}$, provided the spatial tensor $\bar{q}^{a b}$ has the transformation (\ref{eq:Gauge transformation of unnatural spatial metric}).
The unnatural volume is not unique and there is no simple way to determine a viable one besides checking its transformation properties explicitly.
This is relevant in EMG because $\bar{q}_{ab}$ could be for instance the gravitational field $q_{ab}$ (which is different from the emergent spatial metric $\tilde{q}_{ab}$) or a different function of the phase space, and, as long as it has the correct transformation properties, it can be used for densitization.
In the following, we will refer to $\bar{q}_{ab}$ as an auxiliary field.

As a final point, if one considers additional matter fields into the theory that present manifestations independently from the spacetime metric, then one has to make sure that such manifestations of the matter fields are covariant too \cite{EMGPF}.
For example, if the matter field in consideration is described by some spacetime tensor $f$, which need not be a fundamental expression of matter but rather may be an emergent representation of it, then one has to apply the matter covariance condition on this field too:
\begin{equation}\label{eq:Covariance condition - general}
    \delta_\epsilon f |_{\rm O.S.} = \mathcal{L}_\xi f |_{\rm O.S.}
    \,.
\end{equation}
Here, we will be interested in coupling the electromagnetic field, whose observations are associated to the strength tensor $F_{\mu\nu}$, playing the role of $f$ above on which the covariance conditions must be applied.
Because the spacetime metric is emergent, then it is allowed not only to depend on gravity, but also on the matter fields as long as the anomaly-freedom of the constraint algebra and all the covariance conditions are satisfied.
Similarly, the strength tensor or, equivalently, the emergent electric field could have a nontrivial dependence not only on the electromagnetic variables but also on gravity.
We will explore the details of its covariant coupling to EMG in Section~\ref{sec: Emergent electromagnetism} for the four dimensional theory and in Section~\ref{sec:Emergent electromagnetism: Spherical symmetry} for the spherically symmetric model.
Before that, we present the spherically symmetric model of EMG in vacuum, where the modified Hamiltonian constraint and emergent metric can be obtained in explicit form.

\section{Spherically symmetric emergent modified gravity: Vacuum}
\label{sec:Spherical EMG Vacuum}

In the spherically symmetric theory in vacuum, the spacetime metric takes the general form
\begin{equation}
    {\rm d} s^2 = - N^2 {\rm d} t^2 + q_{x x} ( {\rm d} x + N^x {\rm d} t )^2 + q_{\vartheta \vartheta} {\rm d} \Omega^2
    \label{eq:ADM line element - spherical}
    \,.
\end{equation}
The classical spatial metric components can be written in terms of the classical radial and angular densitized triads $E^x$ and $E^\varphi$, respectively,  $q_{xx} = (E^\varphi)^2/E^x$ and $q_{\vartheta \vartheta} = E^x$.
We work in natural units where
\begin{equation}\label{eq:Natural units}
    c=G=\epsilon_0=1\,,
\end{equation}
such that the dimensions of all quantities are suitable powers of length $L$.

The symplectic structure of the canonical theory is
\begin{equation}
    \{ K_x (x) , E^x (y)\}
    = \{ K_\varphi (x) , E^\varphi (y) \}
    = \delta (x-y)
    \ ,
\end{equation}
where $K_x$ and $K_\varphi$ are the radial and angular components of the extrinsic curvature.
Within spherical symmetry, only the Hamiltonian constraint and the radial vector constraint are non-trivial.
The hypersurface deformation algebra (\ref{eq:Hypersurface deformation algebra - HaHa})-(\ref{eq:Hypersurface deformation algebra - HH}) becomes
\begin{eqnarray}
    \{ H_x [ N_1^x] , H_x [ N^x_2 ] \} &=& - H_x [\mathcal{L}_{N^x_2} N^x_1]
    \,,
    \label{eq:Hypersurface deformation algebra - HaHa - spherical}
    \\
    \{ H [ N ] , H_x [ N^x]\} &=& - H [ N^x N' ]
    \label{eq:Hypersurface deformation algebra - HHa - spherical}
    \,, \\
    \{ H [ N_1 ] , H [ N_2 ] \} &=& - H_x [ q^{xx} ( N_2 N_1' - N_1 N_2' )]
    \,,\quad
    \label{eq:Hypersurface deformation algebra - HH - spherical}
\end{eqnarray}
where the prime denotes a derivative with respect to the radial coordinate $x$, and $q^{xx}=E^x/(E^\varphi)^2$ is the inverse of the radial metric component $q_{xx}$.
The algebra (\ref{eq:Hypersurface deformation algebra - HaHa - spherical})-(\ref{eq:Hypersurface deformation algebra - HH - spherical}) holds even in the presence of matter\textemdash when gauge fields are present, typically the third bracket receives a contribution from the respective Gauss constraint, but the U(1) case of electromagnetism is the exception and hence we shall neglect this in the following.
We denote the constraints in vacuum by $H^{\rm grav}$ and $H^{\rm grav}_x$.

In emergent modified gravity we keep the vector constraint in its standard classical form, which in vacuum is given by
\begin{equation}
    H_x^{\rm grav} =
    E^\varphi K_\varphi'
    - K_x (E^x)'
    \,,
    \label{eq:Diffeomorphism constraint - Gravity - spherical}
\end{equation}
and hence the bracket (\ref{eq:Hypersurface deformation algebra - HaHa - spherical}) is unchanged.
On the other hand, we consider a modified Hamiltonian constraint and write it as $\tilde{H}$ to distinguish it from the classical one.
We impose that $\tilde{H}$ is anomaly-free, reproducing the brackets (\ref{eq:Hypersurface deformation algebra - HHa - spherical}) and (\ref{eq:Hypersurface deformation algebra - HH - spherical}) up to a modified structure function $\tilde{q}^{xx}$ not necessarily equal to the classical $q^{xx}$,
\begin{eqnarray}
    \{ H_x [ N_1^x] , H_x [ N^x_2 ] \} &=& - H_x [\mathcal{L}_{N^x_2} N^x_1]
    \,,
    \label{eq:Hypersurface deformation algebra - HaHa - spherical - modified}
    \\
    \{ \tilde{H} [ N ] , H_x [ N^x]\} &=& - \tilde{H} [ N^x N' ]
    \label{eq:Hypersurface deformation algebra - HHa - spherical - modified}
    \,, \\
    \{ \tilde{H} [ N_1 ] , \tilde{H} [ N_2 ] \} &=& - H_x [ \tilde{q}^{xx} ( N_2 N_1' - N_1 N_2' )]
    \,.\quad
    \label{eq:Hypersurface deformation algebra - HH - spherical - modified}
\end{eqnarray}
It follows that the gauge transformations of the lapse and shift become
\begin{eqnarray}
    \delta_\epsilon N &=& \dot{\epsilon}^0 + \epsilon^x N' - N^x (\epsilon^0)'
    \,,
    \label{eq:Off-shell gauge transformations for lapse - EMG- spherical}
    \\
    \delta_\epsilon N^x &=& \dot{\epsilon}^x + \epsilon^x (N^x)' - N^x (\epsilon^x)
    \nonumber\\
    &&
    + \tilde{q}^{xx} \left(\epsilon^0 N' - N (\epsilon^0)' \right)
    \,.
    \label{eq:Off-shell gauge transformations for shift - EMG - spherical}
\end{eqnarray}
To derive an explicit expression for the constraint and the structure function, we define an ansatz for $\tilde{H}$ as function of the phase-space variables and their derivatives up to second order \cite{EMGCov,alonso2021anomaly}.
The condition of anomaly-freedom of the constraint algebra (\ref{eq:Hypersurface deformation algebra - HaHa - spherical - modified})-(\ref{eq:Hypersurface deformation algebra - HH - spherical - modified}) places strong restrictions on this ansatz and it determines $\tilde{q}^{xx}$.
The inverse of the modified structure function $\tilde{q}_{xx}=1/\tilde{q}^{xx}$ is then replaced in the line element (\ref{eq:ADM line element - spherical}), defining the emergent spacetime
\begin{equation}
    {\rm d} s^2 = - N^2 {\rm d} t^2 + \tilde{q}_{x x} ( {\rm d} x + N^x {\rm d} t )^2 + \tilde{q}_{\vartheta \vartheta} {\rm d} \Omega^2
    \label{eq:ADM line element - spherical - EMG}
    \,.
\end{equation}
Unlike the radial component, the modified angular component $\tilde{q}^{\vartheta \vartheta}$ cannot be derived from the constraint algebra because the angular components of the vector constraint trivialize due to the underlying spherical symmetry.
The modified angular structure function must then be chosen using phenomenological considerations, and we may here take it as an undetermined function of the radial triad, $\tilde{q}^{\vartheta \vartheta}=\tilde{q}^{\vartheta \vartheta}(E^x)$.

Having identified the emergent metric, the last step is to impose the spacetime covariance condition (\ref{eq:Covariance condition of 3-metric - reduced - EMG}), which in spherical symmetry reduces to the simpler conditions
\begin{eqnarray}
    \frac{\partial (\delta_{\epsilon^0} \tilde{q}^{xx})}{\partial (\epsilon^0)'} \bigg|_{\text{O.S.}}
    = \frac{\partial (\delta_{\epsilon^0} \tilde{q}^{xx})}{\partial (\epsilon^0)''} \bigg|_{\text{O.S.}}
    = \dotsi
    = 0
    \,, \nonumber \\
    \frac{\partial (\delta_{\epsilon^0} \tilde{q}^{\vartheta \vartheta})}{\partial (\epsilon^0)'} \bigg|_{\text{O.S.}}
    = \frac{\partial (\delta_{\epsilon^0} \tilde{q}^{\vartheta \vartheta})}{\partial (\epsilon^0)''} \bigg|_{\text{O.S.}}
    = \dotsi
    = 0
    \,.
    \label{eq:Covariance condition of 3-metric - reduced - spherical}
\end{eqnarray}

As usual in canonical theories, it is possible to use canonical transformations that change the form of the phase-space variables.
In doing so, one may obtain constraints and a structure function that look very different from their expressions prior to the transformations, but both are in fact describing the same system.
This highlights the importance of dealing with and factoring out canonical transformations, so that one is not misled into formulating an apparently different class of modified theories.
In spherical symmetry, we consider the general set of canonical transformations that preserve the form of the vector constraint,
\begin{eqnarray}
    K_\varphi &=& f_c (E^x , \bar{K}_\varphi)\nonumber\\
    E^\varphi &=& \bar{E}^\varphi \left( \frac{\partial f_c}{\partial
                  \tilde{K}_\varphi} \right)^{-1} 
    \nonumber\\
    K_x &=& \frac{\partial (\alpha_c^2 E^x)}{\partial E^x} \bar{K}_x +
            \bar{E}^\varphi \frac{\partial f_c}{\partial E^x} \left(
            \frac{\partial f_c}{\partial \bar{K}_\varphi}
            \right)^{-1}\nonumber\\ 
    \bar{E}^x &=& \alpha_c^2 (E^x) E^x\,,
    \label{eq:Diffeomorphism-constraint-preserving canonical transformations - Spherical - general}
\end{eqnarray}
where the new variables are denoted with a bar.

We begin factoring out this set as follows.
First, any modified angular component of the form
$\tilde{q}_{\vartheta \vartheta} = \alpha_c^{-2} (E^x) E^x$ can be mapped to
its classical form, $\tilde{q}_{\vartheta \vartheta} \to E^x$, by using the
canonical transformation (\ref{eq:Diffeomorphism-constraint-preserving canonical transformations - Spherical - general}) with $f_c=K_{\varphi}$.
This fixes $\alpha_c$ in the set (\ref{eq:Diffeomorphism-constraint-preserving canonical transformations - Spherical - general}).
Second, we may fix the $f_c$ function so as to simplify the differential equations resulting from imposing anomaly-freedom and spacetime covariance.
Such equations can then be solved exactly in the vacuum for the most general constraint ansatz involving up to second-order derivatives of the phase-space variables, obtaining \cite{EMGCov,EMGscalar}

\begin{widetext}
\begin{eqnarray}
    \!\!\!\!
    \tilde{H}^{\rm grav} \!\!\!&=&\!\!\!
    \frac{\bar{\lambda}^2}{\lambda^2} \lambda_0^2 \frac{\sqrt{\tilde{q}_{xx}} \alpha_0^q}{E^x}
    - \frac{E^\varphi}{2} \sqrt{\tilde{q}^{xx}} \Bigg[
    E^\varphi \left(
    \frac{\lambda^2}{\bar{\lambda}^2} \frac{\alpha_{0q}}{E^x}
    + \frac{\alpha_{2q}}{E^x} \left( c_f \frac{\sin^2 (\bar{\lambda} K_\varphi)}{\bar{\lambda}^2} + 2 \frac{\lambda}{\bar{\lambda}} q \frac{\sin(2\bar{\lambda} K_\varphi)}{2 \bar{\lambda}} \right) \right)
    - \frac{((E^x)')^2}{E^\varphi} \frac{\alpha_{2q}}{4 E^x} \cos^2 (\bar{\lambda} K_\varphi)
    \Bigg]
    \nonumber\\
    &&\!\!
    - \frac{\bar{\lambda}}{\lambda} \lambda_0 \frac{\sqrt{E^x}}{2} \Bigg[ E^\varphi \Bigg(
    \frac{\lambda^2}{\bar{\lambda}^2} \frac{\alpha_0}{E^x}
    + 2 \frac{\sin^2 \left(\bar{\lambda} K_\varphi\right)}{\bar{\lambda}^2}\frac{\partial c_{f}}{\partial E^x}
    + 4 \frac{\sin \left(2 \bar{\lambda} K_\varphi\right)}{2 \bar{\lambda}} \frac{\partial}{\partial E^x} \left(\frac{\lambda}{\bar{\lambda}} q\right)
    \nonumber\\
    &&\qquad \qquad \qquad \quad
    + \left( \frac{\alpha_2}{E^x} - 2 \frac{\partial \ln \lambda^2}{\partial E^x}\right) \left( c_f \frac{\sin^2 \left(\bar{\lambda} K_\varphi\right)}{\bar{\lambda}^2}
    + 2 \frac{\lambda}{\bar{\lambda}} q \frac{\sin \left(2 \bar{\lambda} K_\varphi\right)}{2 \bar{\lambda}} \right)
    \Bigg)
    \nonumber\\
    &&\qquad \qquad \quad
    + 4 K_x \left(c_f \frac{\sin (2 \bar{\lambda} K_\varphi)}{2 \bar{\lambda}}
    + \frac{\lambda}{\bar{\lambda}} q \cos(2 \bar{\lambda} K_\varphi)\right)
    \nonumber\\
    &&\qquad \qquad \quad
    - \frac{((E^x)')^2}{E^\varphi} \left(
    \left(\frac{\alpha_2}{4 E^x} - \frac{\partial \ln \lambda}{\partial E^x} \right) \cos^2 \left( \bar{\lambda} K_\varphi \right)
    - \frac{K_x}{E^\varphi} \bar{\lambda}^2 \frac{\sin \left(2 \bar{\lambda} K_\varphi \right)}{2 \bar{\lambda}}
    \right)
    \nonumber\\
    &&\qquad \qquad \quad
    + \left( \frac{(E^x)' (E^\varphi)'}{(E^\varphi)^2}
    - \frac{(E^x)''}{E^\varphi} \right) \cos^2 \left( \bar{\lambda} K_\varphi \right)
    \Bigg]
    \ ,
    \label{eq:Hamiltonian constraint - modified - periodic - vacuum}
\end{eqnarray}
with structure function
\begin{eqnarray}
    \tilde{q}^{x x}
    &=&
    \left(
    \left( c_{f}
    + \left(\frac{\bar{\lambda} (E^x)'}{2 E^\varphi} \right)^2 \right) \cos^2 \left(\bar{\lambda} K_\varphi\right)
    - 2 \frac{\lambda}{\bar{\lambda}} q \bar{\lambda}^2 \frac{\sin \left(2 \bar{\lambda} K_\varphi\right)}{2 \bar{\lambda}}\right)
    \frac{\bar{\lambda}^2}{\lambda^2} \lambda_0^2 \frac{E^x}{(E^\varphi)^2}
    \ .
    \label{eq:Structure function - modified - periodic - vacuum}
\end{eqnarray}
\end{widetext}
where $\lambda_0 , \alpha_0 , \alpha_2 , c_f , q$, and $\lambda$ are undetermined functions of $E^x$, while $\bar{\lambda}$ is a constant.
The classical constraint can be recovered in different limits, but the simplest is given by $\lambda\to \bar{\lambda}$, followed by $\lambda_0 , c_{f} , \alpha_0, \alpha_2\to 1$ and $\bar{\lambda} , q , \alpha_0^q , \alpha_{0 q} , \alpha_{2 q} \to 0$; the cosmological constant can be recovered by instead setting $\alpha_0 \to 1 - \Lambda E^x$.

If we define the barred functions
\begin{eqnarray}
    &&\lambda_0 = \bar{\lambda}_0 \frac{\lambda}{\bar{\lambda}}
    \quad,\quad
    q = \bar{q} \frac{\bar{\lambda}}{\lambda}
    \quad,\quad
    \alpha_0 = \frac{\bar{\lambda}^2}{\lambda^2} \bar{\alpha}_0
    \quad\,,\nonumber\\
    &&
    \alpha_2 = \bar{\alpha}_2 + 4 E^x \frac{\partial \ln \lambda}{\partial E^x}
    \quad,\quad
    \alpha_{0q} = \frac{\bar{\lambda}^2}{\lambda^2} \bar{\alpha}_{0q}
    \quad,
    \label{eq:Redefinitions of lambda - ease notation}
\end{eqnarray}
the Hamiltonian constraint has the simpler form
\begin{widetext}
\begin{eqnarray}
    \tilde{H}^{\rm grav} &=&
    \bar{\lambda}_0^2 \frac{\sqrt{\tilde{q}_{xx}} \alpha_0^q}{E^x}
    - \frac{E^\varphi}{2} \sqrt{\tilde{q}^{xx}} \Bigg[
    E^\varphi \left(
    \frac{\bar{\alpha}_{0q}}{E^x}
    + \frac{\alpha_{2q}}{E^x} \left( c_f \frac{\sin^2 (\bar{\lambda} K_\varphi)}{\bar{\lambda}^2} + 2 \bar{q} \frac{\sin(2\bar{\lambda} K_\varphi)}{2 \bar{\lambda}} \right) \right)
    \nonumber\\
    &&\qquad \qquad \qquad
    - \frac{((E^x)')^2}{E^\varphi} \frac{\alpha_{2q}}{4 E^x} \cos^2 (\bar{\lambda} K_\varphi)
    \Bigg]
    \nonumber\\
    &&
    - \bar{\lambda}_0 \frac{\sqrt{E^x}}{2} \Bigg[ E^\varphi \Bigg( \frac{\bar{\alpha}_0}{E^x}
    + 2 \frac{\sin^2 \left(\bar{\lambda} K_\varphi\right)}{\bar{\lambda}^2}\frac{\partial c_{f}}{\partial E^x}
    + 4 \frac{\sin \left(2 \bar{\lambda} K_\varphi\right)}{2 \bar{\lambda}} \frac{\partial \bar{q}}{\partial E^x}
    \nonumber\\
    &&\qquad \qquad \qquad \quad
    + \frac{\bar{\alpha}_2}{E^x} \left( c_f \frac{\sin^2 \left(\bar{\lambda} K_\varphi\right)}{\bar{\lambda}^2}
    + 2 \bar{q} \frac{\sin \left(2 \bar{\lambda} K_\varphi\right)}{2 \bar{\lambda}} \right)
    \Bigg)
    \nonumber\\
    &&\qquad \qquad \quad
    + 4 K_x \left(c_f \frac{\sin (2 \bar{\lambda} K_\varphi)}{2 \bar{\lambda}}
    + \bar{q} \cos(2 \bar{\lambda} K_\varphi)\right)
    \nonumber\\
    &&\qquad \qquad \quad
    - \frac{((E^x)')^2}{E^\varphi} \left(
    \frac{\bar{\alpha}_2}{4 E^x} \cos^2 \left( \bar{\lambda} K_\varphi \right)
    - \frac{K_x}{E^\varphi} \bar{\lambda}^2 \frac{\sin \left(2 \bar{\lambda} K_\varphi \right)}{2 \bar{\lambda}}
    \right)
    \nonumber\\
    &&\qquad \qquad \quad
    + \left( \frac{(E^x)' (E^\varphi)'}{(E^\varphi)^2}
    - \frac{(E^x)''}{E^\varphi} \right) \cos^2 \left( \bar{\lambda} K_\varphi \right)
    \Bigg]
    \,,
    \label{eq:Hamiltonian constraint - modified - periodic - vacuum - barred}
\end{eqnarray}
and the structure function also simplifies to
\begin{eqnarray}
    \tilde{q}^{x x}
    &=&
    \left(
    \left( c_{f}
    + \left(\frac{\bar{\lambda} (E^x)'}{2 E^\varphi} \right)^2 \right) \cos^2 \left(\bar{\lambda} K_\varphi\right)
    - 2 \bar{q} \bar{\lambda}^2 \frac{\sin \left(2 \bar{\lambda} K_\varphi\right)}{2 \bar{\lambda}}\right)
    \bar{\lambda}_0^2 \frac{E^x}{(E^\varphi)^2}
    \ .
    \label{eq:Structure function - modified - periodic - vacuum - barred}
\end{eqnarray}

The existence of a mass observable can be used as an additional condition to restrict the constraint (\ref{eq:Hamiltonian constraint - modified - periodic - vacuum}).
Imposing this we find that \cite{EMGscalar}
\begin{equation}\label{eq:Vacuum mass observable condition}
    \alpha_0^q=\alpha_{0q}=\alpha_{2q}=0
    \,,
\end{equation}
and hence the constraint (\ref{eq:Hamiltonian constraint - modified - periodic - vacuum}) reduces to
\begin{eqnarray}
    \tilde{H}^{\rm grav} &=&
    - \bar{\lambda}_0 \frac{\sqrt{E^x}}{2} \Bigg[ E^\varphi \Bigg(
    -\frac{\bar{\alpha}_0}{E^x}
    + 2 \frac{\sin^2 \left(\bar{\lambda} K_\varphi\right)}{\bar{\lambda}^2}\frac{\partial c_{f}}{\partial E^x}
    + 4 \frac{\sin \left(2 \bar{\lambda} K_\varphi\right)}{2 \bar{\lambda}} \frac{\partial \bar{q}}{\partial E^x}
    + \frac{\bar{\alpha}_2}{E^x} \left( c_f \frac{\sin^2 \left(\bar{\lambda} K_\varphi\right)}{\bar{\lambda}^2}
    + 2 \bar{q} \frac{\sin \left(2 \bar{\lambda} K_\varphi\right)}{2 \bar{\lambda}} \right)
    \Bigg)
    \nonumber\\
    &&\qquad \qquad \quad
    + 4 K_x \left(c_f \frac{\sin (2 \bar{\lambda} K_\varphi)}{2 \bar{\lambda}}
    + \bar{q} \cos(2 \bar{\lambda} K_\varphi)\right)
    - \frac{((E^x)')^2}{E^\varphi} \left(
    \frac{\bar{\alpha}_2}{4 E^x} \cos^2 \left( \bar{\lambda} K_\varphi \right)
    - \frac{K_x}{E^\varphi} \bar{\lambda}^2 \frac{\sin \left(2 \bar{\lambda} K_\varphi \right)}{2 \bar{\lambda}}
    \right)
    \nonumber\\
    &&\qquad \qquad \quad
    + \left( \frac{(E^x)' (E^\varphi)'}{(E^\varphi)^2}
    - \frac{(E^x)''}{E^\varphi} \right) \cos^2 \left( \bar{\lambda} K_\varphi \right)
    \Bigg]
    \ ,
    \label{eq:Hamiltonian constraint - modified - periodic - grav obs}
\end{eqnarray}
with the same structure function (\ref{eq:Structure function - modified - periodic - vacuum - barred}).
The vacuum mass observable is given by
\begin{eqnarray}\label{eq:Vacuum mass observable - EMG}
    \mathcal{M}
    &=&
    d_0
    + \frac{d_2}{2} \left(\exp \int {\rm d} E^x \ \frac{\bar{\alpha}_2}{2 E^x}\right)
    \left(
    c_f \frac{\sin^2\left(\bar{\lambda} K_{\varphi}\right)}{\bar{\lambda}^2}
    + 2 \bar{q} \frac{\sin \left(2 \bar{\lambda}  K_{\varphi}\right)}{2 \bar{\lambda}}
    - \cos^2 (\bar{\lambda} K_\varphi) \left(\frac{(E^x)'}{2 E^\varphi}\right)^2
    \right)
    \notag\\
    &&
    + \frac{d_2}{4} \int {\rm d} E^x \ \left( \frac{\bar{\alpha}_0}{E^x} \exp \int {\rm d} E^x \ \frac{\bar{\alpha}_2}{2 E^x}\right)
    \,,
\end{eqnarray}
\end{widetext}
where $d_0$ and $d_2$ are undetermined constants with the classical limits $d_0\to0$ and $d_2 \to 1$.

It will be important to note that the gravitational inverse tensor component $q^{xx}$, which classically is the structure function, is not $E^x/(E^\varphi)^2$, but a careful treatment of canonical transformations as depicted above reveals that it is instead \cite{EMGCov}
\begin{equation}
    q^{x x} = \frac{\bar{\lambda}^2}{\lambda^2} \cos^2 (\bar{\lambda} K_\varphi) \frac{E^x}{(E^\varphi)^2}
    \,.
\end{equation}
This function does not obey the condition (\ref{eq:Covariance condition of 3-metric - reduced - spherical}) and thus cannot be used as a component of the true spacetime metric.
On the other hand, the auxiliary function
\begin{equation}
    \bar{q}^{x x} = \frac{\bar{\lambda}^2}{\lambda^2} \frac{E^x}{(E^\varphi)^2}
    \,,
    \label{eq:Alternative spatial metric}
\end{equation}
does satisfy (\ref{eq:Covariance condition of 3-metric - reduced - spherical}), but it cannot be used as a component of the true spacetime metric either because the transformation of the shift (\ref{eq:Off-shell gauge transformations for shift - EMG - spherical}) requires that it be the emergent one obtained from the hypersurface deformation algebra.
However, that the function (\ref{eq:Alternative spatial metric}) satisfies (\ref{eq:Covariance condition of 3-metric - reduced - spherical}) implies that it defines an unnatural volume (\ref{eq:Unnatural volume}) and may be used as an alternative densitization.
In the following, it will be useful to define the function $\beta$ by the relation
\begin{equation}\label{eq:Barred spatial metric}
    \tilde{q}^{xx} = \beta \bar{q}^{xx}
    \,.
\end{equation}
The lambda factors in (\ref{eq:Alternative spatial metric}) could in principle be absorbed into $\beta$, but we keep them in $\bar{q}^{xx}$ instead for reasons that will become clear in the coming sections after we introduce the electromagnetic field.

\section{Symmetry reduced model vs lower dimensional model: The capture of nonlocal physics}
\label{sec:Symm red vs dim red}

It is important to realize that the starting point of our explicit model in section~\ref{sec:Spherical EMG Vacuum} is not necessarily the symmetry reduction of a four dimensional theory, but rather a lower dimensional theory.
More specifically, we are modelling the spherical symmetry reduction of our four dimensional spacetime theory as a two dimensional spacetime model.
This distinction has profound implications.
For instance, consider the modified Hamiltonian constraint (\ref{eq:Hamiltonian constraint - modified - periodic - grav obs}) and take the classical values $c_f,\lambda_0 \to 1$, $\lambda \to \bar{\lambda}$ followed by $\bar{\lambda},q\to0$, in which case its associated structure function (\ref{eq:Structure function - modified - periodic - vacuum - barred}) becomes classical,
\begin{equation}
    \tilde{q}^{xx} \to q^{xx} = \frac{E^x}{(E^\varphi)^2}
    \,,
\end{equation}
but the constraint preserves the modification freedom of the $\alpha_0$ and $\alpha_2$ functions, and hence dynamics are still modified.

This seems to contradict the uniqueness results of \cite{hojman1976geometrodynamics,kuchar1974geometrodynamics}, which state that anomaly-freedom with the classical structure function uniquely determines the constraint of GR.
This result, however, holds only in the four-dimensional system, and the two-dimensional system modeling spherical symmetry is somehow able to circumvent it.
Consequently, the spherically symmetric system enjoys a larger freedom than the full four-dimensional case, allowing the presence of the modification functions $\alpha_0$ and $\alpha_2$, which have long been exploited to model 2D dilation gravity theories\textemdash\,such modifications in the Hamiltonian formalism can be found in \cite{Bojowald_DeformedGR,Tibrewala_Inhomogeneities,Tibrewala_Midisuperspace}, and in the Lagrangian approach with similar modifications can be found in \cite{Kunstatter_New2Ddilaton}\textemdash\,, and hence allowing modified dynamics, though in a more restricted way compared to when one allows for a modified structure function.

Therefore, it is reasonable to assume that this larger freedom of symmetry reduced models compared to the full four-dimensional one in GR can be extrapolated to EMG, in the sense that much of the modification freedom in the constraint (\ref{eq:Hamiltonian constraint - modified - periodic - grav obs}) would not survive in the full four dimensional theory without spherical symmetry.
This, in turn, could be extended to matter coupling such as electromagnetism, such that some modifications we obtain on the matter contributions in a symmetry reduced model would not apply to the four-dimensional theory either.

In light of this, one could argue that the modifications allowed by EMG in the spherically symmetric system are only pathologies of working in a lower dimensional model and hence do not apply to the real world.
But this is a premature conclusion that rests on the assumption that the four-dimensional theory \emph{must} be local.
We will partially drop this assumption in a nuanced way below.

The best known example of a nonlocal theory is lattice field theory, in which space or spacetime is discretized onto a lattice.
This is particularly useful not only for numerical purposes but also for the quantization of gauge theories using the Wilson action.
While most quantization approaches in flat spacetime preserve manifest Poincar\'e invariance at the expense of hindering its gauge symmetry by requiring gauge fixing, lattice quantum field theory (LQFT) preserves manifest gauge symmetry at the expense of losing manifest Poincar\'e invariance.

Electromagnetism on a lattice can be considered a modified theory with the dynamics generated by the Wilson action.
To define this action on the lattice with the correct continuum limit, consider a four dimensional Euclidean space discretized into hypercubes of side length $\ell$.
Each side of a hypercube is called a link, which has an associated vector $\ell^\mu$ denoting its direction and size, while each face is called a plaquette.
Now consider the Wilson line, also called holonomy, of the electromagnetic four-vector potential along a link given by
\begin{equation}\label{eq:Wilson line}
    U_\ell = {\cal P} \exp \left( i \mathfrak{a} \int_\ell A \right)
    \,,
\end{equation}
where ${\cal P}$ denotes path ordering and we have introduced the real, finite constant $\mathfrak{a}$ with dimensions of $L^{-1}$ such that the argument becomes dimensionless\textemdash\,recall that in natural units (\ref{eq:Natural units}) the components $A_\mu$ are dimensionless and hence the one-form $A$ has dimensions of length $L$ in Cartesian-type coordinates.
We may refer to (\ref{eq:Wilson line}) as a holonomy and to $\mathfrak{a}$ as the electromagnetic holonomy parameter.
If one takes a nontrivial closed path or a loop $C$, we may use Stoke's theorem to rewrite the line integral of the connection as a surface integral instead,
\begin{equation}
    \int_{C} A = \int_{\cal A} F
    \,,
\end{equation}
where $F = {\rm d} A$ is the strength tensor and ${\cal A}$ is a two dimensional surface enclosed by $C$\textemdash\,in natural units (\ref{eq:Natural units}), the components $F_{\mu\nu}$ have dimensions of $L^{-1}$ and hence the two-form $F$ has dimensions of $L$ in Cartesian-type coordinates.
If the loop $C$ encloses a single plaquette with its sides given by the succession of links $\ell^\mu\to\ell^\nu\to-\ell^\mu\to-\ell^\nu$, then its associated holonomy takes the form
\begin{equation}
    U_{\mu\nu} = \exp \left(i \mathfrak{a} \ell^2 F_{\mu\nu} \right)
    \,,
\end{equation}
where $\ell^2$ is the coordinate area of the plaquette.
The Wilson action is defined by
\begin{eqnarray}\label{eq:Wilson action - EM}
    S_W &=& \frac{1}{2} \sum_p \sum_{\mu<\nu} \mathfrak{a}^{-2} {\rm Re} \left[ 1 - U_{\mu\nu} \right]
    \nonumber\\
    &=& \frac{1}{2} \sum_p \sum_{\mu<\nu} \frac{1 - \cos \left(\mathfrak{a} \ell^2 F_{\mu\nu} \right)}{\mathfrak{a}^2}
    \,,
\end{eqnarray}
where $p$ denotes the plaquette of a given loop.
In the limit $\ell\to0$ we obtain
\begin{eqnarray}
    S_W &\to& \frac{1}{4} \int {\rm d} x^4\ \sum_{\mu<\nu} F_{\mu\nu} F_{\mu\nu}
    \nonumber\\
    &&
    = \frac{1}{4} \int {\rm d} x^4\ F_{\mu\nu} F_{\alpha\beta} \delta^{\mu \alpha} \delta^{\nu\beta}
    \,,
\end{eqnarray}
where we have taken $\sum_p \ell^4 \to \int {\rm d} x^4$ and $\mathfrak{a}$ cancels out.
After a Wick-rotation to Minkowski spacetime, we recover the Maxwell action,
\begin{equation}
    S_{\rm EM}
    = - \frac{1}{4} \int {\rm d} x^4\ F_{\mu\nu} F^{\mu\nu}
    \,,
\end{equation}
with the indices raised by the Minkoski metric.
This shows that the Wilson action (\ref{eq:Wilson action - EM}) has the correct continuum limit.
However, if one takes the discreteness, and hence nonlocality, of the Wilson action seriously, they imply that the strength tensor has a bounded effect in the action for a nonvanishing, finite link size $\ell$ or area $\ell^2$, as well as a nonvanishing holonomy parameter $\mathfrak{a}$.
On the other hand, Poincar\'e invariance, which is a defining property of Minkowki spacetime, is lost due to the discrete nature of the lattice.
It is important to note that the continuum limit $\ell\to0$ holds for arbitrary, but finite, $\mathfrak{a}$ and hence this new constant is not required to be small.

In the same spirit, loop quantum gravity considers a discretized spacetime such that the links are associated to holonomies and fluxes of the gravitational connection and its conjugate momentum, respectively.
These holonomy-flux variables are then used as the starting point of loop quantization.
Consequently, the Hamiltonian must be modified to be written in terms of these variables\textemdash\,obtaining an gravitational analog of the Wilson action\textemdash\,for it to be a well-defined quantum operator.
In this general relativistic context, the discrete and nonlocal nature introduced by the holonomy-flux modifications to the Hamiltonian implies that the transformations it generates are themselves modified too.
These transformations are directly related to coordinate transformations and it is therefore an open problem whether such modifications preserve general covariance\textemdash\,the latter replacing Poincar\'e invariance of LQFT in flat spacetime.

The introduction of Wilson loops in LQFT and LQG render the theories nonlocal, unless the continuum limit is taken.
Therefore, it is not possible even for EMG, in its current local formulation, to include these modifications in a manifestly covariant way.
However, it is possible to exploit the larger freedom of symmetry reduced models to identify some allowed modifications as lattice effects. As we now show, a particular modification function of the spherically symmetric EMG constraint (\ref{eq:Hamiltonian constraint - modified - periodic - grav obs}) can be interpreted as the length of the links in a spherical lattice, hence modelling holonomy modifications.
In spherically symmetric LQG \cite{bojowald2000symmetry,bojowald2004spherically}, the holonomies are given by
\begin{eqnarray}
    h^x_e [K_x] &=& \exp \left( i \int_e {\rm d} x\ K_x \right)
    \,,
    \label{eq:Radial holonomy}
    \\
    h^\varphi_{v, \lambda} [K_\varphi] &=& \exp \left( i \int_\lambda {\rm d} \theta\ K_\varphi \right)
    = \exp \left( i \lambda K_\varphi (v) \right)
    \,,\quad
    \label{eq:Angular holonomy}
\end{eqnarray}
where $e$ stands for an arbitrary radial curve of finite coordinate length, $v$ stands for an arbitrary point in the radial line, and $\lambda$ is the coordinate length of an arbitrary angular curve on the 2-sphere intersecting the point $v$.
Here, $K_x$ has dimensions of $L^{-1}$ while $K_\varphi$ is dimensionless and hence the arguments of the holonomies are dimensionless without introducing additional parameters unlike the electromagnetic holonomy case which requires the introduction of $\mathfrak{a}$.
While the radial holonomy integration must remain formal as a nonlocal object, the explicit integration in the angular holonomy (\ref{eq:Angular holonomy}) is possible due to the underlying spherical symmetry.
Therefore, if the freedom of coordinate transformations of general covariance is restricted to the $t$-$x$ plane, and hence effectively treated as a two-dimensional system, the angular holonomies may be treated as simply local functions of the variable $K_\varphi (t,x)$.
Because the Hamiltonian constraint must be Hermitian in the quantum theory, or simply real in the classical case, the holonomy modifications must be combined in such a way that $\lambda K_\varphi$ always appears as the argument of trigonometric functions in the constraints.
This is precisely the case in the spherically symmetric EMG constraint (\ref{eq:Hamiltonian constraint - modified - periodic - grav obs}).

Using the spherically symmetric model as our guiding example, we may extrapolate our conclusion to symmetry reduced models in general: The $D$-dimensional modified theory can include nonlocal modifications in the homogenous directions of symmetry when treated as lower $d$-dimensional system where the coordinate transformations are restricted to the $d$-dimensional plane of inhomogeneity.
The modified spherically symmetric models are therefore hybrid systems where the homogenous directions are allowed to include 'nonlocal' modifications, while covariance is imposed only in the inhomogeneous directions, which are treated in a local way.
Consequently, it was possible to include angular, but not raial, holonomy modifications in spherically symmetric EMG.

Just as the concept of the emergent spacetime is paramount in introducing the angular holonomy modifications, it is the purpose of this work to show that the concept of emergent electromagnetism is necessary to introduce modifications capable of modeling those of the Wilson action (\ref{eq:Wilson action - EM}).
In order to do so, we will start with a brief review of classical electromagnetism, which will serve to introduce the concept of the emergent electric field in a familiar context.
We will proceed to evaluate the general modified theory in both the four dimensional and the spherical symmetry reduced cases with the conclusion that the electromagnetic holonomy modifications are allowed only in the latter, as expected from our discussion in this section.


\section{Classical electromagnetism}
\label{sec:Classical electromagnetism}

\subsection{Hamiltonian formalism in curved spacetimes}

In the classical theory, the equations of motion of the electromagnetic field can be derived from the action
\begin{equation}
    S_{\rm EM} 
    = - \frac{1}{4} \int {\rm d}^4 x\ \sqrt{- \det g} \left( F^{\mu \nu} F_{\mu \nu} - \frac{\theta}{2} \epsilon^{\mu \nu \alpha \beta} F_{\mu \nu}F_{\alpha \beta} \right)
    \label{eq:Action EM contribution - Classical}
\end{equation}
where
\begin{equation}
    F_{\mu \nu} = \partial_\mu A_\nu - \partial_\nu A_\mu
\end{equation}
is the electromagnetic strength tensor and $A_\mu$ is the electromagnetic 4-vector potential.

Using the ADM decomposition, the action (\ref{eq:Action EM contribution - Classical}) can be written as
\begin{widetext}
\begin{eqnarray}
    S_{\rm EM}
    &=& \int {\rm d}^4 x\ \Bigg[ - \frac{1}{4} N \sqrt{\det q} \left( q^{a c} s^\alpha_c F_{a \nu} - n^\alpha F_{0 \nu} \right)
    \left( q^{b d} s^\nu_b F_{\alpha d} - n^\nu F_{\alpha 0} \right)
    - \frac{\theta}{2} \sqrt{\det{q}} \epsilon^{a b c} F_{t a} F_{b c}
    \Bigg]
    \nonumber\\
    &=& \int {\rm d}^4 x\ \Bigg[ - \frac{1}{4} N \sqrt{\det q} \left( - 2 q^{a d} F_{0 a} F_{0 d}
    + q^{a c} q^{b d} F_{a b} F_{c d} \right)
    - \frac{\theta}{2} \sqrt{\det{q}} \epsilon^{a b c} F_{t a} F_{b c}
    \Bigg]
    \,,
\end{eqnarray}
\end{widetext}
where $F_{0 \nu} = n^\mu F_{\mu \nu}$, $F_{t\nu} = t^\mu F_{\mu \nu}$, $\epsilon^{0123}=-1/\sqrt{- \det g}$, and $\epsilon^{123}=1/\sqrt{\det q}$.
Time derivatives appear only in the first and last terms through
\begin{eqnarray}
    F_{0 a} &=& N^{-1} \left( F_{t a} - N^b F_{b a} \right)
    \nonumber\\
    &=& N^{-1} \left( \dot{A}_a - \partial_a A_t - N^b F_{b a} \right)
    \,,
    \label{eq:Stength tensor/potential relation - classical}
\end{eqnarray}
and $F_{ta}$, and hence we get the densitized electric field as the conjugate momentum
\begin{eqnarray}
    E^a &=& \frac{\delta S_{\rm EM}}{\delta \dot{A}_a}
    = \sqrt{\det q} \left( q^{a b} F_{0 b}
    - \frac{\theta}{2} \epsilon^{a b c} F_{b c} \right)
    \nonumber\\
    &=:& \tilde{E}^a - \theta B^a
    \,,
    \label{eq:Electric field / strength tensor relation - classical}
\end{eqnarray}
where we have defined
\begin{equation}
    \tilde{E}^a = q^{a b} \sqrt{\det q} F_{0 a}
    \,,
    \label{eq:Emergent electric field - classical}
\end{equation}
and will refer to it as the physical densitized electric field, and $B^a$ is the densitized magnetic vector field
\begin{equation}
    B^a
    = \frac{\sqrt{\det q}}{2} \epsilon^{a b c} F_{b c}
    \,,
    \label{eq:Magnetic field / strength tensor relation}
\end{equation}
such that $B^a B_a = (\det q) F_{ab} F^{ab} / 2$.
Do not confuse the electric momentum components $E^a$ with the gravitational densitized triads $E^x$ and $E^\varphi$ of the spherically symmetric model in the previous Sections. Given the extensive content of this work, it is inevitable to repeat some symbols while remaining close to standard notations. The reader is encouraged to refer to Table~\ref{tab:Notation1} for a summary of the notation of the four-dimensional model and to Table~\ref{tab:Notation2} for that of the spherically symmetric case.

\begin{table}[!h]
\centering
\vspace{0.5cm}
    \caption{Notation for the four-dimensional system of Sections~\ref{sec:Emergent modified gravity}, \ref{Sec:Transformation of the electromagnetic vector potential}, \ref{sec:Em electric and symmetries - classical}, and \ref{sec: Emergent electromagnetism}.}
    \begin{tabular}{|p{6cm}|p{1cm}|}
     \hline
     \multicolumn{2}{|c|}{Gravitational and spacetime variables} \\
     \hline
     Gravitational field & $q_{ab}$ \\
     Lapse and shift & $N$, $N^a$ \\
     Gravitational gauge functions & $\epsilon^0$, $\epsilon^a$ \\
     Emergent spatial metric & $\tilde{q}_{ab}$ \\
     Auxiliary symmetric spatial tensor & $\bar{q}_{ab}$ \\
     \hline
     \multicolumn{2}{|c|}{Electromagnetic variables} \\
     \hline
     Electromagnetic vector potential & $A_a$ \\
     Electromagnetic scalar potential & $A_t$ \\
     Electric momentum & $E^a$ \\
     Densitized magnetic field & $B^a$ \\
     Electromagnetic gauge function & ${\cal A}_t$ \\
     Smearing constants of symmetry generator & ${\cal A}_a$ \\
     Emergent electric field & $\tilde{E}^a$ \\
     \hline
    \end{tabular}
    \label{tab:Notation1}
\end{table}

With the above definitions, the action can be rewritten in the form of a Legendre transformation,
\begin{widetext}
\begin{eqnarray}
    S_{\rm EM}
    &=& \int {\rm d}^4 x\ \Bigg[
    E^a \dot{A}_a
    - N \left( \frac{\left( E^a
    + \frac{\theta}{2} \sqrt{\det{q}} \epsilon^{a b c} F_{b c} \right) \left( E_a
    + \frac{\theta}{2} \sqrt{\det{q}} \tensor{\epsilon}{_a^b^c} F_{b c} \right)}{2 \sqrt{\det q}}
    + \frac{1}{4} \sqrt{\det q} F^{a b} F_{a b} \right)
    \nonumber\\
    &&\qquad\qquad
    - N^d \left( E^a + \frac{\theta}{2} \sqrt{\det{q}} \epsilon^{a b c} F_{b c} \right) F_{d a}
    - E^a \partial_a A_t
    \Bigg]
    \nonumber\\
    &=& \int {\rm d}^4 x\ \Bigg[
    E^a \dot{A}_a
    - N \left( \frac{\tilde{E}^a \tilde{E}_a}{2 \sqrt{\det q}}
    + \frac{\sqrt{\det q} F^{a b} F_{a b}}{4} \right)
    - N^d \tilde{E}^a F_{d a}
    + A_t \partial_a E^a
    \Bigg]
    \nonumber\\
    &=:&
    \int {\rm d}^4 x\ \left( E^a \dot{A}_a - N H^{\rm EM} - N^a H^{\rm EM}_a
    + A_t G^{\rm EM} \right)
    \,,
    \label{eq:Action EM contribution - ADM decomposition - Classical}
\end{eqnarray}
\end{widetext}
where an integration by parts is used to write the last term proportional to $A_t$, neglecting boundary terms.
Here, $H^{\rm EM}$, $H^{\rm EM}_a$, and $G^{\rm EM}$ are, respectively, the electromagnetic contributions to the Hamiltonian, vector, and Gauss constraints given by
\begin{eqnarray}
    \label{eq:Hamiltonian constraint EM contribution - Electromagnetism - classical}
    H^{\rm EM} &=&
    \frac{\tilde{E}^a \tilde{E}_a + B^a B_a}{2 \sqrt{\det q}}
    \,,\\
    \label{eq:Diffeomorphism constraint EM contribution - Electromagnetism - classical}
    H^{\rm EM}_a &=&
    \tilde{E}^b F_{a b}
    \,,\\
    \label{eq:Gauss constraint - Electromagnetism}
    G^{\rm EM} &=& \partial_a E^a = \partial_a \tilde{E}^a
    \,.
\end{eqnarray}
We note that the Gauss constraint can be written either with $\tilde{E}^a$ or $E^a$ because $\partial_a B^a=0$ by its definition (\ref{eq:Magnetic field / strength tensor relation}). We will come back to this ambiguity below.

The configuration variables of the electromagnetic field are given by $A_a$, and the conjugate momenta by the electric field $E^a$, such that
\begin{equation}
    \{ A_a(x) , E^b(y) \} = \delta_a^b \delta^3 (x-y)
    \,.
\end{equation}
The time component $A_t$ has no momenta since $\delta S_{\rm EM} / \delta \dot{A}_t = 0$, and hence it is a non-dynamical variable and appears as a Lagrange multiplier just as the lapse and shift do.

The full constraints $H = H^{\rm grav} + H^{\rm EM}$ and $H_a = H^{\rm grav}_a + H^{\rm EM}_a$, together with the Gauss constraint, satisfy the algebra
\begin{subequations}
\label{eq:Hypersurface deformation - Electromagnetism}
\begin{eqnarray}
    \{ \Vec{H} [ \Vec{N}] , \Vec{H} [ \Vec{M} ] \} \!\!&=&\!\! - \vec{H} [\mathcal{L}_{\Vec{M}} \Vec{N}]
    \,, \\
    \{ H [ N ] , \Vec{H} [ \Vec{N}]\} \!\!&=&\!\! - H [ N^b \partial_b N ]
    \,, \\
    \{ H [ N ] , H [ M ] \} \!\!&=&\!\! - \vec{H} [ q^{a b} ( M \partial_b N - N \partial_b M )]
    ,\qquad
    \label{eq:HH bracket - electromagnetism}
    \\
    \{ \vec{H} [ \vec{N} ] , G^{\rm EM} [ A_t ] \}
    \!\!&=&\!\! 0
    \,, \\
    \{ H [ N ] , G^{\rm EM} [ A_t] \} \!\!&=&\!\! 0
    \,,
    \label{eq:HG bracket - electromagnetism}
    \\
    \{ G^{\rm EM} [{\cal A}_{t1}] , G^{\rm EM} [{\cal A}_{t2}] \} \!\!&=&\!\! 0
    \,.
    \label{eq:GG bracket - electromagnetism}
\end{eqnarray}
\end{subequations}

The full Hamiltonian generating time evolution and gauge transformations is given by
\begin{equation}
    H [N , \vec{N} , A_t] = H[N] + \vec{H} [\vec{N}] - G^{\rm EM} [A_t]
    \,, \label{eq:Full Hamiltonian - Electromagnetism}
\end{equation}
and we denote an arbitrary gauge transformation of some phase-space function $\mathcal{O}$ by
\begin{equation}
    \delta_{\epsilon,{\cal A}_t} \mathcal{O} = \{ \mathcal{O} , H [\epsilon^0 , \vec{\epsilon} , {\cal A}_t] \}
    \,.
\end{equation}
Demanding that the equations of motion are gauge covariant requires that the lapse, the shift, and the electric potential transform as
\begin{eqnarray}
    \delta_{\epsilon,{\cal A}_t} N &=& \dot{\epsilon}^0 + \epsilon^a \partial_a N - N^a \partial_a \epsilon^0
    \,,
    \label{eq:Off-shell gauge transformations for lapse}
    \\
    \delta_{\epsilon,{\cal A}_t} N^a &=& \dot{\epsilon}^a + \epsilon^b \partial_b N^a - N^b \partial_b \epsilon^a
    \nonumber\\
    &&
    + q^{a b} \left(\epsilon^0 \partial_b N - N \partial_b \epsilon^0 \right)
    \,,
    \label{eq:Off-shell gauge transformations for shift}
    \\
    \delta_{\epsilon,{\cal A}_t} A_t &=& \dot{\cal A}_t
    \label{eq:Off-shell gauge transformations for EM scalar potential}
    \,.
\end{eqnarray}

In the transition to a modified theory we will preserve the form of the constraint algebra (\ref{eq:Hypersurface deformation - Electromagnetism}) and the classical form of the vector and Gauss constraint, while modifying only the Hamiltonian constraint and the structure functions $q^{a b}$ as reviewed in the previous section.
The covariance condition (\ref{eq:Covariance condition - general}) must be satisfied for all the spacetime tensor fields of the theory, one of which is the spacetime metric leading to the spacetime covariance condition (\ref{eq:Covariance condition of 3-metric - reduced - EMG});
we have, however, yet to find the correct spacetime tensor field associated to the electromagnetic field on which to impose the general condition (\ref{eq:Covariance condition - general}).
The two obvious options are the electromagnetic one-form $A_\mu$ and the strength tensor $F_{\mu \nu}$.
There are two reasons why the latter is the correct choice and not the former.
First, as we will show, the gauge transformations of $A_\mu$ do not correspond to its Lie derivatives even in the classical theory, while those of $F_{\mu\nu}$ do.
Second, it is $F_{\mu\nu}$ the one that enters observations via the Lorentz force equations and therefore it is the only one whose gauge transformations are required to be manifestly covariant.

\subsection{Transformation of the electromagnetic vector potential}
\label{Sec:Transformation of the electromagnetic vector potential}

Consider the classical electromagnetic vector constraint (\ref{eq:Diffeomorphism constraint EM contribution - Electromagnetism - classical}).
Its action on the spatial one-form given by the vector potential $A_a$ gives
\begin{equation}\label{eq:EM vector potential transformation}
    \delta_{\vec{\epsilon}} A_a = \{A_a , \vec{H}^{\rm EM}[\vec{\epsilon}]\} 
    = \epsilon^b F_{b a}
    = \epsilon^b \left(\partial_b A_a - \partial_a A_b\right)
    \,.
\end{equation}
On the other hand, a Lie derivative generated by a four-vector $\vec{\xi}=\vec{\epsilon}$ with purely spatial components gives
\begin{equation}\label{eq:EM vector potential transformation-Lie derivative}
    {\cal L}_{\vec{\xi}} A_a
    = \epsilon^b \partial_b A_a + A_b \partial_a \epsilon^b
    = \delta_{\vec{\epsilon}} A_a + \partial_a (\epsilon^b A_b)
    \,.
\end{equation}
We conclude that the gauge transformations of the vector potential do not correspond directly to Lie derivatives even in Maxwell's theory, at least at the local level.
The vector potential integrated over a path $\gamma$ in $\Sigma$ is still invariant up to boundary terms: given
\begin{equation}
    A[\gamma] \equiv \int_\gamma A = \int_{s_i}^{s_f} \frac{{\rm d} \gamma^a}{{\rm d} s} A_a {\rm d} s
    \,,
\end{equation}
its first-order spatial gauge transform is
\begin{eqnarray}
    A [\gamma] &\to& A [\gamma] + \delta_{\vec{\epsilon}} (A [\gamma])
    \nonumber\\
    &=&\tilde{A} [\tilde{\gamma}] - \int_s \partial_a (\epsilon^b A_b) {\rm d} x^a
    \nonumber\\
    &=& \tilde{A} [\tilde{\gamma}] - \epsilon^b A_b \big|_{s_i}^{s_f}
    \,,
\end{eqnarray}
where the tilde stands for the transformed quantities ($\tilde{\gamma}$ is simply the smearing curve in the new coordinates associated to the new gauge.)

We also note that, using the same vector constraint contribution (\ref{eq:Diffeomorphism constraint EM contribution - Electromagnetism - classical}), the spatial gauge transformation of the physical electric field is equivalent to the Lie derivative of a densitized spatial vector field,
\begin{eqnarray}
    \delta_{\vec{\epsilon}} \tilde{E}^a
    = \partial_c (\epsilon^c \tilde{E}^a) - \partial_b (\epsilon^a \tilde{E}^b)
    = \mathcal{L}_{\vec{\epsilon}} \tilde{E}^a
    \,,
\end{eqnarray}
but it is not the case for the electric momentum
\begin{eqnarray}
    \delta_{\vec{\epsilon}} E^a
    = \mathcal{L}_{\vec{\epsilon}} E^a 
    + \theta \partial_c (\epsilon^{a b c} \epsilon^d F_{a d})
    \,,
\end{eqnarray}
which receives a contribution from the $\theta$ term.

These two examples show how the gauge transformation of the fundamental phase space variables need not have a direct relationship to Lie derivatives.
However, despite the peculiar transformation properties of the phase-space variables, the theory is covariant because it is derived from the classical action.
This is the case because the physical manifestations of the gravitational and electromagnetic fields, namely, the spacetime metric and the strength tensor are in fact covariant in the usual sense that the gauge transformations do match Lie derivatives of a symmetric and an antisymmetric tensor of type $(0,2)$, respectively.
The covariance conditions must therefore be placed on these two tensors and not on the fundamental phase space variables.
In this classical setting, one can show that in fact $\delta_\epsilon F_{\mu\nu} |_{\rm O.S.} = \mathcal{L}_\xi F_{\mu \nu} |_{\rm O.S.}$:
Notice for instance that upon substitution of the spatial gauge transformation $\delta_{\vec \epsilon} A_a$ into $F_{ab}$ we obtain, using (\ref{eq:EM vector potential transformation-Lie derivative}), $\delta_{\vec \epsilon} F_{ab} = \partial_a \delta_{\vec \epsilon} A_b-\partial_b \delta_{\vec \epsilon} A_a=\partial_a \mathcal{L}_{\vec \xi} A_b-\partial_b \mathcal{L}_{\vec \xi} A_a=\mathcal{L}_{\vec \xi}F_{ab}$.
More generally, however, the covariance condition can be directly imposed on $F_{\mu\nu}$ without going through the transformation of the vector potential and this is the path we will take for our emergent field theory. Due to the complexity of such condition and to avoid repetition, we will perform its detailed analysis in Section~\ref{sec: Emergent electromagnetism}.

\subsection{Emergent electric field and classical symmetries}
\label{sec:Em electric and symmetries - classical}

Let us note that the electromagnetic effects can only be seen through its interaction with other forms of matter.
This is commonly done by analyzing the motion of a particle with an electric charge $e$ and mass $m$ in the presence of an electromagnetic field described by the Lorentz force equation
\begin{equation}\label{eq:Lorentz force}
    m \frac{{\rm d} u^\mu}{{\rm d} \tau} = e \tensor{F}{^\mu_\nu} u^\nu
    \,,
\end{equation}
where $\tau$ is the particle's proper time and $u^\mu$ is its four-velocity.
On the other hand, the components of the strength tensor can be written as
\begin{equation}
    F_{0a} = \frac{q_{a b} \tilde{E}^b}{\sqrt{\det q}}
    \quad , \quad
    F_{a b} = \partial_a A_b - \partial_b A_a
    \,.
\end{equation}
Therefore, it is $\tilde{E}^a$ that acts as the physical electric field, not the canonical momentum $E^a$.
This inequivalence between the phase space coordinate $E^a$ and the strength tensor component $\tilde{E}^a$ is what leads us to distinguish them based on their roles as a fundamental phase space variable and a physical electric field, respectively.
Because the latter can be seen as a composite function of the phase space, we may refer to it as the emergent electric field, while we refer to the former as simply the electric momentum\textemdash\,as we will show in Sec.~\ref{sec: Emergent electromagnetism}, the word \emph{emergent} is fitting because the relation between $\tilde{E}^a$ and $E^a$, including the $\theta$ term, can be derived from covariance conditions.
In the classical theory, experiments mostly deal only with the physical $\tilde{E}^a$ and cannot distinguish it from the fundamental momentum $E^a$.
However, once the quantum effects of the electromagnetic field are taken into account, the $\theta$ term (and therefore the difference between the emergent electric field and the fundamental electric momentum) can play a role in certain quantum phenomena in the context of topological insulators and the quantum Hall effect \cite{Simon,AxionEM}.
It is, therefore, important to study what the possible effects are in a modified electromagnetic theory where the distinction between fundamental and emergent electric fields is taken into account.

As a final point, recall that in EMG the derivative operator for parallel transport in spacetime is singled out as the one compatible with the emergent metric, $\nabla_\mu \tilde{g}_{\alpha \beta}=0$.
The analog in Maxwell electrodynamics is the derivative operator for parallel transport in the U(1) gauge group, which must be compatible with the electric field by satisfying $D_a E^a=\partial_a E^a=0$\textemdash\,recall that the U(1) group is Abelian and hence the electromagnetic potential does not enter this particular equation\textemdash\,, which can be identified as the Gauss constraint.
In electromagnetism with the $\theta$ term, however, the Gauss constraint (\ref{eq:Gauss constraint - Electromagnetism}) has the peculiarity that the U(1) covariant derivative is compatible with both the electric momentum and the emergent electric field, $\partial_a E^a = \partial_a \tilde{E}^a$, because $\partial_a B^a=0$.
If one obtains a modified theory where the relation between the fundamental and emergent electric fields is more complicated, such that $\partial_a E^a \neq \partial_a \tilde{E}^a$, then we have an ambiguity in how we impose the Gauss constraint: either on $E^a$ or on $\tilde{E}^a$.
In analogy with EMG, where the covariant derivative is compatible with the emergent metric, the latter is the natural choice.
Furthermore, defining the spacetime-densitized strength tensor $\bar{F}_{\mu \nu} = \sqrt{-\det g} F_{\mu \nu}$, recall that in the absence of electric charges one has the covariant conservation law
\begin{equation}\label{eq:Covariant conservation of EM current}
    \nabla_\mu \bar{F}^{\mu \nu} = \partial_\mu \bar{F}^{\mu \nu} = 0
    \,.
\end{equation}
The densitized strength tensor with raised indices has the components
\begin{eqnarray}
    \bar{F}^{t a} &=& - \tilde{E}^a
    \,,
    \\
    \bar{F}^{a b} &=& N \sqrt{\det q} F^{a b} + N^a \tilde{E}^b - N^b \tilde{E}^a
    \,,
\end{eqnarray}
where $F^{ab} = q^{a c} q^{b d} F_{c d}$.
The conservation law (\ref{eq:Covariant conservation of EM current}) can be explicitly written as
\begin{equation}
    \dot{\bar{F}}^{t \nu} + \partial_a \bar{F}^{a \nu} = 0
    \,.
\end{equation}
with the component $\nu=t$ being precisely the Gauss constraint on the emergent electric field
\begin{equation}
    \partial_a \tilde{E}^a = 0
    \,.
\end{equation}
It is therefore necessary to apply the Gauss constraint on the emergent electric field, rather than on the electric momentum, for the conservation law to hold for more general expressions of the emergent electric field.

For completeness, we evaluate the component $\nu=b$ of the conservation law,
\begin{equation}\label{eq:Current conservation condition}
    \dot{\tilde{E}}^a = - \partial_b \bar{F}^{a b}
    \,,
\end{equation}
This equation can be seen as the conservation of three Noether currents $(J^a)^\mu$ (labeled by the spatial index) with components
\begin{equation}
    (J^a)^t = \tilde{E}^a
    \quad , \quad (J^a)^b = - \bar{F}^{a b}
    \,.
\end{equation}
Therefore, we have three Noether charges $(J^a)^t = \tilde{E}^a$ given by the components of the emergent electric field.
Their role as Noether charges means that the phase-space functional
\begin{equation}\label{eq:EM symmetry generator}
    \vec{J}^t[\vec{\cal A}] = \int {\rm d}^3 x\ {\cal A}_a (J^a)^t
    = \int {\rm d}^3 x\ {\cal A}_a \tilde{E}^a
    \,,
\end{equation}
with arbitrary constants ${\cal A}_a$, commutes with the constraints,
\begin{equation}\label{eq:EM symmetry generator brackets}
    \{H[N], \vec{J}^t[\vec{\Lambda}] \}
    = \{\vec{H}[N], \vec{J}^t[\vec{\Lambda}] \}
    = \{ G^{\rm EM}[\Lambda], \vec{J}^t[\vec{\Lambda}] \}
    = 0\,.
\end{equation}
The third bracket is trivially satisfied since $\{\tilde{E}^a(x),\tilde{E}^b(y)\}=0$, while the vanishing of the first two brackets implies the conservation law (\ref{eq:Current conservation condition}) through the boundary terms
\begin{equation}
    \{(J^a)^t , H[N,\vec{N}] \}
    = \partial_b (J^a)^b
    \,.
\end{equation}
We recognize the Noether charge (\ref{eq:EM symmetry generator}) as the average physical electric field in the region of integration, weighted by the arbitrary constants ${\cal A}_a$\textemdash\,choosing ${\cal A}_a=\delta_{a,x}$, ${\cal A}_a=\delta_{a,y}$, and ${\cal A}_a=\delta_{a,z}$ implies three Noether charges that reconstruct the average physical electric field.
The conservation law (\ref{eq:Current conservation condition}) implies that, in the absence of electric charges, the average electric field can evolve only in the presence of a nontrivial magnetic field at the boundary:
$$\dot{\vec{J}}^{\,t}[\vec{\cal A}]=\int_{\partial\Sigma}{\rm d}^2x\; \hat{n}_b {\cal A}_a F^{ab}=\int_{\partial\Sigma}{\rm d}^2x\; \hat{n}_b {\cal A}_a \epsilon^{abc}B_c$$
where $\hat{n}_b$ is the unit normal to the boundary of the integration region $\Sigma$. If the boundary conditions are chosen such that the magnetic field is always normal to the boundary, then average physical electric field in the region of interest remains a constant of the motion.

Since we want to retain the classical symmetries in the modified theory for possibly more complicated expressions of the emergent field $\tilde{E}^a$, we must preserve the existence of the symmetry generator (\ref{eq:EM symmetry generator}) .

\subsection{Spherical symmetry}
\label{sec:EM spherical classical}

\subsubsection{Phase space}

We define spherical symmetry for electromagnetism as the scenario where the electromagnetic field cannot generate Lorentz forces in the angular directions.
This immediately implies that the magnetic field and the angular components of the (emergent) electric field must vanish.
Therefore, $F_{a b}=0$\textemdash\,which implies there is no $\theta$ term contribution and hence fundamental and emergent electric fields are identical\textemdash\,and ${\cal E}^\vartheta={\cal E}^\varphi=0$, such that the only non-trivial electromagnetic phase-space pair is
\begin{equation}
    \{A_x (x) , {\cal E}^x (y)\} = \delta (x - y)
    \ ,
\end{equation}
where we absorbed a factor of $\sqrt{4\pi}$ coming from the angular integration into $A_x$ and ${\cal E}^x$ for simplicity.
Here, we denote the electric field by ${\cal E}^a$ so as not to confuse it with the gravitational triads.
The reader is encouraged to refer to Table~\ref{tab:Notation2} for a summary of the notation used in the spherically symmetric model.

\begin{table}[!htb]
\centering
\vspace{0.5cm}
\caption{Notation for the spherically symmetric system of Sections~\ref{sec:Spherical EMG Vacuum}, \ref{sec:EM spherical classical}, \ref{sec:Emergent electromagnetism: Spherical symmetry}, and \ref{eq:BH sol}}
    \begin{tabular}{|p{6.3cm}|p{1.1cm}|}
     \hline
     \multicolumn{2}{|c|}{Gravitational and spacetime variables} \\
     \hline
     Gravitational momenta (densitized triads) & $E^x$, $E^\varphi$ \\
     Gravitational configuration variables & $K_x$, $K_\varphi$ \\
     Lapse and shift & $N$, $N^x$ \\
     Gravitational gauge functions & $\epsilon^0$, $\epsilon^x$ \\
     Emergent spatial metric (radial component) & $\tilde{q}_{xx}$ \\
     Auxiliary variable & $\bar{q}_{xx}$ \\
     \hline
     \multicolumn{2}{|c|}{Electromagnetic variables} \\
     \hline
     Electromagnetic vector potential & $A_x$ \\
     Electromagnetic scalar potential & $A_t$ \\
     Electric momentum & ${\cal E}^x$ \\
     Electromagnetic gauge function & ${\cal A}_t$ \\
     Smearing constant of symmetry generator & ${\cal A}_x$ \\
     Emergent electric field & $\tilde{\cal E}^x$ \\
     \hline
    \end{tabular}
    \label{tab:Notation2}
\end{table}

\subsubsection{Classical constraints and symmetries}

Because $F_{ab}=0$ in spherical symmetry, the electromagnetic vector constraint contribution (\ref{eq:Diffeomorphism constraint EM contribution - Electromagnetism - classical}) trivializes.
The Hamiltonian constraint contribution (\ref{eq:Hamiltonian constraint EM contribution - Electromagnetism - classical}) reduces to
\begin{equation}
    H^{\rm EM} =
    \frac{q_{xx} ({\cal E}^x)^2}{2 \sqrt{\det q}}
    = \frac{\sqrt{q_{xx}} ({\cal E}^x)^2}{2 E^x}
    = \frac{E^\varphi ({\cal E}^x)^2}{2 (E^x)^{3/2}}
    \ ,
    \label{eq:Hamiltonian constraint EM contribution - spherical - classical}
\end{equation}
where $\det q = q_{xx} (E^x)^2$,
while the Gauss constraint (\ref{eq:Gauss constraint - Electromagnetism}) reduces to
\begin{equation}
    G^{\rm EM} = ({\cal E}^x)'
    \ .
    \label{eq:Gauss constraint - spherical}
\end{equation}

In \cite{TibrewalaCharged,alonsobardaji2023Charged} it was suggested that an electromagnetic vector constraint contribution of the form $A_x ({\cal E}^x)'$ is put by hand so that $A_x$ gauge-transforms as a spatial one-form and ${\cal E}^x$ as a scalar.
Such new generator, $D_x=H_x^{\rm grav}+A_x ({\cal E}^x)'=H_x^{\rm grav}+A_x G^{\rm EM}$, results in a linear combination of the original vector constraint and the Gauss constraint and hence still vanishes on shell.
However, this is a mistake because the vector potential's gauge transformation does not correspond to a one-form's coordinate transformation even in the four dimensional Maxwell theory, see equation (\ref{eq:EM vector potential transformation}) which in fact requires $A_x$ be invariant to radial transformations.
This shows that the vector constraint $\vec{H}$ is not identical to a diffeomorphism generator $\vec{D}$ formed by the linear combination of the vector and Gauss constraints.
Furthermore, such diffeomorphism generator does not take place in the hypersurface deformation algebra of the full four dimensional theory and we shall therefore not include this addition to the vector constraint in the spherically symmetric system either.

The classical electro-gravity system in spherical symmetry has the mass observable
\begin{equation}
    \mathcal{M} \equiv \frac{\sqrt{E^x}}{2} \left( 1 + K_\varphi^2 - \left( \frac{(E^x)'}{2 E^\varphi} \right)^2
    - \frac{\Lambda E^x}{3}
    + \frac{({\cal E}^x)^2}{E^x}
    \right)
    \,,
    \label{eq:Electro-gravity mass - spherical - classical}
\end{equation}
replacing the classical limit of (\ref{eq:Vacuum mass observable - EMG}) of the vacuum system.

Furthermore, the electric field ${\cal E}^x$ is an observable too because it commutes with all the constraints.
This implies that ${\cal E}^x$ is a constant of the motion, and due to the Gauss constraint it is also spatially constant.
We therefore identify the dynamical value of ${\cal E}^x$ as the electric charge observable,
\begin{equation}
    Q = {\cal E}^x\,,
    \label{eq:Electro-gravity charge - spherical - classical}
\end{equation}
replacing the role of the symmetry generator of the four dimensional theory (\ref{eq:EM symmetry generator}).

\section{Emergent electromagnetism}
\label{sec: Emergent electromagnetism}

\subsection{The strength tensor and the emergent electric field}

Based on our discussion of the classical theory with the $\theta$ term, we make a distinction between the fundamental electric momentum $E^a$ and the emergent electric field $\tilde{E}^a$. We do not assume an a-priori relation between the two, rather we want to derive it.
We assume that the electromagnetic vector constraint is unmodified, but still dependent on the emergent electric field, hence taking the form (\ref{eq:Diffeomorphism constraint EM contribution - Electromagnetism - classical}).
Also, following the past section on EMG, we will make a distinction between the gravitational field, denoted here by $q_{a b}$, and the inverse of the structure function, denoted by $\tilde{q}_{a b}$, the latter playing the role of the spatial components of the emergent spacetime metric.
As explained before, the covariance conditions single out the emergent spacetime metric as the only covariant spacetime metric tensor candidate and hence EMG is not a bi-metric theory.
However, when restricted to spatial hypersurfaces we do have two different spatial metric candidates, the gravitational one and the emergent one, because both are covariant under spatial diffeomorphisms, but only the latter has a spacetime extension.

Mimicking the classical relations, we will postulate the relation between the emergent electric field and the field strength tensor, given by
\begin{eqnarray}
    F_{0a} &=& \frac{\bar{q}_{a b} \tilde{E}^b}{\sqrt{\det \bar{q}}}
    \ , 
    \label{eq:Emergent strength field tensor relation to electromagnetic field} \\
    F_{a b} &=& \partial_a A_b - \partial_b A_a
    \ ,
    \label{eq:Emergent strength field tensor magnetic relation to electromagnetic vector potential}
\end{eqnarray}
such that the full tensor is given by $F_{\mu \nu}$ with $F_{0 a} = n^\mu s^\nu_a F_{\mu \nu}$ and $F_{a b} = s^\mu_a s^\nu_b F_{\mu \nu}$.
If magnetic monopoles are considered, the relation (\ref{eq:Emergent strength field tensor magnetic relation to electromagnetic vector potential}) must include them, but we will neglect such possibility.
Here we have used the auxiliary tensor field $\bar{q}_{a b}$ because we so far have an ambiguity of what spatial tensor we must use, either the gravitational field $q_{a b}$ or the emergent spatial metric $\tilde{q}_{a b}$, and we use the auxiliary $\bar{q}_{ab}$ whenever the ambiguity is present.
Therefore, we are presented with four different schemes to define the previous relation.
Scheme 1 is given by
\begin{equation}
    F_{0a} = q_{a b} \tilde{E}^b / \sqrt{\det q}
    \,,
    \label{eq:Emergent strength field tensor relation to electromagnetic field - scheme 1}
\end{equation}
scheme 2 is given by
\begin{equation}
    F_{0a} = q_{a b} \tilde{E}^b / \sqrt{\det \tilde{q}}
    \,,
    \label{eq:Emergent strength field tensor relation to electromagnetic field - scheme 2}
\end{equation}
scheme 3 is given by
\begin{equation}
    F_{0a} = \tilde{q}_{a b} \tilde{E}^b / \sqrt{\det q}
    \,, 
    \label{eq:Emergent strength field tensor relation to electromagnetic field - scheme 3}
\end{equation}
and scheme 4 is given by
\begin{equation}
    F_{0a} = \tilde{q}_{a b} \tilde{E}^b / \sqrt{\det \tilde{q}}
    \,.
    \label{eq:Emergent strength field tensor relation to electromagnetic field - scheme 4}
\end{equation}

The possibility of $\bar{q}_{ab}$ being degenerate, since it may not be a strict metric tensor, raises a problem in the formulation of the different schemes.
This is a complicated problem that cannot be fully addressed until the Hamiltonian and the dynamics are known, at which point we could use the potential degeneracy of either scheme to prefer one over the other in an explicit model.
On the other hand, the canonical dynamics rely only on the fundamental phase space and the Hamiltonian, and not necessarily on the emergent fields. Therefore, if the former remain well-defined in the regions with a degenerate emergent metric, the canonical evolution of the fundamental fields can be propagated throughout.
If that is the case, then the emergent spacetime and strength tensor would then be undefined in such regions, but the fundamental gravitational and electromagnetic variables may remain well-defined, although this must be checked in the particular system at hand once the Hamiltonian and the dynamics are known. An example where this is indeed the case is the following. Consider the gravitational collapse of dust coupled to EMG \cite{EMGPF}, whose dynamical solution consists on the collapse of the matter towards a minimum radius surface beyond which the solutions bounces back; in this case, the minimum radius surface has a singular geometry, as indicated by the divergence of some components of the emergent metric as well as the Ricci scalar, but the fundamental phase space variables, composed by the gravitational field and the dust, remain regular and hence the canonical dynamics can be evolved through the singular minimum radius surface.
However, as we will see later, the electromagnetic covariance conditions will single out scheme 4 as the preferred one for the four dimensional theory, and scheme 1 for the spherically symmetric model, leaving no room for speculations on which one to take.
While the discrepancy of the favored scheme between the four dimensional and spherically symmetric theories might seem like an inconsistency, we refer to the discussion of Section~\ref{sec:Symm red vs dim red} to argue that the introduction of nonlocality in the spherically symmetric model, in contrast to the four dimensional theory which is fully local, might make this discrepancy possible.

\subsection{Electromagnetic covariance conditions}

The covariance condition for the strength tensor reads
\begin{equation}
    \delta_\epsilon F_{\mu\nu} |_{\text{O.S.}} = \mathcal{L}_\xi F_{\mu \nu} |_{\text{O.S.}}
    \,,
    \label{eq:Covariance condition - EM}
\end{equation}
where we have suppressed the ${\cal A}_t$ subscript in the gauge transformation operator because we will impose U(1) invariance of the strength tensor in Subsection~\ref{sec:U(1)-gauge-invariance and symmetry generator} and hence cannot contribute to (\ref{eq:Covariance condition - EM}).
We start by performing the ADM decomposition of the right-hand-side,
\begin{widetext}
\begin{eqnarray}
    \mathcal{L}_\xi F_{\mu \nu} &=&
    \xi^\alpha \partial_\alpha F_{\mu \nu} - 2 F_{\alpha [ \mu} \partial_{\nu ]} \xi^\alpha
    \notag\\
    &=&
    \frac{\epsilon^0}{N} \dot{F}_{\mu \nu}
    + \left( \epsilon^c - \frac{\epsilon^0}{N} N^c \right) \partial_c F_{\mu \nu}
    - 2 \left( F_{t [ \mu} - N^c F_{c [ \mu} \right) \partial_{\nu ]} \left( \frac{\epsilon^0}{N} \right)
    - 2 F_{c [ \mu} \left( \partial_{\nu ]} \epsilon^c - \frac{\epsilon^0}{N} \partial_{\nu ]} N^c \right)
    \ ,
\end{eqnarray}
where the non-vanishing components are, explicitly,
\begin{eqnarray}\label{eq:Lie derivative magnetic components}
    \mathcal{L}_\xi F_{a b} &=&
    \frac{\epsilon^0}{N} \left( \dot{F}_{a b}
    + 2 F_{0 [ a} \partial_{b]} N
    - \partial_a (N^c F_{c b}) + \partial_b (N^c F_{c a}) \right)
    - 2 F_{0 [ a} \partial_{b]} \epsilon^0
    + \partial_a (\epsilon^c F_{c b}) - \partial_b (\epsilon^c F_{c a})
    \,,\\
    \mathcal{L}_\xi F_{t b} &=&
    \epsilon^0 \dot{F}_{0 b}
    + F_{0 b} \delta_\epsilon N
    + F_{c b} \delta_\epsilon N^c
    - F_{c b} \tilde{q}^{c a} \left(\epsilon^0 \partial_a N - N \partial_a \epsilon^0 \right)
    \nonumber\\
    &&+ \left( N \epsilon^c
    - \epsilon^0 N^c \right) \partial_c F_{0 b}
    + F_{0 c} \left( N \partial_{b} \epsilon^c
    - \epsilon^0 \partial_{b} N^c \right)
    + N^a \mathcal{L}_\xi F_{a b}
    \,,
    \label{eq:Lie derivative electric components}
\end{eqnarray}
\end{widetext}
where $\delta_\epsilon N$ and $\delta_\epsilon N^c$ are given by (\ref{eq:Off-shell gauge transformations for lapse}) and (\ref{eq:Off-shell gauge transformations for shift}).
Furthermore, using $\delta_\epsilon F_{t a} = N \delta_\epsilon F_{0 a} + N^b \delta_\epsilon F_{b a} + F_{0 a} \delta_\epsilon N + F_{b a} \delta_\epsilon N^b$ and plugging all the above into (\ref{eq:Covariance condition - EM}) we can separate the equations into spatial and normal contributions, the former given by setting $N=0=\epsilon^0$ and the latter by setting $N^a=0=\epsilon^a$.
Without loss of generality (because all components of $(N,N^a)$ and $(\epsilon^0,\epsilon^a)$ are independent of each other) we will treat them separately in the following subsections.
We will also separate (\ref{eq:Covariance condition - EM}) into magnetic ($\delta_\epsilon F_{ab}$) and electric ($\delta_\epsilon F_{0a}$) components for further clarity.

\subsubsection{Magnetic normal conditions: The emergence of the electric field}

Using the Lie derivative expressions (\ref{eq:Lie derivative magnetic components}) and (\ref{eq:Lie derivative electric components}), the normal contribution to the covariance condition (\ref{eq:Covariance condition - EM}) for the magnetic components simplifies to
\begin{eqnarray}
    &&\frac{1}{\epsilon^0} \{ F_{a b} , \tilde{H} [\epsilon^0] \}
    + 2 F_{0 [ a} \frac{\partial_{b]}\epsilon^0}{\epsilon^0}
    \bigg|_{\rm O.S.}
    \nonumber\\
    &&\qquad\qquad\qquad
    =
    \frac{1}{N} \{ F_{a b} , \tilde{H} [N] \}
    + 2 F_{0 [ a} \frac{\partial_{b]} N}{N}
    \bigg|_{\rm O.S.}
    \qquad
    \label{eq:Normal F_ab cov cond - EM - Full}
\end{eqnarray}
It is important to note that the emergent $\tilde{q}^{a b}$ is the one that directly appears in the covariance conditions raising the spatial indices, see (\ref{eq:Lie derivative magnetic components}) and (\ref{eq:Lie derivative electric components}).
This is a direct consequence of the appearance of $\tilde{q}^{a b}$ in the gauge transformation of the shift (\ref{eq:Off-shell gauge transformations for shift}).
This shows that the covariance conditions favor the schemes 3 and 4 of the relation between the emergent electric field and the strength tensor over schemes 1 and 2.
Consequently, in the following we will narrow our focus to schemes 3 and 4.

The gauge transformation of the spatial strength tensor components takes the generic form
\begin{eqnarray}
    \{ F_{a b} , \tilde{H} [\epsilon^0] \} &=& 2 \mathcal{F}_{a b} \epsilon^0
    - 2 \mathcal{F}_{[a} \partial_{b]} \epsilon^0
    - 2 \mathcal{F}_{[a}^{c_1}  \partial_{b]} \partial_{c_1} \epsilon^0
    \nonumber\\
    &&
    - 2 \mathcal{F}_{[a}^{c_1 c_2} \partial_{b]} \partial_{c_1} \partial_{c_2} \epsilon^0
    - \dotsi
    \nonumber\\
    &&
    - 2 \mathcal{F}_{[a}^{c_1 \dotsi c_{n-1}} \partial_{b]} \partial_{c_1} \dotsi \partial_{c_{n-1}} \epsilon^0
    \,,
    \label{eq:Generic normal transformation of strength tensor - magnetic components}
\end{eqnarray}
for some phase space functions $\mathcal{F}$, where $n$ is the highest derivative order considered in the constraint.
Plugging this into the magnetic normal condition (\ref{eq:Normal F_ab cov cond - EM - Full}) and requiring it to hold for arbitrary $N$ and $\epsilon^0$, it implies the series of conditions
\begin{eqnarray}
    \mathcal{F}_{a} &=& F_{0 a}
    \,,
    \label{eq:Magnetic normal condition - first order}
    \\
    \mathcal{F}_{a}^{c_1} &=& \mathcal{F}_{a}^{c_1 c_2} = \dotsi = \mathcal{F}_{a}^{c_1 \dotsi c_{n-1}} = 0
    \,.
    \label{eq:Magnetic normal condition - Higher orders}
\end{eqnarray}

The higher-order terms in the transformation (\ref{eq:Generic normal transformation of strength tensor - magnetic components}) are given by 
\begin{eqnarray}\label{eq:Highest F}
    \mathcal{F}_{a}^{c_1 \dotsi c_{n-1}}
    &=&
    (-1)^{n} \frac{\partial \tilde{H}}{\partial (\partial_{c_1} \dotsi \partial_{c_{n-1}} E^{a})}
    \,,\\
    \label{eq:Second highest F}
    \mathcal{F}_{a}^{c_1 \dotsi c_{n-2}}
    &=&
    (-1)^{n-1} \frac{\partial \tilde{H}}{\partial (\partial_{c_1} \dotsi \partial_{c_{n-2}} E^{a})}
    \\
    &&
    + (-1)^{n} \partial_{c_{n-1}}\left(\frac{\partial \tilde{H}}{\partial (\partial_{c_1} \dotsi \partial_{c_{n-2}} E^{a})}\right)
    \,,\nonumber
\end{eqnarray}
and so on.
The vanishing of the last term in the higher-order condition (\ref{eq:Magnetic normal condition - Higher orders}) requires the vanishing of the right-hand-side of (\ref{eq:Highest F}), which in turn implies the vanishing of the second line in (\ref{eq:Second highest F}); using this, the vanishing of the second highest order term in (\ref{eq:Magnetic normal condition - Higher orders}), $\mathcal{F}_{a}^{c_1\dotsi c_{n-2}}=0$, implies that the first line of the right-hand-side of (\ref{eq:Second highest F}) must vanish independently.
Iterating this argument for the rest of the higher derivative terms up to ${\cal F}^{c_1}_{a}$ results in the series of conditions
\begin{eqnarray}
    &&\frac{\partial \tilde{H}}{\partial (\partial_{c_1} E^{a})} = \frac{\partial \tilde{H}}{\partial (\partial_{c_1} \partial_{c_2} E^{a})} = \dotsi
    \nonumber\\
    &&\qquad\qquad\qquad
    = \frac{\partial \tilde{H}}{\partial (\partial_{c_1} \dotsi \partial_{c_{n-1}} E^{a})} = 0\,.
\end{eqnarray}
We conclude that the Hamiltonian constraint cannot depend on spatial derivatives of the electric momentum.
Using this, the first-order term in (\ref{eq:Generic normal transformation of strength tensor - magnetic components}) is given by
\begin{equation}
    \mathcal{F}_{a}
    =
    \frac{\partial \tilde{H}}{\partial E^{a}}
    \,.
    \label{eq:Generic normal transformation of strength tensor - magnetic components - first order explicit}
\end{equation}
Substitution into the first-order equation (\ref{eq:Magnetic normal condition - first order}) implies
\begin{equation}
    F_{0 a} = \frac{\partial \tilde{H}}{\partial E^{a}}
    \,,
    \label{eq:Magnetic normal condition - simplified}
\end{equation}
which, using (\ref{eq:Emergent strength field tensor relation to electromagnetic field}), determines the expression for the emergent electric field derived from the Hamiltonian constraint,
\begin{equation}
    \tilde{E}^{a}
    =
    \sqrt{\det \bar{q}} \tilde{q}^{a b} \frac{\partial \tilde{H}}{\partial E^{b}}
    \,,
    \label{eq:Magnetic normal condition - Emergent electric field}
\end{equation}
where $\bar{q}_{ab}$ is used in the determinant in accordance to schemes 3 and 4.

Before treating the electric normal condition, it will be useful to first impose the spatial conditions.

\subsubsection{Spatial conditions}

The spatial covariance conditions for the magnetic and electric components respectively become
\begin{eqnarray}
    \{F_{ab} , \vec{H}[\vec{\epsilon}]\} &=& 
    \partial_a (\epsilon^c F_{c b}) - \partial_b (\epsilon^c F_{c a})
    \,,
    \label{eq:Spatial F_ab cov cond - EM - Full}
    \\
    \{F_{0b} , \vec{H}[\vec{\epsilon}]\} &=& 
    \epsilon^c \partial_c F_{0 b}
    + F_{0 c} \partial_{b} \epsilon^c
    \,.
    \label{eq:Spatial F_0a cov cond - EM - Full}
\end{eqnarray}
Using an unmodified (up to the emergent electric field) vector constraint of the form (\ref{eq:Gauss constraint - Electromagnetism}) ($H_c=\tilde{E}^d F_{c d}$), the spatial covariance conditions reduce to
\begin{eqnarray}
    &&\partial_{[a} \left( \int {\rm d}^3 z\ \frac{\delta \tilde{E}^d(x)}{\delta E^{b]}(z)} \epsilon^c(x) F_{d c}(x) \right) =
    \partial_{[a} (\epsilon^c F_{b] c})
    \,,\qquad \\
    &&\tilde{E}^d \left\{ \frac{\tilde{q}_{b d}}{\sqrt{\det \bar{q}}} , \vec{H}[\vec{\epsilon}] \right\}
    + \frac{\tilde{q}_{b d}}{\sqrt{\det \bar{q}}} \left\{ \tilde{E}^d , \vec{H}[\vec{\epsilon}] \right\}
    \nonumber\\
    &&\qquad\qquad\qquad
    = 
    \epsilon^c \partial_c \left(\frac{\tilde{q}_{b d} \tilde{E}^d}{\sqrt{\det \bar{q}}}\right)
    + \frac{\tilde{q}_{c d} \tilde{E}^d}{\sqrt{\det \bar{q}}} \partial_{b} \epsilon^c
    \,.
\end{eqnarray}
The first condition requires $\delta \tilde{E}^a(x)/\delta E^b(y) = \delta^3 (x-y)$, therefore
\begin{eqnarray}
    \frac{\partial \tilde{E}^a}{\partial E^b} &=& \delta^a_b
    \,, \nonumber\\
    \frac{\partial \tilde{E}^a}{\partial (\partial_{c_1} E^b)} &=& \frac{\partial \tilde{E}^a}{\partial (\partial_{c_1} \partial_{c_2} E^b)} = \dotsi = 0
    \,.
    \label{eq:EM covariance condition - spatial diffeomorphism - 1}
\end{eqnarray}
We use this to define the emergent electric field ansatz
\begin{equation}
    \tilde{E}^a = E^a + \theta \tilde{B}^a
    \,,
    \label{eq:Emergent electric field ansatz}
\end{equation}
where $\theta$ is a constant and $\tilde{B}^a$ is independent of the electric momentum $E^a$ but otherwise an undetermined function of the phase-space.
Using that the auxiliary field $\bar{q}_{a b}$ (either it being the emergent or the gravitational spatial metric) is a density of weight two, the second condition becomes
\begin{eqnarray}\label{eq:Emergent E denisitized vector}
    &&\left\{ \tilde{E}^a , \vec{H}[\vec{\epsilon}] \right\}
    =
    \partial_c (\epsilon^c \tilde{E}^a)
    - \partial_{c} (\epsilon^a \tilde{E}^c)
    \,,
\end{eqnarray}
that is, the emergent electric field must be a densitized spatial vector.
Using (\ref{eq:EM covariance condition - spatial diffeomorphism - 1}) this is possible only if
\begin{eqnarray}
    \label{eq:EM covariance condition - spatial diffeomorphism - reduced}
    &&\left\{ \tilde{B}^a (x) , \vec{H}^{\rm grav}[\vec{\epsilon}] \right\}
    \\
    &&\qquad\qquad
    + \int {\rm d}^3  z\ \epsilon^c (z) F_{cd} (z) \left\{ \tilde{E}^a (x) , \tilde{E}^d (z) \right\}
    = 0
    \,.\nonumber
\end{eqnarray}
If the emergent electric field locally commutes with itself, then the function $\tilde{B}^a$, and hence $\tilde{E}^a$, must be independent of the gravitational variables.

\subsubsection{Electric normal condition}

Setting $N^a = 0 = \epsilon^a$ in the covariance condition (\ref{eq:Covariance condition - EM}) we obtain the electric normal condition,
\begin{eqnarray}
    &&\!\!\!\!\!\!
    \frac{1}{\epsilon^0} \{ F_{0 b} , \tilde{H} [\epsilon^0] \}
    - F_{c b} \tilde{q}^{c a} \frac{\partial_a \epsilon^0}{\epsilon^0}
    \bigg|_{\rm O.S.}
    \\
    &&\qquad\qquad=
    \frac{1}{N} \{ F_{0 b} , \tilde{H} [N] \}
    - F_{c b} \tilde{q}^{c a} \frac{\partial_a N}{N}
    \bigg|_{\rm O.S.}
    .\nonumber
    \label{eq:Normal cov cond - EM - Full}
\end{eqnarray}
Using the relations (\ref{eq:Emergent strength field tensor relation to electromagnetic field}), (\ref{eq:Emergent strength field tensor magnetic relation to electromagnetic vector potential}), and (\ref{eq:Magnetic normal condition - simplified}) to write the strength tensor in terms of the emergent electric field and the vector potential, the electric condition becomes
\begin{eqnarray}
    &&\!\!\!\!\!\!
    \left\{ \frac{\partial \tilde{H}}{\partial E^a} , \tilde{H} [\epsilon^0] \right\} \frac{1}{\epsilon^0}
    - F_{a b} \tilde{q}^{b c} \frac{\partial_c \epsilon^0}{\epsilon^0}
    \bigg|_{\rm O.S.}
    \\
    &&\qquad\qquad\qquad=
    \left\{ \frac{\partial \tilde{H}}{\partial E^a} , \tilde{H} [N] \right\} \frac{1}{N}
    - F_{a b} \tilde{q}^{b c} \frac{\partial_c N}{N}
    \bigg|_{\rm O.S.}
    .\nonumber
\end{eqnarray}
Because this must be satisfied for arbitrary $\epsilon^0$ and $N$, it implies the series of equations
\begin{equation}
    \frac{\partial \{ \partial \tilde{H}/\partial E^{a} , \tilde{H} [\epsilon^0]\}}{\partial (\partial_{b_1} \partial_{b_2} \epsilon^0)} =
    \frac{\partial \{ \partial \tilde{H}/\partial E^{a} , \tilde{H} [\epsilon^0]\}}{\partial (\partial_{b_1} \partial_{b_2} \partial_{b_3} \epsilon^0)} = \dotsi =
    0
    \,,
    \label{eq:Electric covariance condition - higher orders - reduced form}
\end{equation}
and
\begin{equation}
    \frac{\partial \left( \{ \partial \tilde{H}/\partial E^{a} , \tilde{H} [\epsilon^0]\} \right)}{\partial (\partial_{c} \epsilon^0)} \tilde{q}_{c b} = 
    \tensor{F}{_a_b}
    \,.
    \label{eq:Electric covariance condition - first order - reduced form}
\end{equation}

\subsection{U(1) invariance and the symmetry generator}\label{sec:U(1)-gauge-invariance and symmetry generator}

U(1) gauge invariance is the notion that physical observables are invariant under the U(1) gauge transformations.
This requires that not only the equations of motion are U(1)-invariant, which is achieved by the commutation of the Gauss constraint with the Hamiltonian and vector constraint, but also that the physical variables, here the emergent metric $\tilde{g}_{\mu \nu}$ and the strength tensor $F_{\mu \nu}$, commute with the Gauss constraint.
We will apply the former below, while the latter readily gives the conditions
\begin{eqnarray}
    \{ \tilde{q}^{a b} , G^{\rm EM} [{\cal A}_t] \} &=& 0
    \,, \label{eq:U(1)-invariance - q}
    \\
    \{ \tilde{E}^{a} , G^{\rm EM} [{\cal A}_t] \} &=& 0
    \,, \label{eq:U(1)-invariance - E}
    \\
    \{ F_{ab} , G^{\rm EM} [{\cal A}_t] \} &=& 0
    \,. \label{eq:U(1)-invariance - F}
\end{eqnarray}
Using the expression (\ref{eq:Emergent electric field ansatz}) we find that the third bracket is realized automatically.
We will come back to the other two brackets later.

Imposing the existence of the symmetry generator $\vec{J}^t[\vec{{\cal A}}]$ given by (\ref{eq:EM symmetry generator}) is, unlike the gauge symmetry, only a symmetry of the equations of motion and hence only requires the brackets (\ref{eq:EM symmetry generator brackets}) using the modified $\tilde{H}$ and $\tilde{E}^a$.
The commutation of the symmetry generator with the Gauss constraint is implied by (\ref{eq:U(1)-invariance - E}).
The commutation of the symmetry generator with the vector constraint is guaranteed by the emergent electric field being a densitized spatial vector (\ref{eq:Emergent E denisitized vector}) and hence discarded as a boundary term.
The commutation with the Hamiltonian constraint is nontrivial and we will come back to it below.

Going back to U(1)-invariance, the vector constraint commutes with the Gauss constraint because of (\ref{eq:U(1)-invariance - E}) and (\ref{eq:U(1)-invariance - F}).
The commutation of the Gauss constraint with itself gives
\begin{eqnarray}
    &&\!\!\!\!\!\!\!\!\!\!
    \{G^{\rm EM} [{\cal A}_{t1}],G^{\rm EM} [{\cal A}_{t2}]\}
    \\
    &&\quad
    = \int {\rm d}^3 x{\rm d}^3 y\ \frac{\partial {\cal A}_{t1} (x)}{\partial x^a}\frac{\partial {\cal A}_{t2} (y)}{\partial y^b} \left\{ \tilde{E}^a (x) , \tilde{E}^b (y) \right\}
    = 0
    \,,\nonumber
\end{eqnarray}
where we have taken integrations by parts to place the spatial derivatives on the ${\cal A}_{ti}$ functions and neglected boundary terms.
This must hold for arbitrary ${\cal A}_{t1} (x)$ and ${\cal A}_{t2} (x)$, which implies that
\begin{equation}\label{eq:Commutative emergent electric field}
    \left\{ \tilde{E}^a (x) , \tilde{E}^b (y) \right\} = 0
    \,.
\end{equation}
Using this, the spatial condition (\ref{eq:EM covariance condition - spatial diffeomorphism - reduced}) requires $\tilde{B}^a$ be independent of the gravitational variables, and it makes the bracket (\ref{eq:U(1)-invariance - E}) trivial.
As with the symmetry generator, the commutation of the Gauss constraint with the Hamiltonian constraint is nontrivial and we will come back to it below.

First, let us reflect on the function $\tilde{B}^a$ in the expression of the emergent electric field (\ref{eq:Emergent electric field ansatz}), which classically reproduces the $\theta$ term.
Because $\tilde{B}^a$ depends on $F_{ab}$ and its derivatives, the only way to have the correct indices while keeping it independent of the gravitational variables is through the combination $\sqrt{\det \bar{q}} \epsilon^{abc} F_{bc}$, which is the densitized magnetic field when $\bar{q}_{ab}$ is the spatial metric and $\epsilon^{abc}$ is the spatial volume element associated to $\bar{q}_{ab}$ with its indices raised by the $\bar{q}^{ab}$.
In the context of EMG and the possibility discussed above of densitizing and raising/lowering indices with either the gravitational or the emergent spatial metric, we seem to have an ambiguity of choice here.
However, as we just concluded, the normal covariance conditions prefer the use of the emergent spatial metric for raising and lowering indices.
We therefore have $\epsilon^{abc}=\tilde{q}^{a a_1} \tilde{q}^{bb_1} \tilde{q}^{cc_1} \epsilon_{a_1b_1c_1}$, such that the combination $\sqrt{\det \bar{q}} \epsilon^{abc} F_{bc}$ is independent of the gravitational variables only if $\sqrt{\det \bar{q}}=\sqrt{\det \tilde{q}}$.
We find that the covariance conditions prefer the use of the scheme 4 where the relation (\ref{eq:Emergent strength field tensor relation to electromagnetic field}) between the emergent electric field and the strength tensor makes use of the emergent spatial metric both for raising and lowering indices and for the densitization.
Hence, in the following we will restrict ourselves to the use of the emergent spatial metric for these two purposes.

Because now it is the spatial metric used in the determinant of the expression (\ref{eq:Magnetic normal condition - Emergent electric field}), and since it satisfies the spacetime covariance condition (\ref{eq:Covariance condition of 3-metric - reduced - EMG}), we can write the electric covariance conditions (\ref{eq:Electric covariance condition - higher orders - reduced form}) and (\ref{eq:Electric covariance condition - first order - reduced form}) in terms of $\tilde{E}^a$ itself.
The transformation of the emergent electric field takes the generic form
\begin{equation}\label{eq:Emergent E generic normal transformation}
    \!\!\!\!
    \{ \tilde{E}^a , \tilde{H} [\epsilon^0] \} = \tilde{\cal E}^a \epsilon^0 + \tilde{\cal E}^{a b} \partial_b \epsilon^0
    + \tilde{\cal E}^{a b_1 b_2} \partial_{b_1} \partial_{b_2} \epsilon^0
    + \dotsi
    \,,
\end{equation}
where the $\tilde{\cal E}$ are some, yet undetermined, functions of the phase space\textemdash\,these are not related to the variables of the spherical symmetric model of previous sections, see Table~\ref{tab:Notation1} to refer to the symbols used in the present section.
In terms of these functions, the vanishing of the higher-order terms (\ref{eq:Electric covariance condition - higher orders - reduced form}) and the first-order equation (\ref{eq:Electric covariance condition - first order - reduced form}) respectively translate into the conditions
\begin{equation}
    \tilde{\cal E}^{a b_1 b_2} =
    \tilde{\cal E}^{a b_1 b_2 b_3} = \dotsi =
    0
    \ ,
    \label{eq:Normal electric covariance condition - reduced - higher orders}
\end{equation}
and
\begin{equation} \label{eq:Normal electric covariance condition - reduced - first order}
    \tilde{\cal E}^{a b} = 
    \sqrt{\det \tilde{q}} \tensor{F}{^a^b}
    \ .
\end{equation}
Using this, the bracket between the symmetry generator (\ref{eq:EM symmetry generator}) and the Hamiltonian constraint implies, after an integration by parts,
\begin{equation}
    \int {\rm d}^3 x\; {\cal A}_a \left( \tilde{\cal E}^a - \partial_b \tilde{\cal E}^{a b} \right) \epsilon^0
    = 0
    \,.
\end{equation}
Because this must hold for arbitrary ${\cal A}_a$ and $\epsilon^0$ we obtain
\begin{equation}\label{eq:Symmetry generator condition}
    \tilde{\cal E}^a = \partial_b \tilde{\cal E}^{a b}
    = \partial_b \left(\sqrt{\det \tilde{q}} \tensor{F}{^a^b}\right)
    \,.
\end{equation}

Finally, the U(1)-invariance of the Hamiltonian constraint is achieved by preserving the bracket $\{ \tilde{H} [ \epsilon^0 ] , G^{\rm EM} [{\cal A}] \}=0$, and we now show explicitly that its vanishing is automatically satisfied by the previous conditions:
\begin{eqnarray}
    &&\!\!\!\!\!\!\!\!\!
    \{ \tilde{H} [ \epsilon^0 ] , G^{\rm EM} [{\cal A}_t] \} =
    - \int {\rm d}^3x\ {\cal A}_t \partial_a \left(\{ \tilde{E}^a , \tilde{H} [\epsilon^0] \}\right)
    \quad
    \\
    &&\qquad\qquad\qquad
    =
    - \int {\rm d}^3x\ {\cal A}_t \partial_a \left(\tilde{\cal E}^a \epsilon^0 + \tilde{\cal E}^{a b} \partial_b \epsilon^0\right)
    \nonumber\\
    &&\qquad\qquad\qquad
    =
    - \int {\rm d}^3x\ {\cal A}_t \partial_a \left( (\partial_b \tilde{\cal E}^{a b}) \epsilon^0 + \tilde{\cal E}^{a b} \partial_b \epsilon^0\right)
    .\nonumber
\end{eqnarray}
This is a boundary term and hence neglected.

The self-commutation of the emergent electric field (\ref{eq:Commutative emergent electric field}) together with the result that it is independent of the gravitational variables, leads us to the conclusion that the brackets (\ref{eq:U(1)-invariance - E}) and (\ref{eq:U(1)-invariance - F}) determine that a local phase-space function $\mathcal{O}$ is U(1)-invariant only if
\begin{eqnarray}
    0 &=& \{{\cal O} (x) , G^{\rm EM} [{\cal A}_t] \} 
    = \int {\rm d}^3 z \frac{\partial {\cal A}_t (z)}{\partial z^a} \frac{\delta {\cal O} (x)}{\delta A_a (z)}
    \nonumber\\
    &=& \frac{\partial {\cal O}}{\partial A_a} \partial_a {\cal A}_t
    - \frac{\partial {\cal O}}{\partial (\partial_b A_a)} \partial_a \partial_b {\cal A}_t
    \nonumber\\
    &&
    + \frac{\partial {\cal O}}{\partial (\partial_{b_1} \partial_{b_2} A_a)} \partial_a \partial_{b_1} \partial_{b_2} {\cal A}_t
    - \dotsi
    \,,
\end{eqnarray}
which must hold for arbitrary ${\cal A}_t (x)$, and hence
\begin{eqnarray}
    \frac{\partial {\cal O}}{\partial A_a} &=& 0
    \,, \\
    \frac{\partial {\cal O}}{\partial (\partial_{b_1} A_a)} &=& - \frac{\partial {\cal O}}{\partial (\partial_a A_{b_1})}
    \,, \\
    \frac{\partial {\cal O}}{\partial (\partial_{b_1} \dotsi \partial_{b_j} \dotsi \partial_{b_n} A_a)} &=& - \frac{\partial {\cal O}}{\partial (\partial_{b_1} \dotsi \partial_{a} \dotsi \partial_{b_n} A_{b_j})}
    .\qquad
\end{eqnarray}
These conditions determine the dependence
\begin{eqnarray}
    {\cal O} &=& {\cal O} (\partial_{b_1} \dotsi \partial_{b_j} \dotsi \partial_{b_n} A_a - \partial_{b_1} \dotsi \partial_a \dotsi \partial_{b_n} A_{b_j})
    \nonumber\\
    &=&
    {\cal O} ( F_{b a} , \partial_{b_1} \dotsi \partial_{b_{j-1}} \partial_{b_{j+1}} \dotsi \partial_{b_n}  F_{b_j a} )
    \,.
    \label{eq:U(1)-invariance restriction of O - EM}
\end{eqnarray}
The same conclusion is true for smeared functions and hence both the structure function and the Hamiltonian constraint must have the dependence (\ref{eq:U(1)-invariance restriction of O - EM}) for them be U(1) invariant.

Higher-order spatial derivatives of the magnetic part of the strength tensor are therefore allowed, but the vector potential cannot appear in any way other than in the form of the magnetic part of the strength tensor\textemdash\,the only way to circumvent this is by modifying the Gauss constraint as it happens when introducing electrically charged matter, a possibility that we leave for future works.

\subsection{The Bianchi identities}

The Bianchi identity of classical electromagnetism
\begin{equation}
    \partial_\alpha F_{\beta \gamma} + \partial_\beta F_{\gamma \alpha} + \partial_\gamma F_{\alpha \beta} = 0
    \,,
\end{equation}
follows from $F = {\rm d} A$ being an exact two-form and hence ${\rm d} F = 0$.
But this identity does not follow so trivially in the canonical formulation where $F_{0a} = \tilde{E}_a / \sqrt{\det q}$ such that $F_{\mu \nu}$ is not manifestly an exact two-form.
The purely spatial part of the Bianchi identity is trivial because $F_{ab} = \partial_a A_b - \partial_b A_a$ is still an exact two-form in the spatial hypersurface, while the case with two time indices vanishes due to the antisymmetry of $F_{\mu \nu}$.
The only non-trivial component reads
\begin{eqnarray}
    \frac{1}{2} \dot{F}_{b c} &=& - \partial_{[b} F_{c] t}
    = - \partial_{[b} \left( N F_{c] 0} + N^d F_{c] d} \right)
    \nonumber\\
    &=& \partial_{[b} F_{|0| c]} N
    - F_{0 [b} \partial_{c]} N
    - \partial_{[b} F_{c] d} N^d
    + F_{d [c} \partial_{b]} N^d
    \nonumber\\
    &=& \partial_{[b} F_{|0| c]} N
    - F_{0 [b} \partial_{c]} N
    + \frac{1}{2} \mathcal{L}_{\vec{N}} F_{b c}
    \,,
\end{eqnarray}
where we used the purely spatial Bianchi identity to obtain the last line.
Using Hamilton's equation of motion in the left-hand-side, the Bianchi identity becomes
\begin{equation}
    \{ F_{a b} , \tilde{H} [N] \}
    = 2 \partial_{[a} F_{|0| b]} N
    - 2 F_{0 [a} \partial_{b]} N
    \,.
    \label{eq:Bianchi condition - ADM}
\end{equation}
Using the generic normal transformation of the spatial strength tensor (\ref{eq:Generic normal transformation of strength tensor - magnetic components}) and requiring the equation (\ref{eq:Bianchi condition - ADM}) be satisfied for arbitrary $N$, we obtain the series of equations
\begin{eqnarray}
    \mathcal{F}_{a b}
    &=& \partial_{[a} F_{|0| b]}
    \,,
    \label{eq:Bianchi condition - zeroth order}
    \\
    \mathcal{F}_{a} &=& F_{0 a}
    \,,
    \label{eq:Bianchi condition - first order}
    \\
    \mathcal{F}_{bc}^{c_1} &=& \mathcal{F}_{bc}^{c_1 c_2} = \dotsi = 0
    \,.
    \label{eq:Bianchi condition - Higher orders}
\end{eqnarray}
The second and third line are precisely the magnetic normal conditions (\ref{eq:Magnetic normal condition - first order}) and (\ref{eq:Magnetic normal condition - Higher orders}), while the first line can be analyzed as follows.
We first compute $\mathcal{F}_{a b} = 2 \partial_{[a} (\partial \tilde{H} / \partial E^{b]}) $, which follows from $\tilde{H}$ being independent of spatial derivatives of $E^a$. Then equation (\ref{eq:Bianchi condition - zeroth order}) is satisfied by use of (\ref{eq:Magnetic normal condition - simplified}).
Therefore, the Bianchi identity is implied by the magnetic normal covariance condition.

\subsection{Summary of the conditions}

We have imposed the electromagnetic covariance conditions, the existence of the symmetry generator, and the U(1)-invariance conditions.
These turned out to be quite lengthy and interrelated.
As a summary, the electromagnetic covariance conditions are given by the spatial conditions (\ref{eq:EM covariance condition - spatial diffeomorphism - 1}) and (\ref{eq:EM covariance condition - spatial diffeomorphism - reduced}), and the normal conditions (\ref{eq:Normal electric covariance condition - reduced - higher orders}) and (\ref{eq:Normal electric covariance condition - reduced - first order}).
The relation between the emergent electric field and the Hamiltonian constraint is given by the expression (\ref{eq:Magnetic normal condition - Emergent electric field}), while the relation between the emergent electric field and the strength tensor is given by (\ref{eq:Emergent strength field tensor relation to electromagnetic field}) under the preferred scheme 4.
We rewrite and simplify here all the conditions derived above for clarity.
In the following, however, we will consider only first-order spatial derivatives of the electromagnetic vector potential, mimicking the order of the classical theory.

The relation between the electromagnetic variables and the strength tensor are given by
\begin{eqnarray}
    F_{0a} &=& \frac{\tilde{q}_{a b} \tilde{E}^b}{\sqrt{\det \tilde{q}}}
    \,, 
    \label{eq:Emergent strength field tensor relation to electromagnetic field - summary} \\
    F_{a b} &=& \partial_a A_b - \partial_b A_a
    \,,
    \label{eq:Emergent strength field tensor magnetic relation to electromagnetic vector potential - summary}
\end{eqnarray}
where the magnetic normal covariance condition favors the explicit use of the emergent spatial metric and also determines
\begin{equation}
    F_{0 a} = \frac{\partial \tilde{H}}{\partial E^a}
    \,,
    \label{eq:dH/dE - summary}
\end{equation}
providing, together with (\ref{eq:Emergent strength field tensor relation to electromagnetic field - summary}), a derivation of the emergent electric field from the Hamiltonian constraint.
The spatial covariance conditions determined that the emergent electric field takes the form
\begin{equation}\label{eq:Em E and B - summary}
    \tilde{E}^a = E^a + \theta B^a
    \,,
\end{equation}
where $\theta$ is a constant and $B^{a}$ is given by
\begin{equation}\label{eq:Densitized magnetic field - summary}
    B^a=\frac{1}{2}\sqrt{\det \tilde{q}} \epsilon^{a b c} F_{b c}
    \,,
\end{equation}
and $\{\tilde{E}^a(x),\tilde{E}^b(y)\}=0$.

U(1)-invariance requires that the Hamiltonian constraint be dependent on the electromagnetic vector potential only through $F_{ab}$ (recall that we will not consider spatial derivatives of $F_{ab}$ here).
Using this, the generic normal transformation of the emergent electric field has the form
\begin{equation}\label{eq:Emergent E generic normal transformation - summary}
    \{ \tilde{E}^a , \tilde{H} [\epsilon^0] \} = \tilde{\cal E}^a \epsilon^0 + \tilde{\cal E}^{a b} \partial_b \epsilon^0
    + \tilde{\cal E}^{a b_1 b_2} \partial_{b_1} \partial_{b_2} \epsilon^0
    + \dotsi
    \,,
\end{equation}
where the coefficients are given by
\begin{eqnarray}
    \tilde{\cal E}^a &=&
    \partial_{b} \tilde{\cal E}^{ab}
    \,,\label{eq:Emergent E generic normal transformation - first-order - summary}
    \\
    \tilde{\cal E}^{ab} &=&
    2 \frac{\partial \tilde{H}}{\partial F_{a b}}
    + 2 \theta \frac{\partial B^a}{\partial F_{b c}} \frac{\partial \tilde{H}}{\partial E^c}
    \,,\label{eq:Emergent E generic normal transformation - second-order - summary}
    \\
    \tilde{\cal E}^{ab_1b_2} &=& \tilde{\cal E}^{ab_1b_3} = \dotsi = 0
    \,, \label{eq:Emergent E generic normal transformation - higher-order - summary}
\end{eqnarray}
obtained by direct computation.
The first and third line are in agreement with the symmetry generator condition (\ref{eq:Symmetry generator condition}) and the electric normal condition (\ref{eq:Normal electric covariance condition - reduced - higher orders}), respectively.

Finally, we recall the first-order term of the electric normal condition (\ref{eq:Electric covariance condition - first order - reduced form}) and we write it here for clarity
\begin{equation}
    \tilde{\cal E}^{ab}
    = \sqrt{\det \tilde{q}}\tensor{F}{^a^b}
    \,.
    \label{eq:Normal electric covariance condition - reduced - first order - summary}
\end{equation}

We can now rewrite the electromagnetic covariance conditions by combining (\ref{eq:dH/dE - summary}), (\ref{eq:Em E and B - summary}), (\ref{eq:Densitized magnetic field - summary}), (\ref{eq:Emergent E generic normal transformation - second-order - summary}), and (\ref{eq:Normal electric covariance condition - reduced - first order - summary}) into the single equation
\begin{equation}
    \frac{\partial \tilde{H}}{\partial F_{a b}}
    = \frac{\sqrt{\det \tilde{q}}}{2} \tensor{F}{^a^b}
    - \frac{\theta}{2} \epsilon^{a b c} \left(E_c + \theta B_c\right)
    \,.
    \label{eq:Normal electric covariance condition - summary}
\end{equation}
Combining (\ref{eq:Emergent strength field tensor relation to electromagnetic field - summary}), (\ref{eq:dH/dE - summary}), and (\ref{eq:Em E and B - summary}) we obtain the equation
\begin{equation}\label{eq:dH/dE simp - summary}
    \frac{\partial \tilde{H}}{\partial E^a} = \frac{\tilde{q}_{a b} \left(E^b + \theta B^b\right)}{\sqrt{\det \tilde{q}}}
    \,.
\end{equation}
Our goal is, therefore, to solve equations (\ref{eq:Normal electric covariance condition - summary}) and (\ref{eq:dH/dE simp - summary}).

In general, U(1)-invariance allows the structure function $\tilde{q}^{ab}$ be dependent on the emergent electric field $\tilde{E}^a$ and the magnetic strength tensor $F_{ab}$, which complicates solving (\ref{eq:Normal electric covariance condition - summary}).
However, in the special case where the structure function is independent of the electric momentum, a substitution of the expressions (\ref{eq:Emergent strength field tensor relation to electromagnetic field - summary}) and (\ref{eq:Em E and B - summary}) into (\ref{eq:dH/dE - summary}) allows for a direct integration, yielding
\begin{equation}
     \tilde{H} = \frac{\tilde{q}_{a b} E^a (\frac{1}{2}E^b + \theta B^b)}{\sqrt{\det \tilde{q}}}
     + \tilde{H}_A
     \,,
\end{equation}
where $\tilde{H}_A$ is an undetermined phase-space function independent of $E^a$, while (\ref{eq:Normal electric covariance condition - summary}) remains complicated.
On the other hand, in the case where the structure function is independent of the magnetic tensor $F_{a b}$, the condition (\ref{eq:Normal electric covariance condition - summary}) can be directly integrated to yield
\begin{equation}
    \tilde{H} = \frac{B_a B^a}{2 \sqrt{\det \tilde{q}}}
    - \theta \frac{B_c (E^c + \frac{1}{2} \theta B^c)}{\sqrt{\det \tilde{q}}}
    + \tilde{H}_E
    \,,
\end{equation}
where $\tilde{H}_{\rm E}$ is independent of $F_{a b}$, while (\ref{eq:dH/dE simp - summary}) remains complicated.
In the case where the structure function is independent of both $E^a$ and $F_{a b}$ the last two results combine into
\begin{equation}
     \tilde{H} = \frac{\tilde{E}_a \tilde{E}^a + B_a B^a}{2 \sqrt{\det \tilde{q}}}
     + \tilde{H}_{\rm grav}
     \,,
     \label{eq:Electromagnetic Hamiltonian in EMFT}
\end{equation}
where $\tilde{H}_{\rm grav}$ is independent of the electromagnetic variables.
The constraint contribution (\ref{eq:Electromagnetic Hamiltonian in EMFT}) has precisely the form of the classical Hamiltonian constraint including the $\theta$ term with the metric replaced by the emergent one.

Therefore, unless the emergent spacetime depends on the electromagnetic variables at the kinematical level, no local modified electromagentic theory exists in four dimensions to the derivative order considered here.

\section{Emergent electromagnetism: Spherical symmetry}
\label{sec:Emergent electromagnetism: Spherical symmetry}

We will show in this section that the spherically symmetric model, unlike the four dimensional theory, allows for modifications and a nontrivial relation between the emergent and fundamental electric fields beyond the $\theta$ term contribution.
We use the same notation of Sections~\ref{sec:Spherical EMG Vacuum} and \ref{sec:EM spherical classical}, where we respectively formulated spherically symmetric EMG and dealt with the spherically symmetric classical electromagnetism. See Table~\ref{tab:Notation2} for a summary of the notation used in the present Section.
We follow the procedure of Section~\ref{sec: Emergent electromagnetism} to implement all the conditions on the modified theory involving the electromagnetic field.
These conditions include electromagnetic covariance, anomaly freedom, existence of a mass and charge observables as the symmetry generators, and U(1) invariance in the simpler setting of spherical symmetry.

\subsection{Electromagnetic covariance}

Imposing spherical symmetry, the spatial covariance conditions (\ref{eq:Spatial F_ab cov cond - EM - Full}) and (\ref{eq:Spatial F_0a cov cond - EM - Full}) reduce to the single equation
\begin{equation}
    \{F_{0x} , \vec{H}[\vec{\epsilon}]\} = (\epsilon^x F_{0 x})'
    \,,
    \label{eq:Spatial F_0x cov cond - EM - Spherical}
\end{equation}
which implies that $F_{0x}$ must be a density of weight one.

The electric normal covariance condition (\ref{eq:Normal F_ab cov cond - EM - Full}) reduces to
\begin{equation}
    \frac{1}{\epsilon^0} \{ F_{0 x} , \tilde{H} [\epsilon^0] \}
    \bigg|_{\rm O.S.} =
    \frac{1}{N} \{ F_{0 x} , \tilde{H} [N] \}
    \bigg|_{\rm O.S.}
    \,,
    \label{eq:Normal F_0x cov cond - EM - Spherical}
\end{equation}
and hence implies the series of conditions
\begin{equation}
    \frac{\partial (\delta_{\epsilon^0} F_{0 x})}{\partial (\epsilon^0)'} \bigg|_{\rm O.S.} = \frac{\partial (\delta_{\epsilon^0} F_{0 x})}{\partial (\epsilon^0)''} \bigg|_{\rm O.S.} = \dotsi = 0
    \,.
    \label{eq:Normal cov cond - EM - Spherical - reduced}
\end{equation}

The relation between the strength tensor electric component and the Hamiltonian constraint (\ref{eq:dH/dE - summary}) is provided in the full four-dimensional theory by the magnetic normal covariance condition, which trivializes in the spherical symmetry reduced model because the magnetic components of the strength tensor are trivial.
Therefore, we simply extrapolate such result and postulate it here:
\begin{equation}
    F_{0 x} = \frac{\partial \tilde{H}}{\partial {\cal E}^x}
    \,.
    \label{eq:Strength tensor/Hamiltonian - spherical}
\end{equation}
This expression is a density of weight one and therefore automatically satisfies (\ref{eq:Spatial F_0x cov cond - EM - Spherical}).
Furthermore, the preference of the scheme 4 to determine the relation between the emergent electric field and the strength tensor (\ref{eq:Emergent strength field tensor relation to electromagnetic field - summary}), using the emergent spatial metric for densitization, was based on the appearance of the structure function in the magnetic normal covariance condition, which is trivial in spherical symmetry.
Therefore, we face an extra freedom in the definition of this relation if we restrict ourselves to the equations of the reduced model.
While the four-dimensional theory clearly favors the relation (\ref{eq:Emergent strength field tensor relation to electromagnetic field - summary}) with the emergent spatial metric, we may take advantage of the 2D model's ambiguity and reconsider the different schemes:
Scheme 1 is given by
\begin{equation}
    F_{0x} = \frac{\bar{q}_{xx}}{E^x \sqrt{\bar{q}_{xx}}} \tilde{\cal E}^x
    = \frac{\lambda}{\bar{\lambda}} \frac{E^\varphi}{(E^x)^{3/2}} \tilde{\cal E}^x
    \,,
    \label{eq:Emergent strength field tensor relation to electromagnetic field - scheme 1 - spherical}
\end{equation}
where we used the definition (\ref{eq:Alternative spatial metric}) for $\bar{q}_{xx}$, scheme 2 is given by
\begin{equation}
    F_{0x} = \frac{\bar{q}_{xx}}{E^x \sqrt{\tilde{q}_{xx}}} \tilde{\cal E}^x
    = \beta^{1/2} \frac{\lambda}{\bar{\lambda}} \frac{E^\varphi}{(E^x)^{3/2}} \tilde{\cal E}^x
    \,, 
    \label{eq:Emergent strength field tensor relation to electromagnetic field - scheme 2 - spherical}
\end{equation}
where we used the definition (\ref{eq:Barred spatial metric}), scheme 3  is given by
\begin{equation}
    F_{0x} = \frac{\tilde{q}_{xx}}{E^x \sqrt{\bar{q}_{xx}}} \tilde{\cal E}^x
    = \beta^{-1} \frac{\lambda}{\bar{\lambda}} \frac{E^\varphi}{(E^x)^{3/2}} \tilde{\cal E}^x
    \,, 
    \label{eq:Emergent strength field tensor relation to electromagnetic field - scheme 3 - spherical}
\end{equation}
and scheme 4 by
\begin{equation}
    F_{0x} = \frac{\tilde{q}_{xx}}{E^x \sqrt{\tilde{q}_{xx}}} \tilde{\cal E}^x
    = \beta^{-1/2} \frac{\lambda}{\bar{\lambda}} \frac{E^\varphi}{(E^x)^{3/2}} \tilde{\cal E}^x
    \,.
    \label{eq:Emergent strength field tensor relation to electromagnetic field - scheme 4 - spherical}
\end{equation}
Combining these expressions with (\ref{eq:Strength tensor/Hamiltonian - spherical}), we may express the emergent electric field as
\begin{equation}
    \tilde{\cal E}^x = \beta^{n/2} \frac{\bar{\lambda}}{\lambda} \frac{(E^x)^{3/2}}{E^\varphi} \frac{\partial \tilde{H}}{\partial {\cal E}^x}
    \,.
    \label{eq:Emergent strength field tensor relation to electromagnetic field - scheme n - spherical}
\end{equation}
with $n=0$ for scheme 1, $n=1$ for scheme 2, $n=-2$ for scheme 3, and $n=-1$ for scheme 4.
The identification of the emergent electric field (\ref{eq:Emergent strength field tensor relation to electromagnetic field - scheme n - spherical}) is necessary to define the Gauss constraint and implement U(1) invariance as discussed in Subsection~\ref{sec:Em electric and symmetries - classical}.

\subsection{General constraint ansatz, anomaly freedom, and covariance}

In imposing anomaly freedom, we start by defining the general ansatz
\begin{eqnarray}
    \tilde{H} &=& a_0
    + ((E^x)')^2 a_{x x}
    + ((E^\varphi)')^2 a_{\varphi \varphi}
    + (E^x)' (E^\varphi)' a_{x \varphi}
    \nonumber\\
    &&
    + (E^x)'' a_2
    + (K_\varphi')^2 b_{\varphi \varphi}
    + (K_\varphi)'' b_2
    + (E^x)' K_\varphi' c_{x \varphi}
    \nonumber\\
    &&
    + (E^\varphi)' K_\varphi' c_{\varphi \varphi}
    + (E^\varphi)'' c_2
    \,,
    \label{eq:Hamiltonian constraint ansatz - Generalized vacuum - Extended}
\end{eqnarray}
for our modified Hamiltonian constraint, where $a_0$, $a_{i j}$, $a_2$,
$b_{\varphi \varphi}$, $b_2$, $c_2$, $c_{i j}$ are all functions of the phase space variables, but not of their derivatives with the only exception that they are allowed to depend on derivatives of ${\cal E}^x$.
We have included terms quadratic in first-order radial derivatives and linear in second-order radial derivatives of all the phase space variables, except $K_x$ because this would break covariance as
demanded by (\ref{eq:Covariance condition of 3-metric - reduced - spherical}).
Also, mirroring the classical constraint, we do not include dependence on $A_x$ of any coefficient in the ansatz, which is related to U(1) invariance, to be discussed below.
The general ansatz (\ref{eq:Hamiltonian constraint ansatz - Generalized vacuum - Extended}) precludes us, in general, from having a simple relation of the full constraint as sum of gravitational and electromagnetic contributions of the form $\tilde{H}=\tilde{H}_{\rm grav}+\tilde{H}_{\rm EM}$, with $\tilde{H}_{\rm grav}$ independent of the electromagnetic variables, if the modification functions are allowed to depend on ${\cal E}^x$. While such separation in the contribution is possible in special cases, we will consider the general constraint ansatz (\ref{eq:Hamiltonian constraint ansatz - Generalized vacuum - Extended}) to encompass a more general class of models.

Because both the Hamiltonian and vector constraints do not include $A_x$, the implementation of anomaly-freedom and spacetime covariance follows almost the exact same procedure as that of \cite{EMGCov,EMGscalar}, with the exception that the modification functions are allowed to depend on the electric momentum ${\cal E}^x$.
Therefore, we may start with the general constraint compatible with anomaly-freedom and spacetime covariance given by (\ref{eq:Hamiltonian constraint - modified - periodic - vacuum}) with structure function (\ref{eq:Structure function - modified - periodic - vacuum}), but we allow the undetermined functions to also depend on ${\cal E}^x$, $({\cal E}^x)'$, and $({\cal E}^x)''$.
Taking this into account the bracket between the modified Hamiltonian constraint with itself gives
\begin{widetext}
\begin{equation}\label{eq:H,H bracket - spherical - EM}
    \{\tilde{H}[N],\tilde{H}[M]\}
    = H_x[\tilde{q}^{xx}(NM'-MN')]
    - \int {\rm d} x\ \bar{\lambda}_0^2 \frac{E^x}{E^\varphi} \cos^2 (\bar{\lambda} K_\varphi) \left( \frac{\sin (2 \bar{\lambda} K_\varphi)}{2 \bar{\lambda}} \frac{\partial c_f}{\partial {\cal E}^x} + \cos (2 \bar{\lambda} K_\varphi) \frac{\partial \bar{q}}{\partial {\cal E}^x}\right) ({\cal E}^x)'
    \,.
\end{equation}
\end{widetext}
This last term is not necessarily anomalous because in the case where the emergent electric field equals the electric momentum $\tilde{\cal E}^x={\cal E}^x$, it is proportional to the Gauss constraint and hence the constraint algebra would remain first-class. Recall that it is typical of gauge fields for the $\{H,H\}$ bracket to include a Gauss constraint contribution, though U(1) is the exception.
However, as we will show, the electromagnetic covariance conditions will require this term to vanish.

It is now straightforward, though a lengthy procedure, to obtain the strength tensor component (\ref{eq:Strength tensor/Hamiltonian - spherical}) which, in particular contains the term $F_{0x}\supset(\partial \ln ( \frac{\bar{\lambda}}{\lambda} \lambda_0) / \partial {\cal E}^x) \tilde{H}$. On-shell, this term does not contribute to either the emergent electric field nor to the covariance condition (\ref{eq:Normal cov cond - EM - Spherical - reduced}) and hence we shall disregard it here.
To easily take this into account in the following, we may simply write instead
\begin{eqnarray}\label{eq:Electric component of strength tensor - spherical}
    &&
    F_{0x} = \bar{\lambda}_0 \frac{\partial (\tilde{H}/\bar{\lambda}_0)}{\partial {\cal E}^x}
    = f
    + C_1 \frac{\partial c_f}{\partial {\cal E}^x}
    + C_2 \frac{\partial^2 c_f}{\partial E^x \partial {\cal E}^x}
    \nonumber\\
    &&\qquad\qquad\qquad\qquad\quad
    + D_1 \frac{\partial q}{\partial {\cal E}^x}
    + D_2 \frac{\partial^2 q}{\partial E^x \partial {\cal E}^x}
    \,,\qquad
\end{eqnarray}
where, for notational ease, we have used the definitions (\ref{eq:Redefinitions of lambda - ease notation}) for the barred functions.
The coefficients $f$, $C_i$ and $D_i$ are complicated phase-space dependent functions, but, as we now show, their explicit expressions will not be relevant, except for $f$ which we write later below.

Now that we have the expression for $F_{0x}$, we can apply the electric covariance condition (\ref{eq:Normal cov cond - EM - Spherical - reduced}).
Only the first and second order terms are nontrivial.
The second order term implies the equation
\begin{equation}
    \frac{\sin (2 \bar{\lambda} K_\varphi)}{2 \bar{\lambda}} \frac{\partial c_f}{\partial {\cal E}^x}
    + \cos (2 \bar{\lambda} K_\varphi) \frac{\partial \bar{q}}{\partial {\cal E}^x}
    = 0
    \,.
\end{equation}
Because $c_f$ and $\bar{q}$ are independent of $K_\varphi$ we conclude that they must also be independent of ${\cal E}^x$.
Taking this into account, the first order term is satisfied too.
We thus find that the whole expression (\ref{eq:Electric component of strength tensor - spherical}), except for $f$, and the extra term in the bracket (\ref{eq:H,H bracket - spherical - EM}) vanish.

The Hamiltonian constraint therefore simplifies to
\begin{widetext}
\begin{eqnarray}
    \tilde{H} \!\!&=&\!\!
    \frac{\bar{\lambda}^2}{\lambda^2} \lambda_0^2 \frac{\sqrt{\tilde{q}_{xx}} \alpha_0^q}{E^x}
    - \frac{E^\varphi}{2} \sqrt{\tilde{q}^{xx}} \Bigg[
    E^\varphi \left(
    \frac{\lambda^2}{\bar{\lambda}^2} \frac{\alpha_{0q}}{E^x}
    + \frac{\alpha_{2q}}{E^x} \left( c_f \frac{\sin^2 (\bar{\lambda} K_\varphi)}{\bar{\lambda}^2} + 2 q \frac{\sin(2\bar{\lambda} K_\varphi)}{2 \bar{\lambda}} \right) \right)
    - \frac{((E^x)')^2}{E^\varphi} \frac{\alpha_{2q}}{4 E^x} \cos^2 (\bar{\lambda} K_\varphi)
    \Bigg]
    \nonumber\\
    &&\!\!
    - \frac{\bar{\lambda}}{\lambda} \lambda_0 \frac{\sqrt{E^x}}{2} \Bigg[ E^\varphi \Bigg(
    \frac{\lambda^2}{\bar{\lambda}^2} \frac{\alpha_0}{E^x}
    + 2 \frac{\sin^2 \left(\bar{\lambda} K_\varphi\right)}{\bar{\lambda}^2}\frac{\partial c_{f}}{\partial E^x}
    + 4 \frac{\sin \left(2 \bar{\lambda} K_\varphi\right)}{2 \bar{\lambda}} \frac{\partial}{\partial E^x} \left(\frac{\lambda}{\bar{\lambda}} q\right)
    \nonumber\\
    &&\qquad \qquad
    + \left( \frac{\alpha_2}{E^x} - 2 \frac{\partial \ln \lambda^2}{\partial E^x}\right) \left( c_f \frac{\sin^2 \left(\bar{\lambda} K_\varphi\right)}{\bar{\lambda}^2}
    + 2 \frac{\lambda}{\bar{\lambda}} q \frac{\sin \left(2 \bar{\lambda} K_\varphi\right)}{2 \bar{\lambda}} \right)
    \Bigg)
    + 4 K_x \left(c_f \frac{\sin (2 \bar{\lambda} K_\varphi)}{2 \bar{\lambda}}
    + \frac{\lambda}{\bar{\lambda}} q \cos(2 \bar{\lambda} K_\varphi)\right)
    \nonumber\\
    &&\qquad \qquad
    - \frac{((E^x)')^2}{E^\varphi} \left(
    \frac{\alpha_2}{4 E^x} \cos^2 \left( \bar{\lambda} K_\varphi \right)
    - \frac{K_x}{E^\varphi} \bar{\lambda}^2 \frac{\sin \left(2 \bar{\lambda} K_\varphi \right)}{2 \bar{\lambda}}
    \right)
    + \left( \frac{(E^x)' (E^\varphi)'}{(E^\varphi)^2}
    - \frac{(E^x)''}{E^\varphi} \right) \cos^2 \left( \bar{\lambda} K_\varphi \right)
    \Bigg]
    ,
    \label{eq:Hamiltonian constraint - modified - periodic}
\end{eqnarray}
with structure function
\begin{equation}
    \tilde{q}^{x x}
    =
    \left(
    \left( c_{f}
    + \left(\frac{\bar{\lambda} (E^x)'}{2 E^\varphi} \right)^2 \right) \cos^2 \left(\bar{\lambda} K_\varphi\right)
    - 2 \frac{\lambda}{\bar{\lambda}} q \bar{\lambda}^2 \frac{\sin \left(2 \bar{\lambda} K_\varphi\right)}{2 \bar{\lambda}}\right)
    \frac{\bar{\lambda}^2}{\lambda^2} \lambda_0^2 \frac{E^x}{(E^\varphi)^2}
    \,.
    \label{eq:Structure function - modified - periodic}
\end{equation}
\end{widetext}
where $\lambda_0 , \alpha_0,\alpha_2, q$, and $\lambda$ are undetermined functions of $E^x$, ${\cal E}^x$, $({\cal E}^x)'$, and $({\cal E}^x)''$, while $c_f$ and the combination $\bar{q}=\frac{\lambda}{\bar \lambda} q$ are undetermined functions of $E^x$ only.

\subsection{Canonical transformations, periodicity, and the holonomy parameter}

In \cite{EMGCov}, the general vacuum constraint was obtained originally with a non-constant $\lambda(E^x)$ as the frequency in the trigonometric functions of $K_\varphi$, but the constraint was not fully periodic.
Only after a canonical transformation defined by $K_\varphi \to \frac{\bar{\lambda}}{\lambda} K_\varphi$ and leaving the vector constraint unchanged does the constraint become periodic at the expense of introducing the fiducial constant parameter $\bar{\lambda}$.
We may wish to recover this result with the electromagnetic coupling, that is, to be able to eliminate the fiducial parameter after reversing the canonical transformation, $K_\varphi \to \frac{\lambda}{\bar{\lambda}} K_\varphi$.
The complete canonical transformation with $\lambda=\lambda (E^x,{\cal E}^x)$ leaving $E^x$ and ${\cal E}^x$ unchanged is given by
\begin{eqnarray}
    K_\varphi &\to& \frac{\lambda}{\bar{\lambda}} K_\varphi
    \quad,\quad
    E^\varphi \to \frac{\bar{\lambda}}{\lambda} E^\varphi
    \,,    \label{eq:Diffeomorphism-constraint-preserving canonical transformations - Spherical - residual after modulo}\\
    A_x &\to& A_x + E^\varphi K_\varphi \frac{\partial \ln \lambda}{\partial {\cal E}^x}
    \quad,\quad
    {\cal E}^x \to {\cal E}^x
    \,, \nonumber\\
    K_x &\to& K_x + E^\varphi K_\varphi \frac{\partial \ln \lambda}{\partial E^x}
    \quad,\quad
    E^x \to E^x
    \,. \nonumber
\end{eqnarray}
However, such canonical transformation changes the vector constraint to the form
\begin{equation}
    H_x = E^\varphi K_\varphi' - K_x (E^x)' + E^\varphi K_\varphi \frac{\partial \ln \lambda}{\partial {\cal E}^x} ({\cal E}^x)'
    \,.
\end{equation}
Invariance of the vector constraint then restricts $\lambda$ to depend on $E^x$ only.
We will therefore only consider $\lambda=\lambda(E^x)$ in the following.
Furthermore, since $\bar{q}=\frac{\lambda}{\bar \lambda} q$ is independent of ${\cal E}^x$, it implies that $q$ itself is independent of ${\cal E}^x$ too.

After performing this canonical transformation and using the unbarred functions (\ref{eq:Redefinitions of lambda - ease notation}), the fiducial $\bar{\lambda}$ disappears from the structure function and the whole constraint.

The emergent electric field now takes the form
\begin{eqnarray}\label{eq:Emergent electric field - spherical - general}
    &&\!\!\!\!\!\!\!\!
    \tilde{\cal E}^x
    = \beta^{n/2} \frac{(E^x)^{3/2}}{E^\varphi} \frac{\partial \tilde{H}}{\partial {\cal E}^x}
    = \beta^{n/2} \frac{\bar{\lambda}_0}{2} E^x \left( - \frac{\partial \bar{\alpha}_0}{\partial {\cal E}^x}
    - c_f \frac{\partial \bar{\alpha}_2}{\partial {\cal E}^x} \right)
    \nonumber\\
    &&
    - \frac{\beta^{(n+1)/2} }{2} E^x \left( \frac{\partial \bar{\alpha}_{0q}}{\partial {\cal E}^x}
    + c_f \frac{\partial \alpha_{2q}}{\partial{\cal E}^x} \right)
    + \beta^{(n-1)/2} \bar{\lambda}_0^2 \frac{\partial \alpha_0^q}{\partial {\cal E}^x}
    \nonumber\\
    &&
    + \frac{\beta^{(n+2)/2}}{2 \bar{\lambda}_0} E^x \frac{\partial \bar{\alpha}_2}{\partial {\cal E}^x}
    + \frac{\beta^{(n+3)/2}}{2 \bar{\lambda}_0} E^x \frac{\partial \alpha_{2q}}{\partial{\cal E}^x}
    \,.
\end{eqnarray}

\subsection{Charge observable I}

Now that we have the expression of the emergent electric field (\ref{eq:Emergent electric field - spherical - general}) we shall impose the existence of the charge observable
\begin{equation}
    J^x[{\cal A}_x] = \int {\rm d} x\ {\cal A}_x \tilde{\cal E}^x
    \,,
\end{equation}
where ${\cal A}_x$ is a constant.
This requires that $J^x[{\cal A}_x]$ Poisson-commutes with the constraints on shell.
It is straightforward to check that $\{J^x[{\cal A}_x] , G[{\cal A}_t]\}=0$ because the expression obtained for $\tilde{\cal E}^x$ commutes with itself.
Also, because $\tilde{\cal E}^x$ is in general a spatial scalar through its dependence on the gravitational variables, the commutation of the symmetry generator with the vector constraint yields
\begin{equation}
    \{J^x[{\cal A}_x] , H_x[\epsilon^x]\}
    = \int {\rm d} x\ {\cal A}_x \epsilon^x (\tilde{\cal E}^x)'
    = G [{\cal A}_x \epsilon^x]
    \,,
\end{equation}
which vanishes on shell.
Finally, we must ensure that the symmetry generator commutes with the Hamiltonian constraint.
To this end, we note that the spacetime covariance condition implies that the transformation of the $\beta$ function takes the form 
\begin{equation}
    \{\beta,\tilde{H}[\epsilon^0]\} = {\cal B} \epsilon^0
    \,,
\end{equation}
where ${\cal B}$ is some phase-space function that can be expanded in powers of $\beta$,
\begin{eqnarray}
    {\cal B} &=&
    {\cal B}_{-1/2} \beta^{-1/2}
    + {\cal B}_0
    + {\cal B}_{1/2} \beta^{1/2}
    + {\cal B}_1 \beta
    \nonumber\\
    &&
    + {\cal B}_{3/2} \beta^{3/2}
    + {\cal B}_{2} \beta^{2}
    + {\cal B}_{5/2} \beta^{5/2}
    \,,
\end{eqnarray}
where the ${\cal B}_i$ coefficients are complicated functions of the phase space.
Using this and the expression for the emergent electric field (\ref{eq:Emergent electric field - spherical - general}) we find that the bracket takes the form
\begin{equation}\label{eq:Symm gen cond}
    \{J^x[{\cal A}_x] , \tilde{H}[\epsilon^0]\}=
    \int {\rm d} x\ {\cal A}_x \mathcal{J} \epsilon^0
    = 0
    \,,
\end{equation}
with no derivatives of $\epsilon^0$ because of the electric covariance condition (\ref{eq:Normal cov cond - EM - Spherical - reduced}), and where ${\cal J}$ is a complicated phase space function to be discussed in detail below.
The appearance of the arbitrary $\epsilon^0$ in the integrand implies that $\mathcal{J}$ must vanish locally.

We may therefore expand $\mathcal{J}$ into derivative terms which should vanish independently.
However, since the symmetry condition is imposed on shell, we must take this into account, and we do so by replacing $K_x=E^\varphi K_\varphi'/(E^x)'$ which comes from solving the vector constraint $H_x=0$.
We need not impose the Hamiltonian constraint because a direct calculation shows that $\mathcal{J}$ does not include $(E^x)''$, hence the vanishing of the Hamiltonian constraint cannot contribute to this expression.
Before doing so, however, we are now in the position to discuss other conditions.

\subsection{U(1) gauge invariance}

The vanishing of $\mathcal{J}$ implies that $\{\tilde{\cal E}^x (x) , \tilde{H}[\epsilon^0]\}=0$ is satisfied locally too.
This in turn implies that U(1) gauge invariance is automatically realized by imposing the symmetry condition,
\begin{eqnarray}
    \{ G [\mathcal{A}_t] , \tilde{H}[\epsilon^0] \}
    &=& \left\{ \int{\rm d} x\ \mathcal{A}_t(x) (\tilde{\cal E}^x (x))' , \tilde{H}[\epsilon^0] \right\}
    \nonumber\\
    &=&
    \int{\rm d} x\ \mathcal{A}_t(x) \left(\{ \tilde{\cal E}^x (x) , \tilde{H}[\epsilon^0] \}\right)'
    \nonumber\\
    &=&
    \int{\rm d} x\ \mathcal{A}_t(x) {\cal J}'
    = 0
    \,,
\end{eqnarray}
and hence we need not consider it independently.

\subsection{Mass observable}

We will now derive the conditions for the existence of a mass observable which exists in the classical system.
We follow the same procedure developed in \cite{EMGscalar} for the modified vacuum system.
It turns out that generalizing the vacuum observable (\ref{eq:Vacuum mass observable - EMG}) to incorporate the electromagnetic degrees of freedom is quite easy as we now show.

A phase space function $\mathcal{D}$ is a Dirac observable if
$\delta_\epsilon \mathcal{D} = \mathcal{D}_H \tilde{H} + \mathcal{D}_x H_x + \mathcal{D}_{\cal E} G$, where
$\mathcal{D}_H$, $\mathcal{D}_x$, and $\mathcal{D}_{\cal E}$ depend on the phase-space variables and on
the gauge function $\epsilon$.
We may therefore consider the dependence $\mathcal{D} (E^x , K_\varphi , (E^x)'/E^\varphi , \beta , \tilde{\cal E}^x)$.
Because the emergent electric field commutes with the constraints locally as imposed by the existence of the charge observable, it is itself a Dirac observable.
Furthermore, any spatial derivatives of $\tilde{\cal E}^x$ arising from the gauge transformation of $\mathcal{D}$ will be proportional to the Gauss constraint.
We therefore conclude that the dependence of the Dirac observable on $\tilde{\cal E}^x$ does not play any crucial role and we may disregard it and simply re-introduce it in the final result where appropriate.
Indeed, the procedure will result in the same vacuum mass observable (\ref{eq:Vacuum mass observable - EMG}) and the same conditions (\ref{eq:Vacuum mass observable condition}), hence we set
\begin{equation}
    \alpha_0^q=\alpha_{0q}=\alpha_{2q}=0
    \,.
\end{equation}
The only exception to the derivation is that the parameters $d_0$ and $d_2$ are not necessarily constant but may, as well as $\bar{\alpha}_0$ and $\bar{\alpha}_2$, depend on $\tilde{\cal E}^x$.
This immediately implies that $\bar{\alpha}_0$ will in fact depend on ${\cal E}^x$ because it is the only undetermined function that can reproduce the classical limit with an electric field given by (\ref{eq:Electro-gravity mass - spherical - classical}).

\subsection{Charge observable II}

Taking the above results into account, the charge observable condition $\mathcal{J}=0$, from (\ref{eq:Symm gen cond}), can be expanded into powers of $\beta$.
Factoring out an overall factor $\beta^{-n/1+1}$, this condition has only integer powers from $\beta^0$ to $\beta^3$.
This can now be expanded into derivative powers of $(E^x)'$ (no spatial derivatives of other variables appear) which must vanish independently.
We first focus on the highest order term $((E^x)')^6$, which implies the equation
\begin{widetext}
\begin{equation}\label{eq:Symm gen 1}
    \left( \frac{(n+2) \bar{\alpha}_2-4}{4 E^x} 
    - (n+1) \frac{\partial \ln \bar{\lambda}_0}{\partial E^x} 
    + \frac{\partial \ln \lambda}{\partial E^x} \right) \frac{\partial \bar{\alpha}_2}{\partial {\cal E}^x}
    - \frac{\partial^2 \bar{\alpha}_2}{\partial E^x \partial {\cal E}^x}
    = 0
    \,.
\end{equation}
Using this, the next order $((E^x)')^4$ implies the equation
\begin{equation}\label{eq:Symm gen 2}
    \left( \frac{n}{2} \left( c_f \frac{\bar{\alpha}_2}{2 E^x} + \frac{\partial c_f}{\partial E^x}\right)
    + \bar{\lambda}^2 \frac{n+2}{2} \frac{\bar{\alpha}_0}{2 E^x} \right)\frac{\partial \bar{\alpha}_2}{\partial {\cal E}^x}
    + \bar{\lambda}^2 \left( \frac{n \bar{\alpha}_2-4}{4 E^x} - (n+1) \frac{\partial \ln \bar{\lambda}_0}{\partial E^x}
    + \frac{\partial \ln \lambda}{\partial E^x} \right) \frac{\partial \bar{\alpha}_0}{\partial {\cal E}^x}
    - \bar{\lambda}^2 \frac{\partial^2 \bar{\alpha}_0}{\partial E^x \partial {\cal E}^x}
    = 0
    \,.
\end{equation}
\end{widetext}
Using this, the next order $((E^x)')^2$ implies the equation
\begin{equation}\label{eq:Symm gen 3}
    n \left( \bar{\lambda}^2 \bar{\alpha}_0 + c_f \bar{\alpha}_2 + 2 E^x \frac{\partial c_f}{\partial E^x} \right)
    \left( \bar{\lambda}^2 \frac{\partial \bar{\alpha}_0}{\partial {\cal E}^x} + c_f \frac{\partial \bar{\alpha}_2}{\partial {\cal E}^x} \right)
    = 0
    \,.
\end{equation}
Using these three equations we find that the zeroth order vanishes automatically and is hence not an independent condition.

Equation (\ref{eq:Symm gen 3}) is the simpler to start with.
It is trivial for scheme 1 where $n=0$, but not for the other schemes. in which case it poses a restriction on the modification functions.
The vanishing of the term in the first parenthesis can be thought of as an equation for $\bar{\alpha}_0$, $\bar{\alpha}_2$ or $c_f$, but all those solutions are incompatible with the classical limit.
The second parenthesis, however, can be solved for $\bar{\alpha}_2$,
\begin{equation}\label{eq:alpha2 schemes not 1}
    \bar{\alpha}_2 = \bar{a}_2 (E^x) - \bar{\lambda}^2 \frac{\bar{\alpha}_0}{c_f}
    \,,
\end{equation}
which is compatible with the classical limit $\bar{\alpha}_2\to1$.
We chose to solve for $\bar{\alpha}_2$ because $\bar{\alpha}_0$ must be able to reproduce the classical dependence on the electric field independently of $\bar{\alpha}_2$.
We will solve the other two equations for the different schemes separately.

\subsubsection{Scheme 1}

We set $n=0$ with which equation (\ref{eq:Symm gen 3}) trivializes, while equations (\ref{eq:Symm gen 1}) and (\ref{eq:Symm gen 2}) simplify into
\begin{equation}\label{eq:Symm gen 1 - scheme 1}
    \left( \frac{ \bar{\alpha}_2-2}{2 E^x} 
    - \frac{\partial \ln \left( \frac{\bar{\lambda}}{\lambda} \bar{\lambda}_0\right)}{\partial E^x} \right) \frac{\partial \bar{\alpha}_2}{\partial {\cal E}^x}
    - \frac{\partial^2 \bar{\alpha}_2}{\partial E^x \partial {\cal E}^x} = 0
    \,,
\end{equation}
\begin{equation}\label{eq:Symm gen 2 - scheme 1}
    \left( \frac{1}{E^x} + \frac{\partial \ln \left( \frac{\bar{\lambda}}{\lambda} \bar{\lambda}_0\right)}{\partial E^x} \right) \frac{\partial \bar{\alpha}_0}{\partial {\cal E}^x}
    - \frac{\bar{\alpha}_0}{2 E^x} \frac{\partial \bar{\alpha}_2}{\partial {\cal E}^x}
    + \frac{\partial^2 \bar{\alpha}_0}{\partial E^x \partial {\cal E}^x}
    = 0
    \,.
\end{equation}
And the emergent electric field takes the form
\begin{equation}\label{eq:Emergent electric field - scheme 1}
    \tilde{\cal E}^x = - \frac{E^x}{2} \frac{\bar{\lambda}}{\lambda} \bar{\lambda}_0 \frac{\partial \bar{\alpha}_0}{\partial {\cal E}^x}
    - \frac{\bar{\lambda}}{\lambda} \frac{c_f - \beta}{2\bar{\lambda}^2} \frac{\partial \bar{\alpha}_2}{\partial {\cal E}^x}
    \,.
\end{equation}

Equations (\ref{eq:Symm gen 1 - scheme 1}) and (\ref{eq:Symm gen 2 - scheme 1}) are hard to solve exactly.
We will therefore present here only one class of solutions defined by simply taking $\bar{\alpha}_2$ independent of ${\cal E}^x$. In this case, (\ref{eq:Symm gen 1 - scheme 1}) trivializes, while (\ref{eq:Symm gen 2 - scheme 1}) simplifies with the general solution
\begin{equation}\label{eq:alpha0 sol}
    \bar{\alpha}_0 = \bar{a}_0 (E^x) - \frac{2}{E^x} \int {\rm d} {\cal E}^x\ \frac{\lambda}{\bar{\lambda}} \frac{{\cal E}^x}{\bar{\lambda}_0} c_E ({\cal E}^x)
    \,,
\end{equation}
where $\bar{a}_0$ and $c_E$ are are undetermined functions of $E^x$ and ${\cal E}^x$, respectively, up to compatibility with the classical limit.
The emergent electric field takes the simple form
\begin{equation}\label{eq:Emergent electric field - scheme 1 - class 1}
    \tilde{\cal E}^x = c_E {\cal E}^x
    \,.
\end{equation}
We therefore identify the classical limit $c_E \to 1$ and $\bar{a}_0 \to 1 - \Lambda E^x$.

\subsubsection{scheme 2}

We set $n=1$, which implies we must adopt (\ref{eq:alpha2 schemes not 1}).
Using this, equations (\ref{eq:Symm gen 1}) and (\ref{eq:Symm gen 2}) become
\begin{eqnarray}\label{eq:Symm gen 1 - scheme 2} 
    c_f \frac{\partial^2 \bar{\alpha}_0}{\partial E^x \partial {\cal E}^x}
    - \left( \frac{c_f (3 \bar{a}_2 - 4) - 3 \bar{\lambda}^2 \bar{\alpha}_0}{4 E^x}
    + \frac{\partial c_f}{\partial E^x}\right.&&
    \nonumber
    \\
    \left.
    - 2 c_f \frac{\partial \ln \bar{\lambda}_0}{\partial E^x}
    + c_f \frac{\partial \ln \lambda}{\partial E^x} \right) \frac{\partial \bar{\alpha}_0}{\partial {\cal E}^x}
    = 0\,,
    &&
\end{eqnarray}
\begin{equation}\label{eq:Symm gen 2 - scheme 2}
    \left( \frac{\partial c_f}{\partial E^x} + c_f \frac{\bar{a}_2}{2 E^x} \right) \frac{\partial \bar{\alpha}_0}{\partial {\cal E}^x} = 0
    \,.
\end{equation}
Requiring $\partial \bar{\alpha}_0 / \partial {\cal E}^x \neq 0$ as necessary for the correct classical limit, the solution to the second equation is
\begin{equation}\label{eq:cf sol scheme 2}
    c_f = - c \exp \int {\rm d} E^x\ \frac{\bar{a}_2}{2 E^x}
    \,
\end{equation}
with constant $c$.
Using the classical value $\bar{a}_2\to1$, this becomes
\begin{equation}
    c_f
    = c / \sqrt{E^x}
    \,.
\end{equation}
Therefore, scheme 2 is incompatible with the classical limit $c_f\to 1$, and we will not consider it any further.

\subsubsection{Scheme 3}

We set $n=-2$, which implies we must adopt (\ref{eq:alpha2 schemes not 1}).
Using this, equations (\ref{eq:Symm gen 1}) and (\ref{eq:Symm gen 2}) become
\begin{eqnarray}\label{eq:Symm gen 1 - scheme 3}
    \Bigg[ \frac{\partial c_f}{\partial E^x}
    + c_f \left( - \frac{1}{E^x} 
    + \frac{\partial \ln \bar{\lambda}_0}{\partial E^x} 
    + \frac{\partial \ln \lambda}{\partial E^x} \right) \Bigg] \frac{\partial \bar{\alpha}_0}{\partial {\cal E}^x}&&
    \nonumber\\
    - c_f \frac{\partial^2 \bar{\alpha}_0}{\partial E^x \partial {\cal E}^x}
    = 0&&
    \,,
\end{eqnarray}
\begin{equation}\label{eq:Symm gen 2 - scheme 3}
    \frac{c_f}{2 E^x} \frac{\partial \bar{\alpha}_0}{\partial {\cal E}^x}
    = 0
    \,.
\end{equation}
The second equation is not compatible with the classical limit $c_f\to 1$ for an ${\cal E}^x$-dependent $\bar{\alpha}_0$.
Therefore, we will not consider scheme 3 any further either.

\subsubsection{Scheme 4}

We set $n=-1$.
We must adopt (\ref{eq:alpha2 schemes not 1}).
Using this, equations (\ref{eq:Symm gen 1}) and (\ref{eq:Symm gen 2}) become
\begin{eqnarray}\label{eq:Symm gen 1 - scheme 4}
    \left( \frac{c_f(\bar{a}_2-4) - \bar{\lambda}^2 \bar{\alpha}_0}{4 E^x}
    + \frac{\partial c_f}{\partial E^x} + c_f \frac{\partial \ln \lambda}{\partial E^x} \right) \frac{\partial \bar{\alpha}_0}{\partial {\cal E}^x}
    &&
    \nonumber\\
    - c_f \frac{\partial^2 \bar{\alpha}_0}{\partial E^x \partial {\cal E}^x}
    = 0
    \,,&&
\end{eqnarray}
\begin{eqnarray}\label{eq:Symm gen 2 - scheme 4}
    \left( \frac{\partial c_f}{\partial E^x} + c_f \frac{\bar{a}_2}{2 E^x} \right) \frac{\partial \bar{\alpha}_0}{\partial {\cal E}^x} = 0
    \,.
\end{eqnarray}
The second equation is precisely (\ref{eq:Symm gen 2 - scheme 2}) of scheme 2 which we showed is incompatible with the classical limit.
Therefore, we will not consider scheme 4 any further either.

Imposing the existence of the charge observable, together with compatibility with the classical limit, forces us to take scheme 1 and, therefore, adopt the $\bar{\alpha}_0$ function and the emergent electric field given by (\ref{eq:alpha0 sol}) and (\ref{eq:Emergent electric field - scheme 1 - class 1}), respectively.
This is in contrast to the four dimensional theory, which requires the use of scheme 4.

\subsection{Canonical transformations II}

As mentioned before, after performing the canonical transformation (\ref{eq:Diffeomorphism-constraint-preserving canonical transformations - Spherical - residual after modulo})\textemdash\,with ${\cal E}^x$-independent $\lambda$\textemdash\,and using the unbarred functions (\ref{eq:Redefinitions of lambda - ease notation}), the fiducial $\bar{\lambda}$ disappears from structure function and the whole constraint, but let us check this in detail for the $\alpha_0$ term given by (\ref{eq:alpha0 sol}) that we just solved for.
This term originally appears in the constraint as
\begin{eqnarray}
    \tilde{H} \!\!&\supset&\!\! - \bar{\lambda}_0 \frac{\sqrt{E^x}}{2} E^\varphi \frac{\bar{\alpha}_0}{E^x}
    \\
    \!\!&=&\!\! -  \lambda_0 \frac{\sqrt{E^x}}{2} E^\varphi \left( \frac{\lambda}{\bar{\lambda}} \frac{a_0}{E^x} - \frac{\lambda}{\bar{\lambda}} \frac{2}{(E^x)^2} \int {\rm d} {\cal E}^x\ \frac{{\cal E}^x}{\lambda_0} c_E ({\cal E}^x) \right)
    \nonumber
\end{eqnarray}
which, after the canonical transformation (\ref{eq:Diffeomorphism-constraint-preserving canonical transformations - Spherical - residual after modulo}), it becomes
\begin{equation}
    \tilde{H} \supset -  \lambda_0 \frac{\sqrt{E^x}}{2} E^\varphi \left( \frac{a_0}{E^x} - \frac{2}{(E^x)^2} \int {\rm d} {\cal E}^x\ \frac{{\cal E}^x}{\lambda_0} c_E ({\cal E}^x) \right)
    \,.
\end{equation}
The cancellation of the fiducial $\bar{\lambda}$ is possible only by the crucial addition of the $\bar{\lambda}/\lambda$ factor in the auxiliary function (\ref{eq:Alternative spatial metric}) and hence in the relation between the emergent electric field and the strength tensor (\ref{eq:Emergent strength field tensor relation to electromagnetic field - scheme n - spherical}).
Notice that in the latter expressions (\ref{eq:Alternative spatial metric}) and (\ref{eq:Emergent strength field tensor relation to electromagnetic field - scheme n - spherical}), the $\bar{\lambda}/\lambda$ factors disappear too under the canonical transformation (\ref{eq:Diffeomorphism-constraint-preserving canonical transformations - Spherical - residual after modulo}).

\subsection{General constraint}

We have completed the implementation of anomaly freedom, spacetime and electromagnetic covariance, and U(1) invariance, as well as the existence of the symmetry generators corresponding to the mass and charge observables characteristic of the spherically symmetric system.
We summarize our results in this section.
In the following, it will be useful to define the quantity
\begin{equation}\label{eq:Emergent calQ}
    {\cal Q} ({\cal E}^x,E^x) \equiv 2 \int_0 {\rm d} {\cal E}^x\ \frac{c_E ({\cal E}^x) {\cal E}^x}{\lambda_0}
    \,.
\end{equation}
We will refer to it as the charge function, which has the classical limit ${\cal Q} \to ({\cal E}^x)^2$ as $\lambda_0 \to 1$ and $c_E \to 1$.

\subsubsection{Periodic variables}

The resulting general constraint is given by
\begin{widetext}
\begin{eqnarray}
    \! \tilde{H} \!\!&=&\!\!
    - \bar{\lambda}_0 \frac{\sqrt{E^x}}{2} \Bigg[ E^\varphi \Bigg(
    \frac{\bar{a}_0}{E^x}
    - \frac{\bar{\cal Q}}{(E^x)^2}
    + 2 \frac{\sin^2 \left(\bar{\lambda} K_\varphi\right)}{\bar{\lambda}^2}\frac{\partial c_{f}}{\partial E^x}
    + 4 \frac{\sin \left(2 \bar{\lambda} K_\varphi\right)}{2 \bar{\lambda}} \frac{\partial \bar{q}}{\partial E^x}
    + \frac{\bar{\alpha}_2}{E^x} \left( c_f \frac{\sin^2 \left(\bar{\lambda} K_\varphi\right)}{\bar{\lambda}^2}
    + 2 \bar{q} \frac{\sin \left(2 \bar{\lambda} K_\varphi\right)}{2 \bar{\lambda}} \right)
    \Bigg)
    \nonumber\\
    &&\qquad \qquad \quad
    + 4 K_x \left(c_f \frac{\sin (2 \bar{\lambda} K_\varphi)}{2 \bar{\lambda}}
    + \bar{q} \cos(2 \bar{\lambda} K_\varphi)\right)
    - \frac{((E^x)')^2}{E^\varphi} \left(
    \frac{\bar{\alpha}_2}{4 E^x} \cos^2 \left( \bar{\lambda} K_\varphi \right)
    - \frac{K_x}{E^\varphi} \bar{\lambda}^2 \frac{\sin \left(2 \bar{\lambda} K_\varphi \right)}{2 \bar{\lambda}}
    \right)
    \nonumber\\
    &&\qquad \qquad \quad
    + \left( \frac{(E^x)' (E^\varphi)'}{(E^\varphi)^2}
    - \frac{(E^x)''}{E^\varphi} \right) \cos^2 \left( \bar{\lambda} K_\varphi \right)
    \Bigg]
    \,,
    \label{eq:Hamiltonian constraint - final - EM}
\end{eqnarray}
with structure function
\begin{equation}
    \tilde{q}^{x x}
    =
    \left(
    \left( c_{f}
    + \left(\frac{\bar{\lambda} (E^x)'}{2 E^\varphi} \right)^2 \right) \cos^2 \left(\bar{\lambda} K_\varphi\right)
    - 2 \bar{q} \bar{\lambda}^2 \frac{\sin \left(2 \bar{\lambda} K_\varphi\right)}{2 \bar{\lambda}}\right)
    \bar{\lambda}_0^2 \frac{E^x}{(E^\varphi)^2}
    \,.
    \label{eq:Structure function - final - EM}
\end{equation}
\end{widetext}
where $\bar{\lambda}$ is a constant and the rest of the barred parameters are abbreviations for
\begin{eqnarray}
    &&\!\!\!\!\!\!\!\!
    \lambda_0 = \bar{\lambda}_0 \frac{\lambda}{\bar{\lambda}}
    \quad,\quad
    q = \bar{q} \frac{\bar{\lambda}}{\lambda}
    \quad,\quad
    \Lambda_{0} = \frac{\bar{\lambda}^2}{\lambda^2} \bar{\Lambda}_0
    \quad,\quad
    a_0 = \frac{\bar{\lambda}^2}{\lambda^2} \bar{a}_0
    \,,
    \nonumber\\
    &&\!\!\!\!\!\!\!\!
    {\cal Q} = \frac{\bar{\lambda}^2}{\lambda^2} \bar{\cal Q}
    \quad,\quad
    \alpha_2 = \bar{\alpha}_2 + 4 E^x \frac{\partial \ln \lambda}{\partial E^x}
    \,.
    \label{eq:Redefinitions of lambda - ease notation - final - EM}
\end{eqnarray}
The parameters $\lambda$, $c_f$, $q$, $a_0$, and $\alpha_2$ are undetermined functions of $E^x$, while $c_E$ is an undetermined function of ${\cal E}^x$, and only $\lambda_0$ is an undetermined function of both $E^x$ and ${\cal E}^x$.

The only nontrivial strength tensor component is given by
\begin{equation}\label{eq:Strenght tensor - final}
    F_{0 x} = \frac{\partial \tilde{H}}{\partial {\cal E}^x} = \frac{\lambda}{\bar{\lambda}} \frac{E^\varphi}{(E^x)^{3/2}} c_E {\cal E}^x
    + \frac{\partial \ln \lambda_0}{\partial {\cal E}^x} \tilde{H}
    \,,
\end{equation}
where the second term may be ignored on shell.
The emergent electric field is given by
\begin{equation}\label{eq:Emergent electric field - final}
    \tilde{\cal E}^x \big|_{\rm O.S.} = \frac{\bar{\lambda}}{\lambda} \frac{(E^x)^{3/2}}{E^\varphi} \frac{\partial \tilde{H}}{\partial {\cal E}^x} = \frac{\lambda_0}{2} \frac{\partial {\cal Q}}{\partial {\cal E}^x}
    = c_E {\cal E}^x
    \,,
\end{equation}
and hence $\tilde{\cal E}^x$ depends solely, but nontrivially, on ${\cal E}^x$
If $\lambda_0$ is independent of $E^x$, then ${\cal Q}$ is independent of it too because of its definition (\ref{eq:Emergent calQ}), and in this case (\ref{eq:Emergent electric field - final}) implies that ${\cal Q}$ could be expressed in terms of $\tilde{\cal E}^x$.

The classical limit can be obtained in different ways, but the most straightforward is given by $\lambda \to \bar{\lambda}$,  followed by $\bar{\lambda} , q \to 0$, $\lambda_0 , \alpha_2, c_f , c_E \to 1$, and $a_0 \to 1 - \Lambda E^x$ with cosmological constant $\Lambda$.

The Gauss constraint is given by
\begin{equation}
    G = (\tilde{\cal E}^x)'
    \,,
\end{equation}
which implies that, on shell, the emergent electric field is constant $\tilde{\cal E}^x=Q$, taking the value of the electric charge.
Furthermore, $\tilde{\cal E}^x$ is a symmetry generator since it commutes with all the constraints because none of them depends on the electric vector potential $A_x$.

The mass observable is given by
\begin{widetext}
\begin{eqnarray}\label{eq:Vacuum mass observable - final - EM}
    \mathcal{M}
    &=&
    d_0
    + \frac{d_2}{2} \left(\exp \int {\rm d} E^x \ \frac{\bar{\alpha}_2}{2 E^x}\right)
    \left(
    c_f \frac{\sin^2\left(\bar{\lambda} K_{\varphi}\right)}{\bar{\lambda}^2}
    - \cos^2 (\bar{\lambda} K_\varphi) \left(\frac{(E^x)'}{2 E^\varphi}\right)^2
    + 2 \bar{q} \frac{\sin \left(2 \bar{\lambda}  K_{\varphi}\right)}{2 \bar{\lambda}}
    \right)
    \notag\\
    &&
    + \frac{d_2}{4} \int {\rm d} E^x \ \left( \frac{\bar{\alpha}_0}{E^x} \exp \int {\rm d} E^x \ \frac{\bar{\alpha}_2}{2 E^x}\right)
    \,,
\end{eqnarray}
where $d_0$ and $d_2$ are undetermined functions of the emergent electric field $\tilde{\cal E}^x$, with the classical limit defined by $d_0\to0$ and $d_2 \to 1$.
If we take the classical values $d_0=0$, $d_2=1$, $a_0=1-\Lambda E^x$, and $\alpha_2=1$, together with an $E^x$-independent $\lambda_0$, then the mass observable simplifies to
\begin{equation}\label{eq:Mass observable - final - EM - simplified}
    \mathcal{M}
    =
    \frac{\sqrt{E^x}}{2}
    \Bigg(
    1 - \frac{\Lambda}{3} E^x
    + \frac{{\cal Q}}{E^x}
    + \frac{\bar{\lambda}^2}{\lambda^2} \left( c_f \frac{\sin^2\left(\bar{\lambda} K_{\varphi}\right)}{\bar{\lambda}^2}
    - \cos^2 (\bar{\lambda} K_\varphi) \left(\frac{(E^x)'}{2 E^\varphi}\right)^2
    + 2 \frac{\lambda}{\bar{\lambda}} q \frac{\sin \left(2 \bar{\lambda}  K_{\varphi}\right)}{2 \bar{\lambda}}
    \right)
    \Bigg)
    \,.
\end{equation}
In this simpler case, the structure function (\ref{eq:Structure function - final - EM}) can be written in terms of the mass and the charge function,
\begin{equation}\label{eq:Structure function in mass observable - final - EM - simplified}
    \tilde{q}^{xx}
    =
    \left[ c_f + \lambda^2 \left( 1 - \frac{2 \mathcal{M}}{\sqrt{E^x}}
    - \frac{\Lambda E^x}{3}
    + \frac{{\cal Q}}{E^x} \right) \right] \frac{\bar{\lambda}^2}{\lambda^2} \lambda_0^2 \frac{E^x}{(E^\varphi)^2}
    \,.
\end{equation}
\end{widetext}
This simple choice still allows for a nontrivial emergent electric field. While the strength tensor (\ref{eq:Strenght tensor - final}), which is responsible for the Lorentz force, depends on the emergent electric field only, the gravitational attraction due to the presence of an electric field is affected by the distinction between the emergent and fundamental electric fields because ${\cal Q}\neq({\cal E}^x)^2$ in general.

Furthermore, the structure function and hence the spacetime metric may depend on the electric momentum ${\cal E}^x$ through the global factor $\lambda_0$, even at the kinematical level.
This feature of the spacetime metric directly depending on matter degrees of freedom kinematically was first discussed in \cite{EMGscalar} in the context of scalar matter coupling, and our procedure here shows that it extends to the electric field.

The periodicity of the constraint (\ref{eq:Hamiltonian constraint - final - EM}) on the gravitational variable $K_\varphi$ makes it suitable for loop quantization.
However, this periodicity is specific of the phase space coordinates adopted here and depend on the fiducial $\bar{\lambda}$.
This reference parameter can be completely eliminated by a canonical transformation, as discussed in the previous Subsection, at the expense of breaking periodicity.
While the periodic version is preferred for loop quantization, the non-periodic version can result in simpler equations of motion in certain cases.
The two versions, however, describe the same system because they differ by a simple canonical transformation and hence it is desirable to have both.

\subsubsection{Non-periodic variables}

Taking the canonical transformation (\ref{eq:Diffeomorphism-constraint-preserving canonical transformations - Spherical - residual after modulo})\textemdash\,with ${\cal E}^x$-independent $\lambda$\textemdash\,and using the definitions (\ref{eq:Redefinitions of lambda - ease notation}), the constraint (\ref{eq:Hamiltonian constraint - final - EM}) becomes
\begin{widetext}
\begin{eqnarray}
    \tilde{H}
    &=& - \sqrt{E^x} \frac{\lambda_0}{2} \bigg[ E^\varphi \bigg( c_{f 0}
    + \frac{a_0}{E^x}
    - \frac{\cal Q}{(E^x)^2}
    + 2 \frac{\sin^2 \left(\lambda K_\varphi\right)}{\lambda^2}\frac{\partial c_{f}}{\partial E^x}
    \nonumber\\
    &&
    + 4 \frac{\sin \left(2 \lambda K_\varphi\right)}{2 \lambda} \frac{1}{\lambda} \frac{\partial \left(\lambda q\right)}{\partial E^x}
    + \left(\frac{\alpha_2}{E^x} - 2 \frac{\partial \ln \lambda^2}{\partial E^x}\right) \left( c_f \frac{\sin^2 \left(\lambda K_\varphi\right)}{\lambda^2}
    + 2 q \frac{\sin \left(2 \lambda K_\varphi\right)}{2 \lambda} \right)
    \nonumber\\
    &&
    + \left(4 \frac{K_x}{E^\varphi} + \frac{\partial \ln \lambda^2}{\partial E^x} 2 K_\varphi \right) \left(c_f \frac{\sin (2 \lambda K_\varphi)}{2 \lambda}
    + q \cos(2 \lambda K_\varphi)\right)
    \bigg)
    \nonumber\\
    &&
    + \frac{((E^x)')^2}{E^\varphi} \bigg(
    - \frac{\alpha_2}{4 E^x} \cos^2 \left( \lambda K_\varphi \right)
    + \frac{\sin \left(2 \lambda K_\varphi \right)}{2 \lambda} \frac{\lambda^2}{2} K_\varphi \frac{\partial \ln \lambda^2}{\partial E^x}
    + \frac{K_x}{E^\varphi} \lambda^2 \frac{\sin \left( 2 \lambda K_\varphi \right)}{2 \lambda} \bigg)
    \nonumber\\
    &&
    + \left(\frac{(E^x)' (E^\varphi)'}{(E^\varphi)^2}
    - \frac{(E^x)''}{E^\varphi}\right) \cos^2 \left( \lambda K_\varphi \right)
    \bigg]
    \,,
    \label{eq:Hamiltonian constraint - final - EM - non-periodic}
\end{eqnarray}
with the associated  structure function 
\begin{equation}
    \tilde{q}^{x x} =
    \left(
    \left( c_f
    + \lambda^2 \left( \frac{(E^x)'}{2 E^\varphi} \right)^2
    \right)
    \cos^2 \left( \lambda K_\varphi \right)
    - 2 q \lambda^2 \frac{\sin (2 \lambda K_\varphi)}{2 \lambda}
    \right) \lambda_0^2
    \frac{E^x}{(E^\varphi)^2}
    \,.
    \label{eq:Structure function - final - EM - non-periodic}
\end{equation}
\end{widetext}
The only nontrivial strength tensor component is given by
\begin{equation}\label{eq:Strenght tensor - final - non-periodic}
    F_{0 x} \big|_{\rm O.S.} = \frac{\partial \tilde{H}}{\partial {\cal E}^x} = \frac{E^\varphi}{(E^x)^{3/2}} \lambda_0 \frac{\partial {\cal Q}}{\partial {\cal E}^x}
    \,.
\end{equation}
and the emergent electric field by
\begin{equation}\label{eq:Emergent electric field - final - non-periodic}
    \tilde{\cal E}^x \big|_{\rm O.S.} = \frac{(E^x)^{3/2}}{E^\varphi} \frac{\partial \tilde{H}}{\partial {\cal E}^x} = \frac{\lambda_0}{2} \frac{\partial {\cal Q}}{\partial {\cal E}^x}
    \,.
\end{equation}

If we again consider the simpler case of $E^x$-independent $\lambda_0$ and the classical values $d_0=0$, $d_2=1$, $a_0=1-\Lambda E^x$, and $\alpha_2=1$, we obtain the mass observable
\begin{widetext}
\begin{equation}\label{eq:Mass observable - final - EM - simplified - charge obs - non-periodic}
    \mathcal{M}
    =
    \frac{\sqrt{E^x}}{2}
    \left(
    1 - \frac{\Lambda}{3} E^x + \frac{\cal Q}{E^x} + c_f \frac{\sin^2\left(\lambda K_{\varphi}\right)}{\lambda^2}
    - \cos^2 (\lambda K_\varphi) \left(\frac{(E^x)'}{2 E^\varphi}\right)^2
    + 2 q \frac{\sin \left(2 \lambda  K_{\varphi}\right)}{2 \lambda}
    \right)
    \,,
\end{equation}
and the structure function (\ref{eq:Structure function - final - EM - non-periodic}) can be written as
\begin{equation}\label{eq:Structure function in mass observable - final - EM - simplified - charge obs - non-periodic}
    \tilde{q}^{xx}
    =
    \left( c_f + \lambda^2 \left( 1 - \frac{2 \mathcal{M}}{\sqrt{E^x}}
    - \frac{\Lambda}{3} E^x
    + \frac{\cal Q}{E^x} \right) \right) \lambda_0^2 \frac{E^x}{(E^\varphi)^2}
    \,.
\end{equation}
\end{widetext}

\subsection{The role of emergence for modified electromagnetism}

If one neglects the possibility of the emergent electric field being different from the fundamental electric momentum, then we must take $\tilde{\cal E}^x={\cal E}^x$ such that (\ref{eq:Emergent electric field - final - non-periodic}) requires that $c_E=1$.
In this case, the only possible electromagnetic modification is through the global factor $\lambda_0$.
Attempting to reproduce an important modification such as that of the Wilson action (\ref{eq:Wilson action - EM}) would result in a global factor that does not respect the vacuum limit.
However, choosing nontrivial $c_E({\cal E}^x)$ and $\lambda_0$ we are able to reproduce a similar modification by requiring that (\ref{eq:Emergent calQ}) becomes
\begin{equation}\label{eq:Wilson action mod - spherical}
    {\cal Q} = 2 \frac{1-\cos (\mathfrak{a} \tilde{\cal E}^x)}{\mathfrak{a}^2}
    \,,
\end{equation}
with a real constant $\mathfrak{a}$.

Using the natural units (\ref{eq:Natural units}), in spherical symmetry, the vector potential components $A_x$ and $A_t$ are dimensionless, the strength tensor component $F_{0x}$ has dimensions of $L^{-1}$, while the electric charge and the densitized electric field ${\cal E}^x$ have dimensions of $L$.
The last one is in contrast to the four dimensional case in Cartesian-type coordinates, where the dimensions of the densitized electric field are $L^{-1}$. This discrepancy is owed to the dimensions acquired by the densitization in spherical coordinates.
The constant $\mathfrak{a}$, therefore, has dimensions of $L^{-1}$, similar to the electromagnetic holonomy parameter in the four-dimensional model (\ref{eq:Wilson line}).

Substituting (\ref{eq:Wilson action mod - spherical}) into the expression for the emergent electric field (\ref{eq:Emergent electric field - final - non-periodic}) we obtain
\begin{equation}
    \tilde{\cal E}^x = \lambda_0 \frac{\sin (\mathfrak{a} \tilde{\cal E}^x)}{\mathfrak{a}} \frac{\partial \tilde{\cal E}^x}{\partial {\cal E}^x}
\end{equation}
While solving for the general relation $\tilde{\cal E}^x ({\cal E}^x)$ is complicated, we may not need it for the purposes of studying several physical effects that involve only ${\cal Q}$ and $\tilde{\cal E}^x$.
The classical values ${\cal Q} \to (\tilde{\cal E}^x)^2$ and $\tilde{\cal E}^x\to{\cal E}^x$ are recovered in the limit given by $\lambda_0\to1$ and $\mathfrak{a} \to 0$.
Alternatively, keeping the constant $\mathfrak{a}$ fixed, the charge function (\ref{eq:Wilson action mod - spherical}) approximates its classical value ${\cal Q}\to(\tilde{\cal E}^x)^2= Q^2$ for small enough charges such that $\mathfrak{a} Q^2\ll1$.

The charge function ${\cal Q}$ is bounded from below by zero and from above by
\begin{equation}\label{eq:Max value of charge function - Wilson}
    {\cal Q}_{\rm max} = \frac{4}{\mathfrak{a}^2}
    \,.
\end{equation}
Because ${\cal Q}$ appears directly in the spacetime metric through (\ref{eq:Structure function in mass observable - final - EM - simplified - charge obs - non-periodic}), at least for the simpler case of $E^x$-independent $\lambda_0$, its boundedness may have important geometric effects.

Lastly, we note that the chosen modification (\ref{eq:Wilson action mod - spherical}) will result in the following term in the Hamiltonian constraint,
\begin{equation}\label{eq:Wilson action mod in H - EM - spherical}
    \tilde{H} \supset \lambda_0 \frac{E^\varphi}{(E^x)^{3/2}} \frac{1-\cos (\mathfrak{a} \tilde{\cal E}^x)}{\mathfrak{a}^2}
    \,.
\end{equation}
Because the emergent electric field is directly related to the strength tensor component $F_{0x}$, a comparison with the Wilson action (\ref{eq:Wilson action - EM}) offers an interpretation for $\mathfrak{a}$ as the electromagnetic holonomy parameter.
This is possible despite our treatment of the $t-x$ part of the manifold as a continuum.
The fact that $\tilde{\cal E}^x=Q$ is a constant on shell in the spherically symmetric model might be the reason this modification is possible without full homogeneity in the $t$ and $x$ directions.

It is also worth pointing out that while (\ref{eq:Wilson action mod in H - EM - spherical}) resembles the respective Wilson action term in (\ref{eq:Wilson action - EM}), $F_{0x}$ depends not only on $\tilde{\cal E}^x$ but also on the gravitational variables due to densitization, through (\ref{eq:Strenght tensor - final - non-periodic}), which would in principle have to appear in the argument of the cosine in (\ref{eq:Wilson action mod in H - EM - spherical}) for a full match.
While the relation to the Wilson action is not perfect, we may still consider this modification to model the boundedness effects implied by the Wilson action.
We stress, however, that the concept of the emergent electric field is more general than, and not contingent on, lattice effects, such that the modifications are still applicable beyond this particular interpretation.
We will test the effects of modification (\ref{eq:Wilson action mod - spherical}) explicitly in the next section, where we obtain a nonsingular, charged black hole solution.

\section{Black hole solution}
\label{eq:BH sol}

In this section, we will consider the electrovac constraints (\ref{eq:Hamiltonian constraint - final - EM}) and (\ref{eq:Hamiltonian constraint - final - EM - non-periodic}) and take the classical values $c_f = \alpha_2 = 1$, $a_0 = 1-\Lambda E^x$, $q=0$, and constant $\lambda_0$ for simplicity, where $\Lambda$ is the cosmologocial constant.
We, however, consider arbitrary $\lambda (E^x)$ and $c_E ({\cal E}^x)$, though we will be mainly concerned with monotonically decreasing $\lambda (E^x)$ that are asymptotically vanishing or constant.
We follow the procedure of \cite{ELBH} almost step-by-step to generate the global structure of the black hole solution.

In comparing the relation between the different gauges, we will make use of the coordinate transformation of the strength tensor, whose only nontrivial component in the Eulerian frame is $F_{0 x}$.
This component corresponds, in the observer's frame, to $F_{t x}=t^\mu F_{\mu x}=N F_{0x}$, which is obtained by use of $t^\mu=N n^\mu+N^x s^\mu_x$.
The coordinate transformation from $x^\mu = (t,x)$ to $X^\mu = (T,X)$ is given by
\begin{eqnarray}\label{eq:Strength tensor coord transf}
    &&\!\!\!\!\!\!\!\!
    F_{TX}(X,T) = \frac{\partial x^\mu}{\partial T} \frac{\partial x^\nu}{\partial X} F_{\mu\nu} (t(T,X),x(T,X))
    \\
    &&\!\!\!\!\!\!\!\!
    = \left[ \left(\frac{\partial T}{\partial t} \frac{\partial X}{\partial x}\right)^{-1}
    - \left(\frac{\partial T}{\partial x} \frac{\partial X}{\partial t}\right)^{-1} \right] F_{tx} (t(T,X),x(T,X))
    \nonumber
\end{eqnarray}

\subsection{Schwarzschild gauge: Static exterior}
\label{sec:Schwarzschild exterior}

In this subsection we use the non-periodic version of the constraint (\ref{eq:Hamiltonian constraint - final - EM - non-periodic}) because it gives the simpler equations of motion in the Schwarzschild gauge, defined by
\begin{equation}
    N^x = 0 \quad,\quad E^x = x^2
    \,.
\end{equation}
This is a partial gauge fixing and we retain a residual gauge freedom for the Lagrange multiplier $A_t$ given by the electric potential.

The consistency equation $\dot{E}^x = 0$ implies the condition
\begin{equation}
    \left( 1
    + \lambda^2 \left(\frac{(E^x)'}{2 E^\varphi}\right)^2 \right) \frac{\sin (2 \lambda K_\varphi)}{2 \lambda}
    = 0
    \,,
\end{equation}
which is solved, consistently with the classical limit, by $K_\varphi = 0$.
This, in turn, implies the consistency equation $\dot{K}_\varphi = 0$, which we will use to solve for the lapse below.

We now impose the on-shell conditions.
The vanishing of the Gauss constraint $G=0$ determines $\tilde{\cal E}^x = Q$ with $Q$ being the constant electric charge, and hence
\begin{equation}
    \tilde{\cal E}^x = c_E ({\cal E}^x) {\cal E}^x = Q
    \,.
\end{equation}
The vanishing of the vector constraint $H_x=0$ implies that $K_x = 0$. Finally, the vanishing of the Hamiltonian constraint $\tilde{H}=0$ implies the equation
\begin{equation}
    \left( x^2 - \Lambda x^4 - {\cal Q} \right) (E^\varphi)^2 - 3 x^4 + x^5 (\ln (E^\varphi)^2)'
    = 0
    \,,
\end{equation}
where $\mathcal{Q}$ is given by (\ref{eq:Emergent calQ}), which has the general solution
\begin{equation}
    E^\varphi = x \left(1 - \frac{c_\varphi}{x} - \frac{\Lambda}{3} x^2 + \frac{{\cal Q}}{x^2}\right)^{-1/2}
    \,,
\end{equation}
where $c_\varphi$ is a constant.
Substituting all these results into the mass observable (\ref{eq:Mass observable - final - EM - simplified - charge obs - non-periodic}), and rewriting it as $\mathcal{M} = M$, determines
\begin{equation}
    c_\varphi = 2 M
    \,.
\end{equation}

Lastly, $\dot{K}_\varphi=0$ implies the classical value for the lapse, up to a scaling constant $\mu$, since $K_\varphi = 0$ eliminates all $\lambda$ dependence. Therefore,
\begin{equation}
    N = \frac{1}{\mu \lambda_0} \sqrt{1 - \frac{2 M}{x} - \frac{\Lambda x^2}{3} + \frac{{\cal Q}}{x^2}}
    \,,
\end{equation}
where in the following we shall fix $\mu=1/\lambda_0$ for simplicity\textemdash\,note that $\mu$ can be interpreted as a rescaling of the time coordinate, so there is no loss of generality in this choice.

The structure function (\ref{eq:Structure function in mass observable - final - EM - simplified - charge obs - non-periodic}) can be rewritten as
\begin{widetext}
\begin{equation}
    \tilde{q}^{x x} =
    \lambda_0^2 \left( 1 + \lambda^2 \left( 1 - \frac{2 M}{x}
    - \frac{\Lambda x^2}{3}
    + \frac{{\cal Q}}{x^2} \right) \right) \left(1 - \frac{2 M}{x} - \frac{\Lambda}{3} x^2 + \frac{{\cal Q}}{x^2}\right)
    \,,
    \label{eq:Structure function - non-periodic version - Schwarzschild}
\end{equation}
where $\lambda$ is an arbitrary function of $E^x=x^2$.
The emergent line element is then given by
\begin{equation}
    {\rm d} s^2 =
    - \left(1 - J(x)\right) {\rm d} t^2
    + \left( 1 + \lambda^2 \left( 1 - J(x) \right) \right)^{-1}
    \left(1 - J(x)\right)^{-1} \frac{{\rm d} x^2}{\lambda_0^2}
    + x^2 {\rm d} \Omega^2
    \label{eq:Spacetime metric - modified - Schwarzschild}
\end{equation}
\end{widetext}
where, for brevity, we have defined
\begin{equation}\label{eq:J function}
    J (x) = \frac{2 M}{x} + \frac{\Lambda x^2}{3} - \frac{{\cal Q}}{x^2}
    \,.
\end{equation}
The only nontrivial strength tensor component is given by
\begin{equation}\label{eq:Strength tensor static Sch}
    F_{tx} = N F_{0 x} = N \sqrt{\frac{(E^\varphi)^2}{E^x}} \frac{\tilde{\cal E}^x}{E^x}
    = \frac{Q}{x^2}
    \,.
\end{equation}

For completeness, the equation of motion for the electric vector potential is given by
\begin{eqnarray}
    \dot{A}_x &=& \{A_x , \tilde{H}[N] + H_x[N^x] - G[A_t]\}
    \nonumber\\
    &=& \left(\frac{1}{x^2}
    + A_t' \right) Q
    \,,
\end{eqnarray}
where $A_t$ is the electric scalar potential.
Fixing the residual gauge freedom, we may choose the potential $A_t = 1/x$, such that $\dot{A}_x=0$ and hence $A_x = A_x (x)$ is time independent, but otherwise undetermined up to boundary conditions.
The explicit expression of $A_x$ is not necessary for most applications since it does not appear in the spacetime metric nor in the strength tensor.

In the weak gravity regime, such that ${\cal Q}/x^2,\Lambda x^2, M/x,\lambda\ll1$ we get an approximately Minkowskian spacetime.
Using (\ref{eq:Strength tensor static Sch}), the Lorentz force equation (\ref{eq:Lorentz force}) is identical to that of special relativity, revealing that a nontrivial charge function ${\cal Q}$ has no electrodynamic effects on test particles.
Therefore, the effects of a modified charge function ${\cal Q}$ are purely gravitational and can only be seen in the strong gravity regime.
On the other hand, the dynamical effects of a nontrivial charge function on the electromagnetic field itself, for instance on electromagnetic waves, cannot be studied in spherical symmetry where no local, propagating degrees of freedom exist; therefore, this must be studied on systems with less degree of symmetry or by introducing electromagnetic perturbations.

A new coordinate singularity appears in the spacetime metric (\ref{eq:Spacetime metric - modified - Schwarzschild}) at the radial coordinate $x_\lambda$ solving the equation
\begin{equation}\label{eq:New coord sing Schwarzschild}
    S(x_\lambda) \equiv 1 + \lambda(x_\lambda)^2 \left( 1 - J(x_\lambda) \right) = 0
    \,.
\end{equation}

The case of constant $\lambda = \bar{\lambda}$ and non-emergent electric field, such that $\mathcal{Q}=Q^2$, was previously obtained in \cite{alonsobardaji2023Charged} in the Schwarzschild gauge with no underlying theory for the electromagnetic field.
However, while, in this special case, the model of \cite{alonsobardaji2023Charged} reproduces the correct metric, it is important to note that it differs kinematically to the general model presented here because it adds an electromagnetic contribution to the vector constraint, a feature we argued in Subsection~\ref{sec:EM spherical classical} is conceptually wrong, stemming from an incomplete discussion of electromagnetic covariance.
In the following, we will proceed to obtain a solution in other gauges, identify the coordinate transformations relating them all, and sew them together to arrive at the global structure.

\subsection{Schwarzschild gauge: Homogeneous interior}

Seeking a homogeneous solution for the black hole's interior, we may set all spatial derivatives to zero and label the time coordinate by $t_{\rm h}$ and the spatial coordinate by $x_{\rm h}$.
The homogeneous gauge is defined by the partial gauge fixing
\begin{equation}\label{eq:Homogeneous partial gauge fixing}
    N^x = 0 \quad,\quad N' = 0
    \,,
\end{equation}
which preserves homogeneity.

Completing the spacetime gauge fixing by choosing $E^x = t_{\rm h}^2$ the equations of motion can be solved\textemdash\,following exactly the procedure of \cite{ELBH}, but including the charge function in the $J$ function (\ref{eq:J function}) almost trivially\textemdash\,, obtaining
\begin{equation}
    \frac{\sin^2 ( \lambda K_\varphi)}{\lambda^2} = J(t_{\rm h}) - 1
    \,,
\end{equation}
for the angular curvature,
\begin{equation}
    E^\varphi = \frac{t_{\rm h}}{\mu} \sqrt{\left( 1 - \lambda^2 \left(J(t_{\rm h})-1\right) \right) \left(J(t_{\rm h}) - 1\right)}
    \,,
\end{equation}
for the angular triad with constant $\mu$, and
\begin{equation}
    N
    = \frac{\left( 1-\lambda^2\left(J(t_{\rm h})-1\right)\right)^{-1/2}}{\lambda_0} \left(J(t_{\rm h})-1\right)^{-1/2}
    \,,
\end{equation}
for the lapse.
The radial curvature $K_x$ is obtained by solving the Hamiltonian constraint $\tilde{H}=0$, but the expression is long and not necessary for our application here.

The structure function is given by
\begin{equation}
    \tilde{q}^{x x} = \mu^2 \lambda_0^2 \left(J(t_{\rm h})-1\right)^{-1}
    \,,
\end{equation}
and we therefore obtain the emergent spacetime metric
\begin{eqnarray}
    {\rm d} s^2 &=&
    - \left( 1 - \lambda^2 \left(J(t_{\rm h}) - 1\right)
    \right)^{-1} \left(J(t_{\rm h}) - 1\right)^{-1} \frac{{\rm d} t_{\rm h}^2}{\lambda_0^2}
    \nonumber\\
    &&
    + \left(J(t_{\rm h}) - 1\right) {\rm d} x_{\rm h}^2
    + t_{\rm h}^2 {\rm d} \Omega^2
    \,,
    \label{eq:Spacetime metric homogeneous - modified - Schwarzschild}
\end{eqnarray}
where we have again taken $\mu=1/\lambda_0$.
This matches the static Schwarzchild metric (\ref{eq:Spacetime metric - modified - Schwarzschild}) up to the label swap $x \to t_{\rm h}$, $t \to x_{\rm h}$.
The only nontrivial strength tensor component is given by
\begin{equation}\label{eq:Strength tensor hom Sch}
    F_{t_{\rm h} x_{\rm h}}=N F_{0_{\rm h} x_{\rm h}} = N \sqrt{\frac{(E^\varphi)^2}{E^x}} \frac{\tilde{\cal E}^x}{E^x}
    = \frac{Q}{t_{\rm h}^2}
    \,.
\end{equation}

For completeness, the equation of motion for the electric vector potential is
\begin{equation}
    \dot{A}_x = \{A_x , \tilde{H}[N] + H_x[N^x] - G[A_t]\} = \frac{Q}{t_{\rm h}^2}
    \,,
\end{equation}
where we have used the partial gauge fixing $A_t'=0$, and hence
\begin{equation}
    A_x = c_a - \frac{Q}{3 t_{\rm h}^3}
    \,,
\end{equation}
where $c_a$ is a constant and may be chosen according to initial conditions.

\subsection{Gullstrand-Painlev\'e gauge}

\subsubsection{Schwarzschild to Gullstrand-Painlev\'e coordinate transformation}
The metric (\ref{eq:Spacetime metric - modified - Schwarzschild}) is static, and, thus, has a timelike Killing vector $\partial_t$.
Therefore, a timelike geodesic observer with 4-velocity $u^\mu$ has a Killing conserved energy given by $\varepsilon=-u_t = - g_{t\nu}u^\nu$.
Furthermore, if the observer is at rest at $x=x_0$, we have $u^t |_{x= x_0}=\tilde{g}^{tt} u_t |_{x=x_0}=-\tilde{g}^{tt}|_{x=x_0}\varepsilon$.
Timelike normalization, $-1=\tilde{g}_{\mu\nu} u^\mu u^\nu|_{x=x_0}=\tilde{g}_{tt}(u^t)^2|_{x=x_0}$, then implies
\begin{equation}
    \varepsilon^2 = 1 - J(x_0)
    \,.
\end{equation}
The covector field $u_\mu$ defines the 1-form
\begin{widetext}
\begin{equation}\label{eq:Sch to GP coord transf}
    {\rm d} t_{\rm GP} = - u_\mu d x^\mu
    =
    \varepsilon {\rm d} t
    + \frac{s}{\lambda_0} \sqrt{J(x) - J(x_0)} \left( 1 + \lambda^2 \left( 1 - J(x) \right)
    \right)^{-1/2}
    \left(1 - J(x)\right)^{-1} {\rm d} x
    \,,
\end{equation}
where $s = \pm 1$, and we have obtained $u_x$ from the normalization $\tilde{g}^{\mu \nu} u_\mu u_\nu=-1$.
The explicit integration will depend on the choice of $\lambda$.
Substituting ${\rm d} t = \varepsilon^{-1} {\rm d} t_{\rm GP} + \varepsilon^{-1} u_x {\rm d} x$ in (\ref{eq:Spacetime metric - modified - Schwarzschild}) we obtain the emergent metric in Gullstrand-Painlev\'e (GP) coordinates
\begin{equation}\label{eq:Spacetime metric - GP}
    {\rm d} s^2_{\rm GP}
    =
    - {\rm d} t_{\rm GP}^2
    + x^2 {\rm d} \Omega^2
    + \frac{1}{\lambda_0^2\varepsilon^{2}} \left( 1 + \lambda^2 \left( 1 - J(x) \right)
    \right)^{-1}
    \left( {\rm d} x 
    + s \lambda_0 \sqrt{J(x)-J(x_0)} \sqrt{1 + \lambda^2 \left( 1 - J(x) \right)} {\rm d} t_{\rm GP} \right)^2
    \,.
    \nonumber
\end{equation}
\end{widetext}

Since the homogeneous Schwarzschild metric (\ref{eq:Spacetime metric homogeneous - modified - Schwarzschild}) is identical to the static one up to the label swap of the $t$ and $x$ coordinates, it follows that it is related to the GP metric (\ref{eq:Spacetime metric - GP}) by the same coordinate transformation up to a label swap $t_{\rm GP}\to X_{\rm GP} = x_{\rm h}$, $x \to T_{\rm GP}$.

\subsubsection{Gullstrand-Painlev\'e gauge}

The GP metric (\ref{eq:Spacetime metric - GP}) covers both the static exterior and the homogeneous interior in a single coordinate chart.
Furthermore, this solution can be obtained directly from solving the equations of motion in the appropriate GP gauge defined as
\begin{equation}
    N = 1 \quad,\quad E^x = x^2
    \,,
\end{equation}
up to the choice of an electric potential.
The result of solving the equations of motion is, see \cite{ELBH} for details,
\begin{equation}
    E^\varphi = x/ \varepsilon
    \,,
\end{equation}
for the angular triad with constant $\varepsilon$,
\begin{equation}
    \frac{\sin^2 (\lambda K_\varphi)}{\lambda^2}
    =
    \left( 1 + \lambda^2 \varepsilon^2 \right)^{-1} \left( J(x) + \varepsilon^2 - 1 \right)
    \,,
\end{equation}
for the angular curvature, and
\begin{eqnarray}
    N^x
    &=&
    - \lambda_0 \left( 1
    + \lambda^2 \varepsilon^2 \right) \frac{\sin (2 \lambda K_\varphi)}{2 \lambda}
    \\
    &=&
    - \lambda_0 \sqrt{J(x) + \varepsilon^2 - 1} \sqrt{ 1 + \lambda^2 \left( 1 - J(x) \right)}
    \,,\nonumber
\end{eqnarray}
for the shift.
The radial curvature is obtained from solving the vector constraint, yielding $K_x = E^\varphi K_\varphi' / (2 x)$.

The resulting metric is therefore given by (\ref{eq:Spacetime metric - GP}) by identifying $\varepsilon^2 = \left(1 - J(x_0)\right)$.
The only nontrivial strength tensor component is given by
\begin{equation}
    F_{t_{\rm GP} x}^{\rm (GP)}=N F_{0 x}^{\rm (GP)} = N \sqrt{\frac{(E^\varphi)^2}{E^x}} \frac{\tilde{\cal E}^x}{E^x}
    = \frac{Q}{\varepsilon x^2}
    \,.
\end{equation}
Using (\ref{eq:Strength tensor coord transf}) and (\ref{eq:Sch to GP coord transf}), the transformation of the strength tensor (\ref{eq:Strength tensor static Sch}) from the static Schwarzschild coordinates yields
\begin{equation}
    F_{t_{\rm GP} x}^{\rm (GP)}
    = \left(\frac{\partial t_{\rm GP}}{\partial t}\right)^{-1} F_{tx}^{\rm Sch} (t(t_{\rm GP},x))
    = \frac{Q}{\varepsilon x^2}
    \,,
\end{equation}
which is in agreement with the canonical derivation.

\subsection{Internal time gauge: Homogeneous interior}
\label{sec:Internal time}

For this subsection, we will use the periodic version of the constraint given by (\ref{eq:Hamiltonian constraint - final - EM}) because it gives the simpler equations of motion in the following gauge.

Once again, we seek a homogeneous solution for the black hole's interior and hence set all spatial derivatives to zero, label the time coordinate by $t_{\varphi}$ and the spatial coordinate by $x_{\rm h}$, and impose the partial gauge fixing (\ref{eq:Homogeneous partial gauge fixing}) to preserve homogeneity under time evolution.

However, instead of completing the gauge by setting $E^x=t_{\rm h}^2$, we choose the angular curvature to define the time coordinate by $K_\varphi = - t_{\varphi}$.
In this gauge, the radial triad is given by the solution to the equation
\begin{equation}
    \frac{\sin^2 \left(\bar{\lambda} t_\varphi\right)}{\lambda(E^x)^2}
    = J\left(\sqrt{E^x}\right) - 1
    \,,
    \label{eq:Curvature-triad - Homogeneous - modified - periodic}
\end{equation}
the angular triad is given by
\begin{equation}
    E^\varphi = - \lambda_0 \sqrt{E^x} \frac{\bar{\lambda}^2}{\lambda^2} \frac{\sin \left(2 \bar{\lambda} t_\varphi\right)}{2\bar{\lambda}}
\end{equation}
and the lapse by
\begin{widetext}
\begin{equation}
    N
    =
    \frac{2 \sqrt{E^x}}{\lambda_0} \frac{\bar{\lambda}}{\lambda} \left( 1 - \Lambda E^x - \frac{\cal Q}{E^x}
    + \left( 1 - 2 \frac{\partial \ln \lambda^2}{\partial \ln E^x}\right) \frac{\sin^2 \left(\bar{\lambda} t_\varphi\right)}{\lambda^2}
    \right)^{-1}
    \,.
    \label{eq:Lapse homogeneous - modified - Internal time}
\end{equation}

The resulting line element is given by
\begin{equation}
    {\rm d} s^2 =
    - 4 E^x \frac{\bar{\lambda}^2}{\lambda^2}
    \left( 1 - \Lambda E^x - \frac{{\cal Q}}{E^x} + \left( 1 - 2 \frac{\partial \ln \lambda^2}{\partial \ln E^x} \right) \frac{\bar{\lambda}^2}{\lambda^2} \frac{\sin^2 \left(\bar{\lambda} t_\varphi\right)}{\bar{\lambda}^2} \right)^{-2} \frac{{\rm d} t_\varphi^2}{\lambda_0^2}
    + \frac{\bar{\lambda}^2}{\lambda^2} \frac{\sin^2 \left(\bar{\lambda} t_\varphi\right)}{\bar{\lambda}^2} {\rm d} x_{\rm h}^2
    + E^x {\rm d} \Omega^2
    \,.
    \label{eq:Spacetime metric homogeneous - modified - Internal time}
\end{equation}
This is related to the homogeneous Schwarzschild metric (\ref{eq:Spacetime metric homogeneous - modified - Schwarzschild}) by the coordinate transformation defined by
\begin{eqnarray}\label{eq:Schwarzchild-Internal time transformation - Homogeneous}
    &&\frac{\sin^2 \left(\bar{\lambda} t_\varphi\right)}{\lambda(t_{\rm h}^2)^2}
    =
    \frac{2 M}{t_{\rm h}} - 1 + \frac{\Lambda t_{\rm h}^2}{3} - \frac{\cal Q}{t_{\rm h}^2}
    \,,
    \\
    &&{\rm d} t_{\rm h}
    =
    - 2 \sqrt{E^x} \frac{\bar{\lambda}^2}{\lambda^2}
    \frac{\sin \left(2 \bar{\lambda} t_{\varphi}\right)}{2 \bar{\lambda}}
    \left( 
    1 - \Lambda E^x - \frac{{\cal Q}}{E^x}
    + \left( 1 - 2 \frac{\partial \ln \lambda^2}{\partial \ln E^x} \right) \frac{\bar{\lambda}^2}{\lambda^2} \frac{\sin^2 \left(\bar{\lambda} t_\varphi\right)}{\bar{\lambda}^2} \right)^{-1} {\rm d} t_\varphi
    = N \frac{\lambda}{\bar{\lambda}} \frac{E^\varphi}{\sqrt{E^x}} {\rm d} t_\varphi
    \nonumber
\end{eqnarray}
\end{widetext}

The only nontrivial strength tensor component is given by
\begin{equation}
    F_{t_\varphi x}^{(\varphi)}=NF_{0_\varphi x}^{(\varphi)} = N \frac{\lambda}{\bar{\lambda}} \frac{E^\varphi}{\sqrt{E^x}} \frac{Q}{E^x}
    \,,
\end{equation}
where the variables are given by the above solutions, yielding a long expression.
Using (\ref{eq:Strength tensor coord transf}) and (\ref{eq:Schwarzchild-Internal time transformation - Homogeneous}), the transformation of the strength tensor (\ref{eq:Strength tensor hom Sch}) from the homogeneous Schwarzschild coordinates yields
\begin{equation}
    F_{t_\varphi x}^{(\varphi)}
    = \frac{\partial t_{\rm h}}{\partial t_\varphi} F_{t_{\rm h} x_{\rm h}}^{\rm (Sch)} (t_{\rm h}(t_\varphi))
    = N \frac{\lambda}{\bar{\lambda}} \frac{E^\varphi}{\sqrt{E^x}} \frac{Q}{E^x}
    \,,
\end{equation}
which is in agreement with the canonical derivation.

Equation (\ref{eq:Schwarzchild-Internal time transformation - Homogeneous}) shows that the internal time $t_\varphi = \pi / (2\bar{\lambda})$ corresponds to the surface of the coordinate singularity of the Schwarzschild metric since it is identical to equation (\ref{eq:New coord sing Schwarzschild}).
However, the metric in the internal time coordinates (\ref{eq:Spacetime metric homogeneous - modified - Internal time}) remains regular at this surface for certain values of $\Lambda$ and ${\cal Q}$, which we investigate in more detail below.
Furthermore, this is a surface of reflection symmetry \cite{EMGPF}.

\subsection{Global structure}

\subsubsection{Asymptotic flatness}

Setting the cosmological constant to zero, the function $J(x)$ in the general metric (\ref{eq:Spacetime metric - modified - Schwarzschild}) vanishes asymptotically.
Restricting ourselves to the case of asymptotically constant $\lambda(x)$, we recover an asymptotically flat spacetime by fixing the global factor to
\begin{equation}\label{eq:Global factor for asymptotic flatness}
    \lambda_0 = 1 / \sqrt{1+\lambda_\infty^2}\,,
\end{equation}
where $\lambda_\infty=\lim_{x\to\infty}\lambda(x)$.
In the following we assume this value.

\subsubsection{Horizons}

The metric (\ref{eq:Spacetime metric - modified - Schwarzschild}) with static Schwarzschild coordinates predicts the classical radial coordinates for the horizons because they solve the same equation
\begin{equation}\label{eq:Horizons eq}
    1 - \frac{2 M}{x} - \frac{\Lambda x^2}{3} + \frac{{\cal Q}}{x^2} = 0
    \,,
\end{equation}
up to the difference between ${\cal Q}$ and the squared electric charge $Q^2$.
A nontrivial emergent electric field therefore changes the position of the horizon.
Taking a small cosmological constant, we may neglect it at smaller scales where the solutions to the horizon equation is given by
\begin{equation}\label{eq:RN horizons}
    x_\pm \approx M \left(1 \pm \sqrt{1 - \frac{{\cal Q}}{M^2}} \right)
    \,,
\end{equation}
which are real only if ${\cal Q}\le M^2$.
The case ${\cal Q}=M^2$ corresponds to an extremal black hole, while ${\cal Q}> M^2$ is superextremal.
The positive root $x_+$ is the black hole's horizon beyond which the metric turns homogeneous, while the negative root $x_-$ corresponds to a potential inner horizon\textemdash\,we will show later that this horizon does not exist when a reflection symmetry surface appears.
At the larger scales, the cosmological constant dominates and we may neglect the mass and the charge functions, obtaining the cosmological horizon
\begin{equation}
    x_\Lambda \approx \sqrt{\frac{3}{\Lambda}}
    \,.
\end{equation}
The static region is therefore restricted to $x_+<x<x_\Lambda$.
Beyond these values we must adopt the homogeneous metric by simply swapping the time and radial coordinates or use the GP metric.

\subsubsection{Reflection symmetry surfaces}

The new coordinate singularity of the metric (\ref{eq:Spacetime metric - modified - Schwarzschild}) is located at the radius $x_\lambda^{(i)}$, where the superindex in $x_\lambda^{(i)}$ labels the multiple possible values, solving the equation
\begin{equation}\label{eq:Ref surface eq}
    S(x) \equiv 1 + \lambda^2 \left( 1 - \frac{2 M}{x} - \frac{\Lambda x^2}{3} + \frac{{\cal Q}}{x^2} \right) = 0
    \,,
\end{equation}
where $\lambda$ is an arbitrary function of $x$.
The term in parenthesis is precisely the left-hand-side of the horizon equation (\ref{eq:Horizons eq}).
Since this term must be negative for (\ref{eq:Ref surface eq}) to hold, we conclude that the new coordinate singularities, if present, appear only in the homogeneous regions.

Furthermore, a closer look at the mass observable expression (\ref{eq:Mass observable - final - EM - simplified}) shows that equation (\ref{eq:Ref surface eq}) corresponds to the surface where the angular curvature takes the maximal value $K_\varphi = - \pi / (2 \bar{\lambda})$.
In the present case, where the modification function $q$ vanishes, this is a reflection symmetry surface in the sense that, defining $K_\varphi = \pm \pi / (2 \tilde{\mu}) + \delta$, the structure function $\tilde{q}^{x x}$ is invariant under $\delta \to - \delta$, while the Hamiltonian constraint is invariant under the combined operation $\delta \to - \delta$, $K_x \to - K_x$.
The reflection-symmetry surface, defined by $\left( K_\varphi = \pm \pi / (2 \tilde{\mu}) , K_x = 0 \right)$ in the phase space, manifests itself in dynamical solutions as a reflection symmetry surface of the spacetime \cite{EMGPF}.
This is all consistent with our solution in the internal time gauge of the Subsection~\ref{sec:Internal time}.

It will be useful to define $x_\lambda^{(-)}$ as the largest solution to (\ref{eq:Ref surface eq}) lower than the outer horizon coordinate $x_+$, and we also define $x_\lambda^{(+)}$ as the lowest solution to (\ref{eq:Ref surface eq}) larger than the cosmological horizon coordinate $x_\Lambda$.
If $x_\lambda^{(-)}$ exists such that $x_-<x_\lambda^{(-)}<x_+$, then the inner horizon would not exist because $x_\lambda^{(-)}$ represents a minimum radius of the spacetime.
On the other hand, if the solution $x_\lambda^{(-)}$ does not exist or it is lower than the inner horizon, then no minimum radius exists because the region $x<x_-$ is not homogeneous and the radial coordinate can go all the way to $x\to 0$ where a geometric singularity is met, as shown from a computation of the Ricci scalar below.

\subsubsection{Physical singularity and the maximal extension}

The Ricci scalar of the static metric (\ref{eq:Spacetime metric - modified - Schwarzschild}) is given by
\begin{widetext}
\begin{eqnarray}\label{eq:Ricci scalar}
    R &=& \frac{\lambda_0^2}{x^6} \Bigg(
    \frac{2 x^4}{\lambda_0^2} + 4 x^6 \Lambda
    - 2 x^4
    + \lambda^2 \left( \frac{2 \Lambda}{3}  x^4 \left(4 \mathcal{Q} - 12 M x+9 x^2\right)
    - 2 \Lambda^2 x^8 \right)
    + 2 \lambda^2 \left( \mathcal{Q}^2 + 3 M^2 x^2 + \mathcal{Q} \left(x^2 - 4 M x\right) - x^4 \right)
    \nonumber\\
    &&
    - x \left( - \mathcal{Q} +3 M x+\Lambda  x^4-2 x^2\right) \left( 2 M x + \frac{\Lambda}{3} x^4 - \left(\mathcal{Q} + x^2 \right)\right) \frac{\partial \lambda^2}{\partial x}
    \Bigg)
    \,.
\end{eqnarray}

At the minimum radius,
\begin{equation}
    R = \frac{\lambda_0^2}{x^6} \Bigg(
    \frac{2 x^4}{\lambda_0^2}
    + 2 \lambda^2 \left( \mathcal{Q}^2 + 3 M^2 x^2 + \mathcal{Q} \left(x^2 - 4 M x\right) - x^4 \right)
    - x \left( - \mathcal{Q} + 3 M x -2 x^2\right) \left( 2 M x - \left(\mathcal{Q} + x^2 \right)\right) \frac{\partial \lambda^2}{\partial x}
    \Bigg) \bigg|_{x_\lambda^{(i)}}
    \,.
\end{equation}
\end{widetext}
For monotonically decreasing $\lambda(x)$, this generically diverges at $x\to 0$ and hence the existence of a minimum radius, implying no inner horizon, is needed to resolve the singularity.
At large scales, the asymptotically constant $\lambda (x)$ case develops a singularity as $x\to \infty$ and, therefore, a maximum radius is needed to avoid it, while the asymptotically vanishing $\lambda(x)$ case has no such singularity issues at large scales.

\begin{figure}[!htb]
    \centering
    \includegraphics[trim=5.4cm 0cm 5.6cm 0cm,clip=true,width=0.99\columnwidth]{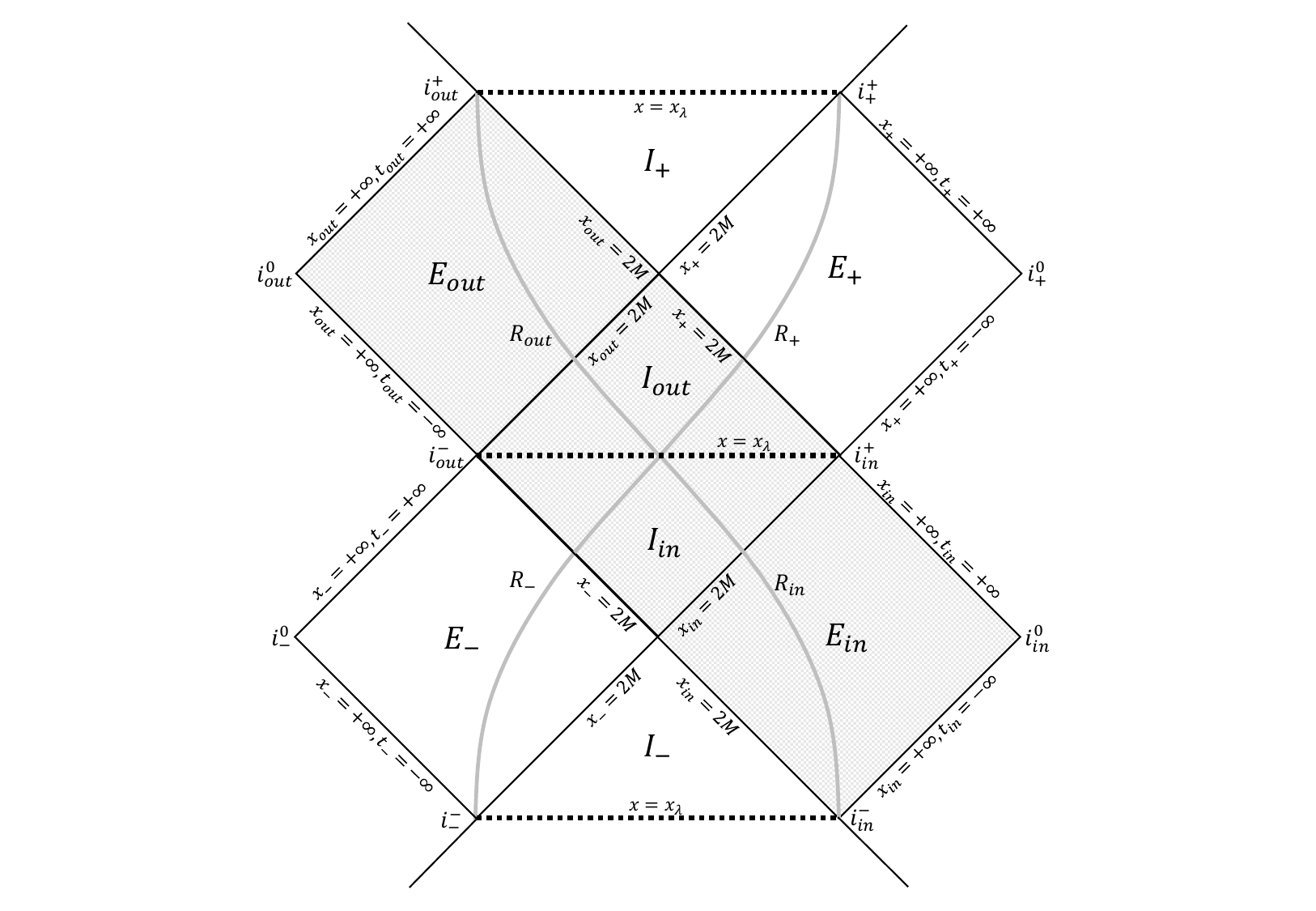}
    \caption{Maximal extension of the vacuum solution for the case with only one solution for the coordinate $x_\lambda^{(-)}$ of the hypersurface of reflection symmetry.
    The the shaded region $E_{\rm in}\cup I_{\rm in}\cup I_{\rm out}\cup E_{\rm out}$ is the wormhole solution obtained by sewing two interior Schwarzschild regions by the reflection symmetry surface $x=x_\lambda^{(-)}$.
    The gray lines $R_{i}$ denote geodesics falling from a remote past to a far future. Credit: Ref.~\cite{ELBH}.}
    \label{fig:Holonomy_KS_Vacuum_Wormhole-Periodic}
  \end{figure}

\begin{figure}[!htb]
    \centering
    \includegraphics[trim=1.65cm 0cm 2cm 0cm,clip=true,width=0.99\columnwidth]{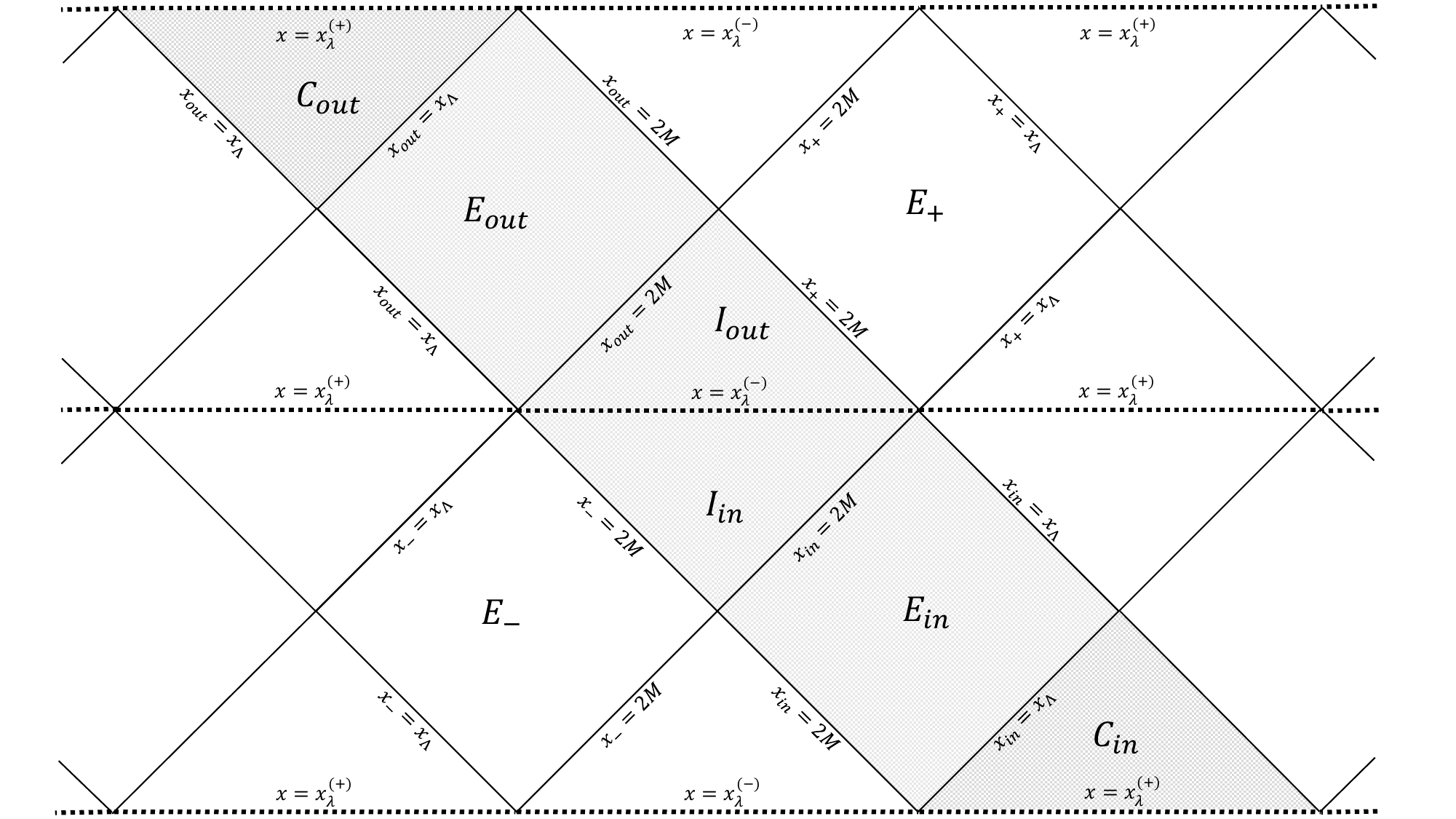}
    \caption{Maximal extension of the wormhole solution in vacuum on a bounded de Sitter background.
    The shaded region $E_{\rm in}\cup I_{\rm in}\cup I_{\rm out}\cup E_{\rm out}$ is the wormhole solution obtained by the sewing process. Credit: Ref.~\cite{ELBH}.}
    \label{fig:Holonomy_KS_Vacuum_Wormhole-Periodic-Cosmo-MaxRadius}
  \end{figure}

\begin{figure}[!htb]
    \centering
    \includegraphics[trim=1.65cm 0cm 2cm 0cm,clip=true,width=0.99\columnwidth]{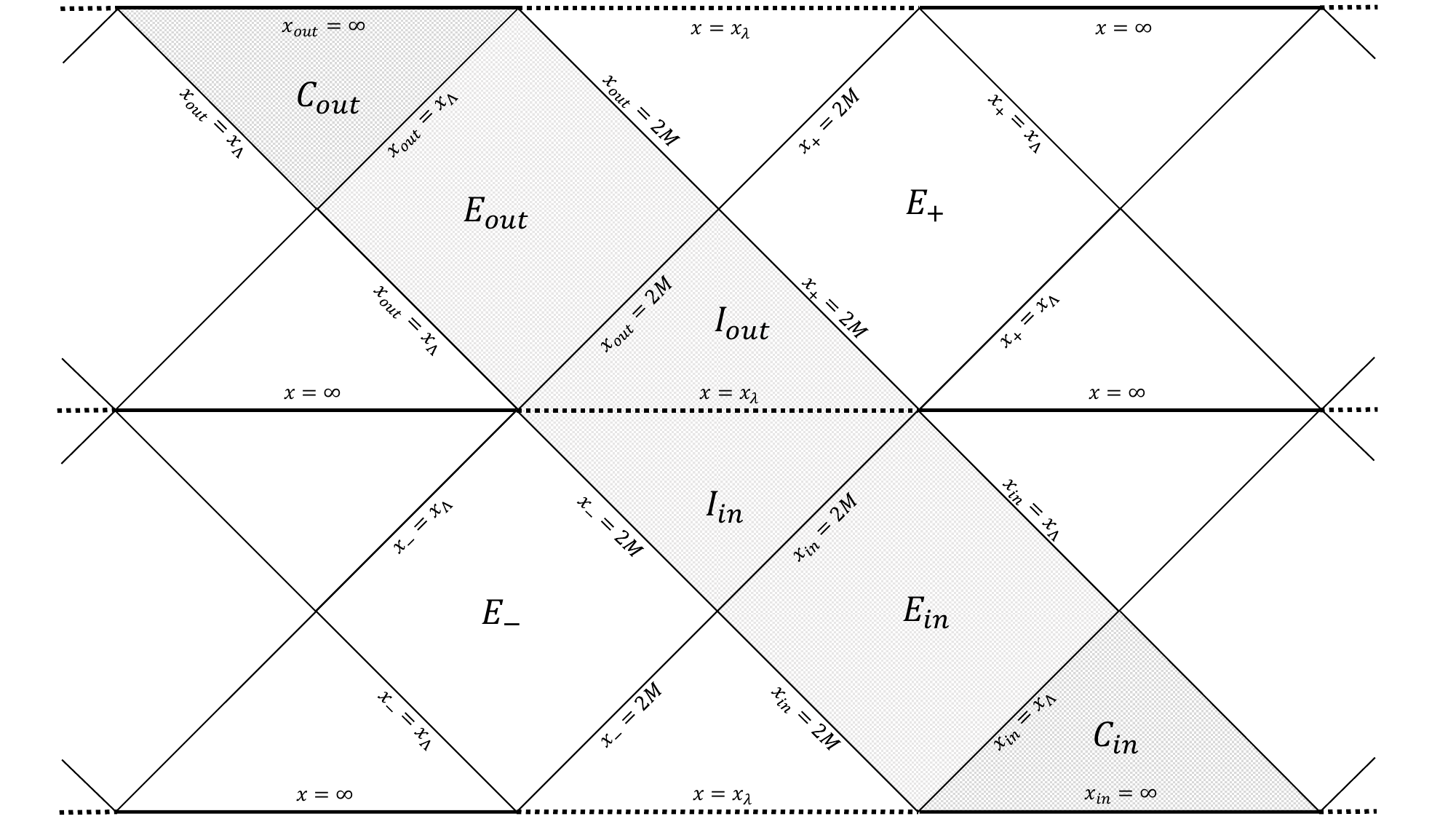}
    \caption{Maximal extension of the wormhole solution in vacuum on an unbounded de Sitter background.
    The shaded region $E_{\rm in}\cup I_{\rm in}\cup I_{\rm out}\cup E_{\rm out}$ is the wormhole solution obtained by the sewing process. Credit: Ref.~\cite{ELBH}.}
    \label{fig:Holonomy_KS_Vacuum_Wormhole-Periodic-Cosmo-NO MaxRadius}
\end{figure}

We can now deduce the global structure of the spacetime as follows.
We start with a static Schwarzschild region with metric (\ref{eq:Spacetime metric - modified - Schwarzschild}).
The GP region with metric (\ref{eq:Spacetime metric - GP}) covers the static Schwarzschild region and the homogeneous Schwarzschild regions, one for $x< x_{\rm H}$ and another for $x> x_\Lambda$.
We can then sew all three charts together into a black hole spacetime that ends at the minimum radius $x_{\lambda}^{(-)}$, rather than meeting an inner horizon $x_-$, and at the maximum radius $x_{\lambda}^{(+)}$ for asymptotically constant $\lambda(x)$ or at $x\to\infty$ for asymptotically vanishing $\lambda(x)$.
Furthermore, the internal time homogenous region with metric (\ref{eq:Spacetime metric homogeneous - modified - Internal time}) covers two homogeneous Schwarzschild regions.
Thus, it can be used to sew together the GP chart with other two by their extremized radii into a single wormhole spacetime.
The generic global structure for monotonically decreasing $\lambda(x)$ is therefore similar to that of \cite{alonso2022nonsingular,Alonso_Bardaji_2022,alonsobardaji2023Charged,ELBH}.
See Figures~\ref{fig:Holonomy_KS_Vacuum_Wormhole-Periodic}, \ref{fig:Holonomy_KS_Vacuum_Wormhole-Periodic-Cosmo-MaxRadius}, and \ref{fig:Holonomy_KS_Vacuum_Wormhole-Periodic-Cosmo-NO MaxRadius} for detailed conformal diagrams of the resulting spacetimes.
The case where $x_-<x_\lambda^{(-)}<x_+$ does not exist has a conformal diagram similar to the Reissner-Nordtr\"om spacetime with an inner horizon and a singularity at $x\to0$, although a maximum radius would appear in the asymptotically constant $\lambda(x)$ case.

We will discuss in detail the cases of constant $\lambda = \bar{\lambda}$ and asymptotically vanishing $\lambda=\Delta/x^2$ in the next two Subsections.

\subsection{Constant holonomy parameter}

Spacetimes with different global structures may result from the relation between different values of the mass, cosmological constant, and electric charge as studied in \cite{alonsobardaji2023Charged} for the simplest case of constant $\lambda=\bar{\lambda}$ and non-emergent electric field.
In particular, the dynamical solutions beyond the minimum and maximum radii, if they exist, imply independent spacetime regions.
Here, we are not interested in those exotic spacetimes, but only in the realistic black hole case described above and, in particular, the effects of having a non-trivial emergent electric field.

Using a constant $\lambda=\bar{\lambda}$, the global factor that determines asymptotic flatness for vanishing cosmological constant is $\lambda_0=1/\sqrt{1+\bar{\lambda}^2}$.
Therefore, we obtain the line element
\begin{eqnarray}\label{eq:Spacetime metric - mu0}
    &&\!\!\!\!\!\!
    {\rm d} s^2 =
    - \left(1 - J(x)\right) {\rm d} t^2
    \\
    &&\quad
    + \left( 1 - \frac{\bar{\lambda}^2}{1+\bar{\lambda}^2} J(x) \right)^{-1}
    \left(1 - J(x)\right)^{-1} {\rm d} x^2
    + x^2 {\rm d} \Omega^2
    \,.\nonumber
\end{eqnarray}

At larges scales, the cosmological constant dominates and the solution $x_{\bar \lambda}^{(+)}$ to (\ref{eq:Ref surface eq}) is given by
\begin{equation}
    x_{\bar{\lambda}}^{(+)} \approx \sqrt{\frac{3}{\Lambda} \frac{1+\bar{\lambda}^2}{\bar{\lambda}^2}}
    \,.
\end{equation}
This is a maximum-radius surface, beyond which the spacetime starts to shrink back, avoiding the singularity of (\ref{eq:Ricci scalar}) at $x \to \infty$.
This solution does not appear in asymptotically vanishing $\lambda(x)$, nor do any singularities at large scales.
Therefore, it may be seen as a pathology of the constant parameter $\lambda=\bar{\lambda}$.
See \cite{ELBH} for a more detailed discussion on asymptotic issues of the $\lambda=\bar{\lambda}$ case and implications of a maximum radius.

At smaller scales, we may neglect the cosmological constant, such that the solution to $S(x_{\bar \lambda}^{(-)})=0$ in equation (\ref{eq:Ref surface eq}) is given by
\begin{equation}
    x_{\bar \lambda}^{(-)} \approx \frac{M \bar{\lambda}^2}{1+\bar{\lambda}^2} \left(1 + \sqrt{1-{\cal Q} \frac{1+\bar{\lambda}^2}{M^2 \bar{\lambda}^2}}\right)
    \,,
\end{equation}
where we only considered the positive root because its existence would imply a reflection symmetry surface, so that the negative root would not be part of the spacetime.
This expression is real only if
\begin{equation}\label{eq:Charge bound - mu0}
    {\cal Q} \le M^2 \bar{\lambda}^2/(1+\bar{\lambda^2})
\end{equation}
and is, therefore, bounded by $M \bar{\lambda}^2/(1+\bar{\lambda}^2)\leq x_{\bar \lambda}^{(-)}\leq 2M \bar{\lambda}^2/(1+\bar{\lambda}^2)$ with the upper bound realized by ${\cal Q}=0$\textemdash\,here we assume ${\cal Q}$ is nonnegative.
Since we expect $\bar{\lambda}\ll1$, the bound (\ref{eq:Charge bound - mu0}), if respected, implies that no (super)extremal black holes exist.

Requiring the reality of $x_{\bar \lambda}^{(-)}$, and hence the bound (\ref{eq:Charge bound - mu0}), the inner horizon coordinate is bounded by
\begin{equation}
    0\le x_- 
    \lesssim M \bar{\lambda^2}/2
    \,,
\end{equation}
which is below the lower bound of $x_{\bar \lambda}^{(-)}$.
We conclude that the inner horizon does not exist for $\bar{\lambda}\ll1$.
On the other hand, the outer horizon is bounded by
\begin{equation}
    2 M \left(1- \bar{\lambda^2}/4\right)\lesssim x_+\le 2 M
\end{equation}

Since it makes little sense to have electric charge with no mass, we define the critical value for the charge function as the limiting case of the inequality (\ref{eq:Charge bound - mu0}) where the mass is Planckian, $M \to m_{\rm Pl}$:
\begin{equation}\label{eq:Charge function critical - mu0}
    \mathcal{Q}_c = \frac{m_{\rm Pl}^2 \bar{\lambda}^2}{1+\bar{\lambda}^2}
    \,.
\end{equation}
This represents the maximum value of the charge function for which no geometric singularity develops by allowing the appearance of a minimum radius.

Adopting the Wilson modification (\ref{eq:Wilson action mod - spherical}) for the charge function, we may equate its maximum value (\ref{eq:Max value of charge function - Wilson}) to the critical one (\ref{eq:Charge function critical - mu0}) and hence determine
\begin{equation}\label{eq:Nonsingulat a - mu0}
    \mathfrak{a} = \frac{2}{m_{\rm Pl}} \sqrt{\frac{1+\bar{\lambda}^2}{\bar{\lambda}^2}}\,.
\end{equation}
Therefore, the Wilson modification, which is possible only if the emergent electric field differs from the fundamental one, guarantees the singularity resolution by providing a minimum radius.
This, in turn, implies that no inner horizon and no (super)extremal black holes exist.
In the limit $\bar{\lambda}\to 0$, $\mathfrak{a}$ diverges.
This, however, is not necessarily problematic because a minimum radius does not exist in such limit, and hence no singularity resolution is possible in that limit. Since this choice of $\mathfrak{a}$ depends on the existence of a minimum radius, it becomes ill-defined when the latter is removed and no singularity resolution is possible.
The choice (\ref{eq:Nonsingulat a - mu0}) is always finite if $\bar{\lambda}$ is finite too; therefore, we conclude that this choice for the electromagnetic holonomy parameter is contingent on the choice of a finite gravitational holonomy parameter $\bar{\lambda}$.
Also, recall that in Subsection~\ref{sec:Symm red vs dim red} we found that the continuum limit of the full theory can be taken for arbitrary electromagnetic holonomy parameter $\mathfrak{a}$ and is, therefore, not necessarily small in the full theory either.
A more faithful characterization of the continuum limit of a Wilson-type action (\ref{eq:Wilson action - EM}), discussed in Subsection~\ref{sec:Symm red vs dim red}, is provided by the combination $\mathfrak{a}\bar{\lambda}^2$ in the $\bar{\lambda}\to0$ limit.
Here, $\bar{\lambda}$ plays the role of the size of the angular links of a lattice.
Indeed,
\begin{equation}
    \mathfrak{a}\bar{\lambda}^2 = \frac{2\bar{\lambda}}{m_{\rm Pl}} \sqrt{1+\bar{\lambda}^2}
    \,,
\end{equation}
vanishes in the $\bar{\lambda}\to0$ limit.
Furthermore, as pointed out in Subsection~\ref{sec:Schwarzschild exterior}, the Lorentz force equation remains unchanged in the weak gravity regime regardless of the charge function ${\cal Q}$.
Therefore, all laboratory observations, independent of the effects of nontrivial ${\cal Q}$, remain valid for all values of $\mathfrak{a}$.
The effects of the electromagnetic holonomy parameter on charged test particles are, therefore, only observable in the strong gravity regime.

\subsection{Decreasing holonomy parameter}

A similar result follows for other choices of $\lambda$.
For instance, the function $\lambda = \sqrt{\Delta}/x$ with constant $\Delta$, sometimes used in models of LQG with $\Delta$ being the area gap \cite{rovelli1995discreteness}, requires a unit global factor $\lambda_0=1$ to recover asymptotic flatness for vanishing cosmological constant.
We therefore obtain the line element
\begin{eqnarray}\label{eq:Spacetime metric - mubar}
    &&\!\!\!\!\!\!\!\!
    {\rm d} s^2 =
    - \left(1 - J(x)\right) {\rm d} t^2
    \\
    &&
    + \left( 1 + \frac{\Delta}{x^2}\left(1-J(x)\right) \right)^{-1}
    \left(1 - J(x)\right)^{-1} {\rm d} x^2
    + x^2 {\rm d} \Omega^2
    \,.\nonumber
\end{eqnarray}
With this holonomy parameter, equation (\ref{eq:Ref surface eq}), which determines the location of the symmetry surfaces, is a complicated fourth-order polynomial.
However, it is clear that at large scales, where the cosmological constant dominates, no maximum-radius surface develops and, furthermore, no singularity appears in the Ricci scalar (\ref{eq:Ricci scalar}) at large scales.
In fact, (\ref{eq:Ref surface eq}) is asymptotically positive $S(\infty)\to1-\Delta\Lambda /3$ for small $\Delta\Lambda$\textemdash\,using the observed value of $\Lambda$ and a Planck sized $\Delta$, we have $\Delta\Lambda\sim 10^{-122}$.

Taking the derivative of (\ref{eq:Ref surface eq}) yields
\begin{equation}
    S'(x) = -\frac{2 \Delta}{x^3}  \left(x^2 - 3 M x + 2 {\cal Q}\right)
    \,.
\end{equation}
The extrema $S'=0$ occur at $x\to \infty$ and at
\begin{equation}\label{eq:Extrema of singularity equation}
    z_\pm = \frac{3M}{2} \left(1\pm\sqrt{1-\frac{8 {\cal Q}}{9 M^2}}\right)
    \,,
\end{equation}
which exist only if ${\cal Q} < 9 M^2/8$.
The second derivative yields
\begin{equation}
    S''(x)=\frac{2\Delta}{x^6} \left(3x^2 - 12 M x + 10{\cal Q}\right)\,,
\end{equation}
which, evaluated at $z_\pm$, becomes
\begin{equation}
    S''(x) |_{z_\pm}
    = \frac{4\Delta}{z_\pm^5} \left(\frac{3 M}{2} - z_\pm\right)\,.
\end{equation}
We find that $x=z_+$ is a maximum of $S(x)$ while $x=z_-$ and $x\to \infty$ are minima.

Therefore, a minimum radius exists only if $S(z_-)\le 0$ and $z_->x_-$.
The critical value of ${\cal Q}$ that solves the equation $S(z_-)=0$ when $M=m_{\rm Pl}$, which has only one real solution, is given by
\begin{equation}\label{eq:Charge function critical - mubar}
    {\cal Q}_c = \Sigma^{-1/3} \left( \frac{1}{4} (3-\Delta\Lambda)^{-1} \left[\Delta - \Sigma^{1/3}\right]^2 - 9 m_{\rm Pl}^2 \Delta \right) 
\end{equation}
where
\begin{widetext}
\begin{equation}
    \Sigma = \Delta \left[ (3-\Delta\Lambda)^2\left( 6 m_{\rm Pl} \left(9 m_{\rm Pl}^2 + \frac{2 \Delta}{3-\Delta\Lambda}\right)^{3/2} + 162 m_{\rm Pl}^4\right)
    - \Delta^2
    - 90 (3-\Delta\Lambda) \Delta m_{\rm Pl}^2
    \right]
    \,.
\end{equation}
\end{widetext}
This critical value vanishes for $\Delta\to0$, and it has a maximum value of ${\cal Q}_c=m_{\rm Pl}^2$ at large enough $\Delta$ where $\Sigma\to0$ and beyond which ${\cal Q}_c$ becomes complex.
When $\Lambda=0$, such maximum value is attained for
\begin{equation}
    \Delta_{\rm max} = 108 m_{\rm Pl}^2\,.
\end{equation}

At this critical value, the minimum radius coordinate $x_\Delta = z_-|_{{\cal Q}\to{\cal Q}_c}$, given by (\ref{eq:Extrema of singularity equation}),
is larger than the inner horizon coordinate (\ref{eq:RN horizons}) only if ${\cal Q}_c < M^2$ when neglecting the cosmological constant.
Since ${\cal Q}_c\le m_{\rm Pl}^2$, this is always the case for realistic (semi)classical black holes.

Adopting the Wilson-like modification (\ref{eq:Wilson action mod - spherical}) for the charge function, we may equate its maximum value (\ref{eq:Max value of charge function - Wilson}) to the critical one (\ref{eq:Charge function critical - mubar}) to determine
\begin{equation}\label{eq:Critical EM holonomy parameter - mubar}
    \mathfrak{a} = \frac{2 \Sigma^{1/6}}{\sqrt{\frac{1}{4} (3-\Delta\Lambda)^{-1} \left[\Delta - \Sigma^{1/3}\right]^2 - 9 m_{\rm Pl}^2 \Delta}}
    \,.
\end{equation}
Therefore, as in the constant holonomy parameter case, the introduction of the Wilson modification via the emergent electric field resolves the singularity by providing a minimum radius.
This guarantees that no inner horizon and no (super)extremal black holes exist.
Also, since ${\cal Q}_c\to0$ as $\Delta\to0$, $\mathfrak{a}$ diverges in such limit as $\mathfrak{a} \propto \Delta^{-1/3}$.
As argued at the end of the previous Subsection and in Subsection~\ref{sec:Symm red vs dim red}, the electromagnetic holonomy parameter is not necessarily small and it can get a large value for small $\Delta$ without contradicting laboratory observations in the weak gravity regime, where only the Lorentz force, which is independent of $\Delta$ and ${\cal Q}$, is relevant.
A more faithful characterization of the continuum limit of a Wilson-type action (\ref{eq:Wilson action - EM}) is provided by the combination $\mathfrak{a}\Delta$ in the $\Delta\to0$ limit, here $\Delta$ playing the role of the squared size of the links of a lattice.
Indeed,
\begin{equation}
    \mathfrak{a}\Delta = \frac{2 \Delta \Sigma^{1/6}}{\sqrt{\frac{1}{4} (3-\Delta\Lambda)^{-1} \left[\Delta - \Sigma^{1/3}\right]^2 - 9 m_{\rm Pl}^2 \Delta}}
\end{equation}
vanishes in the $\Delta\to0$ limit.
Furthermore, in the limit $\Delta\to0$, the minimum radius does not exist and singularity resolution is not possible, hence why the choice (\ref{eq:Critical EM holonomy parameter - mubar}) becomes ill-defined in such limit.
This leads us to conclude, once again, that the choice of the electromagnetic holonomy parameter (\ref{eq:Critical EM holonomy parameter - mubar}) is contingent on the existence of a finite gravitational holonomy parameter.

Taking $\Delta\sim m_{\rm Pl}^2$ and the observed cosmological constant such that $\Delta \Lambda\sim 10^{-122}$, we obtain
\begin{eqnarray}
    \Sigma \sim 3 \times 10^{3} m_{\rm Pl}^6
    \,,
\end{eqnarray}
and
\begin{eqnarray}\label{eq:Critical EM holonomy parameter - mubar - estimate}
    \mathfrak{a}
    \sim \frac{3}{m_{\rm Pl}}
    \,.
\end{eqnarray}
The introduction of the charge function suppresses the gravitational effect of electric charges because they become Planckian, ${\cal Q}_{\rm max}/x^2\sim m_{\rm Pl}^2/x^2$.
For reference, the electron's charge is $e\approx 4\pi m_{\rm Pl}/\sqrt{137}$.
The maximum charge effect ${\cal Q}_{\rm max}\sim (4/3) m_{\rm Pl}^2$ is therefore equivalent to the classical gravitational effect of only a few electrons.

\section{Conclusions}
\label{sec:Conclusions}

Our canonical methods have revealed how new physics can arise from a proper distinction between the canonical and mechanical momenta.
In our nomenclature, we refer to the former as fundamental fields, which make up the phase space, and to the latter as emergent, which are directly involved with observed phenomena and whose nontrivial dependence on the fundamental fields is derived from covariance conditions.
This is the central idea of emergent modified gravity, which makes it possible to obtain a post-Einsteinian gravitational theory where the gravitational field is fundamental while the spacetime is emergent.
The extension of this idea to the electromagnetic theory, such that there is a fundamental electric momentum different from the emergent electric field that directly enters the strength tensor and hence observations, similarly provides a pathway to a modified electromagnetic theory.

The relation between the fundamental and emergent electric fields is derived from the imposition of covariance conditions on the strength tensor, which is the one that directly enters the Lorentz force equation and is hence involved with the observed phenomena.
Both spacetime and electromagnetic covariance conditions are necessary for the modified theory be generally covariant and, besides revealing the relation between the fundamental and emergent fields, it implies strong restrictions on the allowed modifications.
In fact, if the four-dimensional emergent field theory is required to be fully local, then the expression of the emergent electric field is determined by the covariance conditions as identical to the fundamental electric field, up to a contribution from the topological $\theta$ term, and hence local modified electromagnetism is not viable, unless the emergent spacetime is allowed to depend on the electromagnetic variables at a kinematical level.

However, this result can be circumvented if one introduces nonlocal terms such as Wilson loops in lattice field theory.
This would require a generalization of the local covariance conditions presented here to take into account nonlocal transformations.
While a rigorous analysis of covariance in such a setting is challenging, we have proposed the use of symmetry reduced models to take into account nonlocal effects in the homogeneous directions of symmetry while preserving local covariance in the inhomogeneous directions.
We have applied this procedure to the spherically symmetric model, which allows emergent modified gravity to model gravitational \emph{angular} holonomies \cite{EMG,EMGCov,alonso2022nonsingular,ELBH}, and have shown that, unlike the four-dimensional local theory, it allows for the emergent electric field to have a nontrivial relation to the fundamental one.
This freedom in the spherically symmetric model can be used to reproduce a term similar to that of the Wilson action.

In this modified theory, we have found an electrically charged black hole solution where the gravitational angular holonomy parameter is allowed to be scale dependent and the emergent electric field be different from the fundamental one, extending the results of \cite{alonso2022nonsingular,ELBH,alonsobardaji2023Charged}.
The generic global structure of such solution is that of a wormhole with a finite minimum radius and no inner horizon for small electric charges.
For large enough electric charges, the minimum radius does not exist, the inner horizon reappears, and a singularity develops at $x\to0$, unless an electromagnetic modification, suggested for instance by the Wilson action in lattice quantum field theory, is introduced.
In such case, the electric charge is unbounded, but its spacetime effects are bounded from above by lattice effects.
This maximum value on the spacetime effects of the electric charge leads to a robust resolution of the singularity by ensuring the existence of a minimum radius and, furthermore, implies that no inner horizon can develop and hence no (super)extermal black holes can exist.
This result shows how the concept of the emergent electric field opens the way to covariant modified electromagnetic theories with major physical implications.

\section*{Acknowledgements}

The author thanks Martin Bojowald and Marius J\"urgensen for useful discussions and going through a draft version of this work.
This work was supported in part by NSF grant PHY-2206591.

\end{document}